\documentclass[twocolumn]{aastex631}

\newcommand{\amcvn}{{AM Cvn}} 
\newcommand{\escet}{{ES Cet}} 
\newcommand{\hmcnc}{{HM Cnc}} 
\newcommand{\sdss}{{SDSS J0651}} 
\newcommand{\vul}{{V407 Vul}} 
\newcommand{\ztf}{{ZTF J1539}} 

\newcommand{\vbonly}{{\tt VB only}}
\newcommand{\vbgalaxy}{{\tt VB with Galaxy}} 
\newcommand{\vbglobal}{{\tt VB Global Fit}} 

\newcommand{\sangria}{LDCv2.1a}

\begin{document}

\title{Have any LISA verification binaries been found?}
\shorttitle{Have any verification binaries been found?}
\author{Tyson B. Littenberg}
\affiliation{NASA Marshall Space Flight Center, Huntsville, Alabama 35812, USA}
\author{Ananthu K. Lali}
\affiliation{Space Science Department, University of Alabama in Huntsville, 320 Sparkman Drive, Huntsville, Alabama 35899, USA}



\begin{abstract}

Some electromagnetically observed ultra-compact binaries will be strong gravitational wave sources for space-based detectors like the Laser Interferometer Space Antenna (LISA).
These sources have historically been referred to as ``verification binaries'' under the assumption that they will be exploited to assess mission performance.
This paper quantitatively interrogates that scenario by considering targeted analyses of known galactic sources in the context of a full simulation of the galactic gravitational wave foreground.
We find that the analysis of the best currently known LISA binaries, even making maximal use of the available information about the sources, is susceptible to ambiguity or biases when not simultaneously fitting to the rest of the galactic population.  
While galactic binaries discovered electromagnetically in advance of, or during, the LISA survey are highly valuable multimessenger systems, the need for a global treatment of the galactic gravitational wave foreground calls into question whether or not they are the best sources for data characterization.

\end{abstract}



\section{Introduction}
The Laser Interferometer Space Antenna (LISA) will open the mHz band of the gravitational wave (GW) spectrum for direct observations. 
Binary systems with minutes- to hours-long orbital periods emit GWs in the LISA band and are detectable from galactic scales for stellar mass objects to cosmological scales for massive black holes.
At these frequencies a population of galactic ultra compact binaries (UCBs), primarily double white dwarf systems, are expected to be the most numerous sources of GWs observable by LISA.
Population synthesis simulations predict tens of millions of sources in the Milky Way galaxy, tens of thousands of which will be individually resolvable by LISA after several years of observations \citep{Nelemans,Cornish_2017}.

At the GW frequencies detectable by LISA the binary separations for UCBs are sufficiently large that the orbital evolution of the binaries is small, i.e. $\dot{P} T \ll P_0$ where $\dot{P}$ is the first time derivative of the orbital period, $T$ is the elapsed time of the LISA observations, and $P_0$ is the orbital period of the binary at the start of observations.
Due to these relatively small ``chirp'' rates the UCBs are persistent sources in the mHz band and will thus be continuously monitored by LISA. 
As a consequence of UCBs' continuous signal, abundance, and spatial distribution concentrated in the galactic plane and bulge, the majority of sources will blend together to form an irreducible astrophysical foreground, or ``confusion noise,'' which is predicted to exceed the instrument noise below a few mHz and thus limit the sensitivity of LISA at those frequencies \citep{Bender_1997}.

Another consequence of the persistence and abundance of UCBs is the possibility of identifying LISA sources through electromagnetic (EM) surveys in advance of LISA operations.
Several dozen such binaries have been discovered at the time of this writing, many of which were found serendipitously~\citep{Roelofs_2010,2011ApJ...737L..23B, 2017ApJ...847...10B,2019Natur.571..528B}.  
Of these currently-known binaries, ${\sim}20$ will be detectable by LISA within its primary mission lifetime \citep{10.1093/mnras/stad1288, kupfer2024lisa}, and that number is expected to steadily increase as new wide-field and high-cadence surveys come on line (e.g, \citet{Burdge_2020,10.1093/mnras/stad2290}).

Having guaranteed multimessenger sources in the LISA data is a tantalizing opportunity and the existence of these systems has long served as a foundation of the mission's science case.  
GW and EM of UCBs constrain different combinations of the physical parameters of the binary systems allowing degeneracies in one observation to be broken by the other, as demonstrated in \citet{Shah2012} and \citet{10.1093/mnras/stad2579}.
As is true across all of multimessenger astronomy, joint observations are greater than the sum of their parts. 
The discovery of new sources, and further characterization of the already-known binaries, is critical precursor science for the LISA gravitational wave survey.

These multimessenger binaries have an elevated level of interest and importance to the LISA community beyond the prospects of understand the astrophysics and dynamics of UCBs.
Such systems are typically referred to as ``verification binaries'' stemming from the concept that detection, or lack thereof, of the GW signal would be used as part of the verification and validation process of the LISA science data and/or analyses.
However, despite this persistent nomenclature there have been few quantitative studies of \emph{how} the known binaries would be used for instrument characterization, apart from the notable exceptions of \citet{PhysRevD.106.022003} and \citet{shah2023optimal}.

In this paper we investigate the known binaries' utility as verification sources for LISA in the context of the broader UCB population in the galaxy.
Our findings contrast with the canonical view that the currently know binaries will be useful \emph{in isolation} as verification sources, and instead require careful modeling of the entire UCB population near in frequency to the known binaries.  
The currently known binaries all occupy densely populated regions of the LISA frequency band and, based on population synthesis models of the galaxy \citep{Toonen}, have a high probability of overlapping other sources.  

\begin{figure}
\plotone{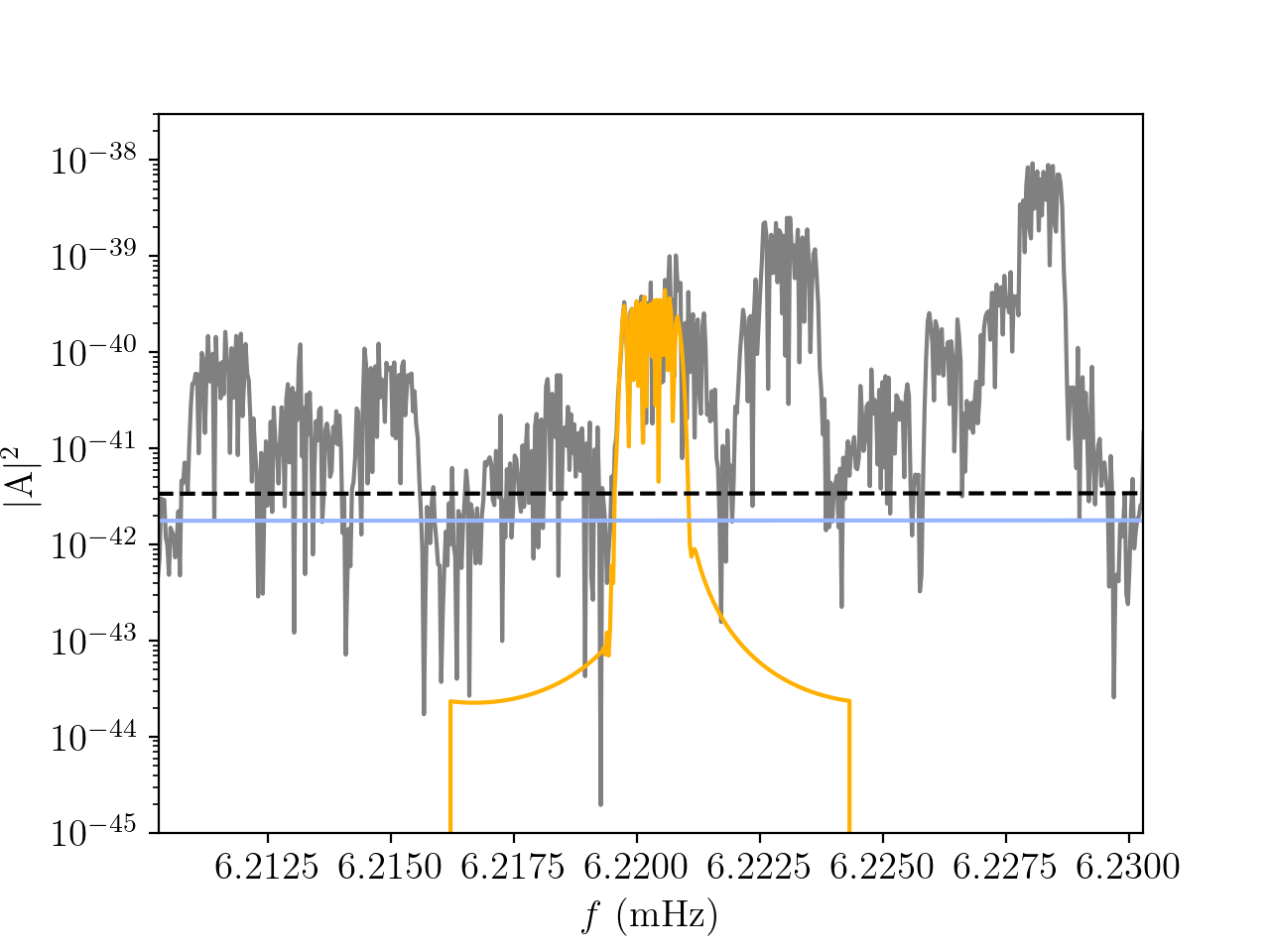}
\caption{Power spectra of the {\hmcnc} waveform (orange) in context with a simulated galaxy (gray) used for the {\sangria} dataset. The blue line indicates the simulated instrument noise level while the black dashed line is the noise level inferred by a global analysis of the data, slightly elevated by unresolved binaries in this year-long data simulation.}
\label{fig:hmcnc_waveforms}
\end{figure}

Figure~\ref{fig:hmcnc_waveforms} shows a particularly egregious example in the form of the power spectrum of simulated LISA data, where the signal from {\hmcnc}, shown in orange, is overlapping other similarly strong signals present in a galactic binary population (gray). The blue line is the simulated instrument noise level, while the black dashed line is the inferred level, slightly higher than the simulation due to unresolved binaries in this 1 year observation time. The simulated data were produced as part of the LISA Data Challenges (LDCs)~\citep{Sangria}.

In this study we will show that analysis methods which search only for the known binaries, even when using maximally informative prior information about the sources, will be dominated by systematic errors from the mis-modeling.
These biases are either due to ignored covariances with neighboring sources or, worse, misclassifying the combined signal from several binaries as coming only from the targeted EM-discovered source. 
To rectify the situation we suggest two avenues of future study.  The first is to develop more concrete protocol for using galactic binaries as verification sources, and to test these procedures in context of the full astrophysical population of GW sources that will be present in the LISA data.  
The second is to consider other sources predicted to be in the galactic populations, particularly at higher frequency where the source density is low and the probability of source confusion becomes negligible. 
While less will be known about these systems \emph{a priori}, it is possible that the extra degrees of freedom needed to model the GW-only binaries are balanced by the relative simplicity of measuring isolated signals.
 
\section{Known Binaries}
For our study we will focus on the highest signal to noise ratio (S/N) UCBs currently identified electromagnetically. 
The rationale for only considering the most prominent binaries is simple: We want sources which will be detectable after relatively short integration times with LISA.
Known UCBs detectable within several ${\sim}$weeks of observing time could be candidates for ``verification'' studies in the early stages of the mission operations, such as using their positive identification as part of the initial data validation, or continuously monitoring the inferred parameters of the binaries for use as calibration sources.

Typical UCB models for LISA data are parameterized with eight quantities:  The initial gravitational wave frequency and it's first time derivative ($f_0,\dot{f}$), the sky location in barycentric ecliptic coordinates ($\theta,\phi$), the gravitational wave amplitude ($\mathcal{A}$), and three angles describing the orientation of the binary to the observer: inclination, polarization, and initial phase ($\iota,\psi,\varphi_0$). 
The sources used in this study and their assumed parameters are found in Table~\ref{tab:parameters}. Not listed in the table are the phase and polarization parameters ($\psi$ and $\varphi_0$) which are randomly drawn from a uniform distribution over the interval $U[0,2\pi]$. The gravitational wave amplitude is derived from the binary parameters as $\mathcal{A} = 2\mathcal{M}^{5/3} (\pi f_0)^{2/3}/d_L$ where the chirp mass $\mathcal{M}\equiv(m_1 m_2)^{3/5} (m_1 + m_2)^{-1/5}$ and $m_i$ is the mass of the $i$'th object in the binary. 

\begin{table*}[t]
\begin{center}
\begin{tabular}{c|c|c|c|c|c|c|c|c}
Name & $f_0$ (mHz) & $\dot{f}$ (Hz/s) & $\cos\theta$ & $\phi$ (rad) & $m_1$ ($M_\odot$) & $m_2$ ($M_\odot$) & $\cos\iota$ & $d_L$ (kpc) \\
\hline
\amcvn & 1.944144722 & 0 & 0.6080 & 2.9737 & 0.68 & 0.13 & 0.7 & 0.302 \\
\escet & 3.22471421 & $-8.3190 \times 10^{-18}$ & -0.3475 & 0.4295 & 0.80 & 0.16 & 0.5 & 1.726 \\
\hmcnc & 6.220278731 & $-1.9555 \times 10^{-17}$ & -0.0820 & 2.1021 & 0.55 & 0.27 & 0.8 & 5 \\
\sdss & 2.613673417 & $3.3473 \times 10^{-17}$ & 0.1011 & 1.7686 & 0.49 & 0.25 & 0.1 & 0.993 \\
\vul & 3.51250011 & $-1.9556 \times 10^{-17}$ & 0.7288 & 5.1486 & 0.80 & 0.18 & 0.5 & 1.7 \\
\ztf & 4.821699107 & $2.7585 \times 10^{-16}$ & 0.9147 & 3.5783 & 0.65 & 0.17 & 0.1 & 2.34 
\end{tabular}
\caption{Source parameters for verification binary simulations.}
\label{tab:parameters}
\end{center}
\end{table*}

The parameters for the known binaries match the simulations provided by the LISA Data Challenge group in dataset {\sangria}~\citep{Sangria}. 
For the current state of the art of known binary parameters see ~\citep{kupfer2024lisa}.

\section{Method}
To study the identification of the loudest currently known binaries we conduct numerical experiments on simulated data $D$ with existing prototype LISA analysis pipelines~\citep{ldasoft} which use Markov Chain Monte Carlo (MCMC) methods to produce samples from the posterior distribution function of the binary parameters.
For the special case of analyzing the known sources, the sampler {\tt vbmcmc} does not search over the orbital period or sky location of the binary, i.e. the prior $p(\bf{x}\rightarrow (f_0,\cos\theta,\phi))=\delta(\bf{x}-\bf{x}^{\rm EM})$, where $\bf{x}^{\rm EM}$ are the parameter values inferred from the EM observations.  
The instrument noise is modeled as a constant over the narrow bandwidth of the analysis.
Priors over the remaining parameters are uniform for the orientation parameters and the frequency derivative $(\cos\iota,\psi,\varphi_0,\dot{f})$ while the prior on $\mathcal{A}$ depends on the signal-to-noise of the binary as described in \cite{PhysRevD.101.123021}.
Furthermore, the analysis assumes the existence of the binary and is not performing the model selection necessary when optimally fitting to the full galaxy signal (handled by a transdimensional MCMC in the blind analysis).
In other words, the prior probability that the verification binaries are in the model is 1.
For each experiment we compare three different analyses
\begin{itemize}
\item {\bf \vbonly:} The simulated data contain only the known binaries, and only the targeted ``verification binary'' search is used.
\item {\bf \vbgalaxy:} The simulated data contain a full galaxy simulation, and only the targeted ``verification binary'' search is used.
\item {\bf \vbglobal:} The simulated data contain a full galaxy simulation, and the targeted ``verification binary`` search is jointly run with the ``blind'' search for the remaining galactic binaries.
\end{itemize}
The {\vbonly} demonstration is to calibrate expectations, as it is the ideal (and perhaps erroneously envisioned) scenario of being able to measure the known binaries independently of other sources possibly present in the data.  
The {\vbgalaxy} analysis serves as the cautionary tale of what could result from a targeted search for the known binaries in isolation, without considering possible contamination from other galactic sources.
Finally the {\vbglobal} analysis is to demonstrate the robust inferences for the known binaries but at the expense of having to use a complete model for the galactic binaries (and instrument noise) thus making the analysis of the known binaries no simpler than the full global fit.
For each analysis we compare results for different LISA observing scenarios, all of which use observational integration times that are short compared to the full mission lifetime.
These are to serve as a proxy for different scenarios for using known binaries for validation of the science data or analysis pipelines early in the mission operations. 
\begin{itemize}
\item {\bf Scenario 1}: Compare results from each analysis for 3 month observation times to test consistency between the different methods. This serves to calibrate our expectations of how well the parameters are constrained, and what biases we might expect from the different analyses.
\item {\bf Scenario 2}: Compare results for a single analysis after 1 week, 1 month, and 3 months, to test the self-consistency of results as more data are acquired. This is a verification binary ``monitoring'' approach where we would expect steadily improving precision of the inferred parameters.
\item {\bf Scenario 3}: Compare results from a single analysis on four subsequent week-long segments, all within the first month of observing. This is a ``jackknife'' test of the parameter inferences in different, independent observing epochs.
\end{itemize}

We will focus on two marginalized projections of the full posterior distribution function $p(\mathcal{A},\cos\iota|D)$ and $p(\psi,\varphi_0|D)$.
The former is effectively testing the consistency of the amplitude recovery, as ${\mathcal{A}}$ is the overall GW amplitude and $\cos\iota$ encodes the relative amplitude between the different GW polarizations.
The latter monitors the phase consistency, with $\varphi_0$ controlling the overall GW phase and $\psi$ setting phase differences of the different GW polarizations.

\section{Results}
\subsection{Scenario 1}

We begin by comparing the marginalized posterior distribution functions of the amplitude and phase parameters for each of the six binaries studied in this work for a three month observing time.  
In all figures comparing analysis methods the color coding conventions will be identical. The magenta, green, and orange colors correspond to the {\vbglobal}, {\vbonly}, and {\vbgalaxy} methods, respectively.  
In each figure panel the dashed lines indicate the true parameter values of the simulated source.  Contours of the posteriors denote the 1 and 2$\sigma$ intervals.  

\begin{figure}
\gridline{\fig{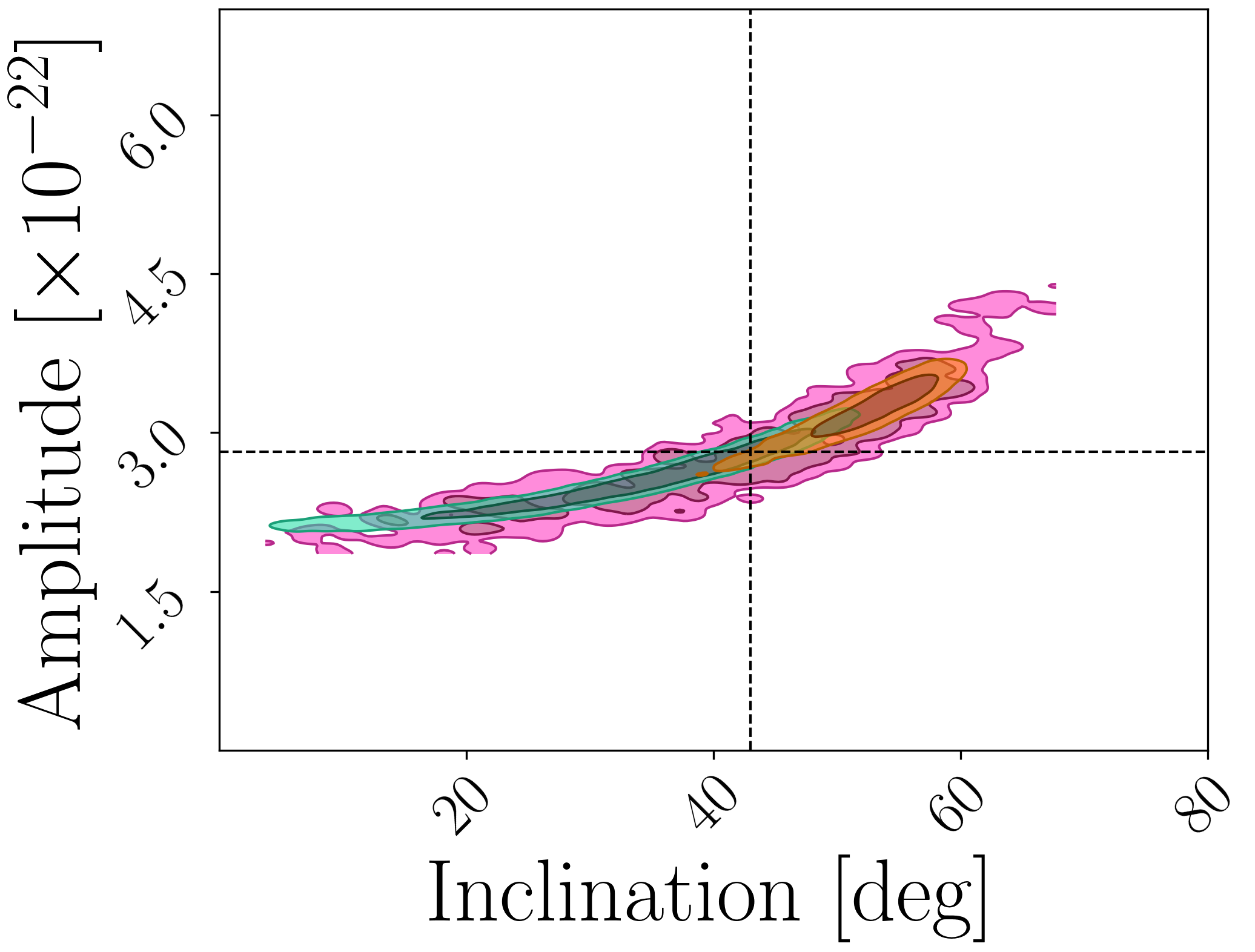}{0.15\textwidth}{{\amcvn}}
          \fig{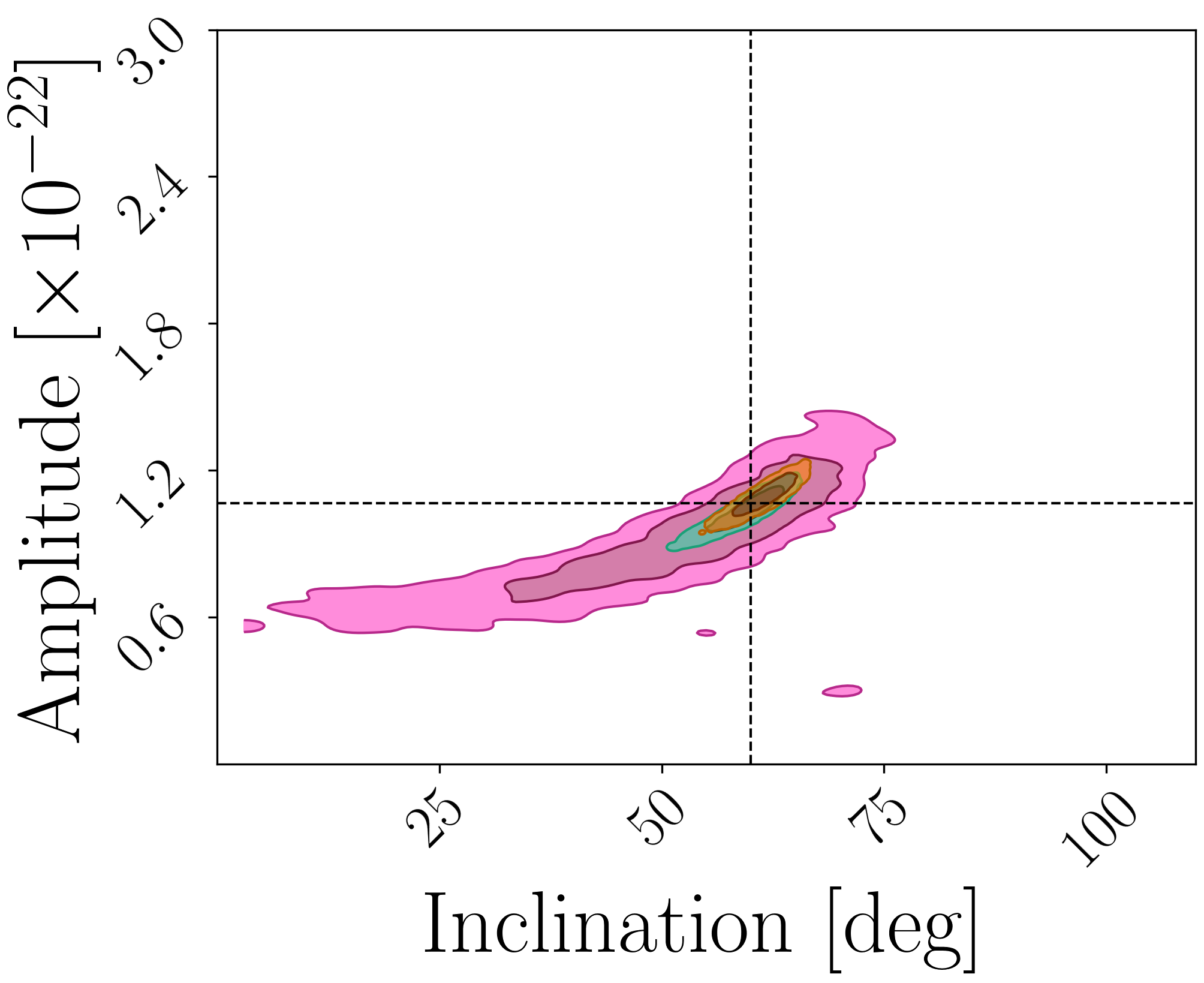}{0.15\textwidth}{{\escet}}
          \fig{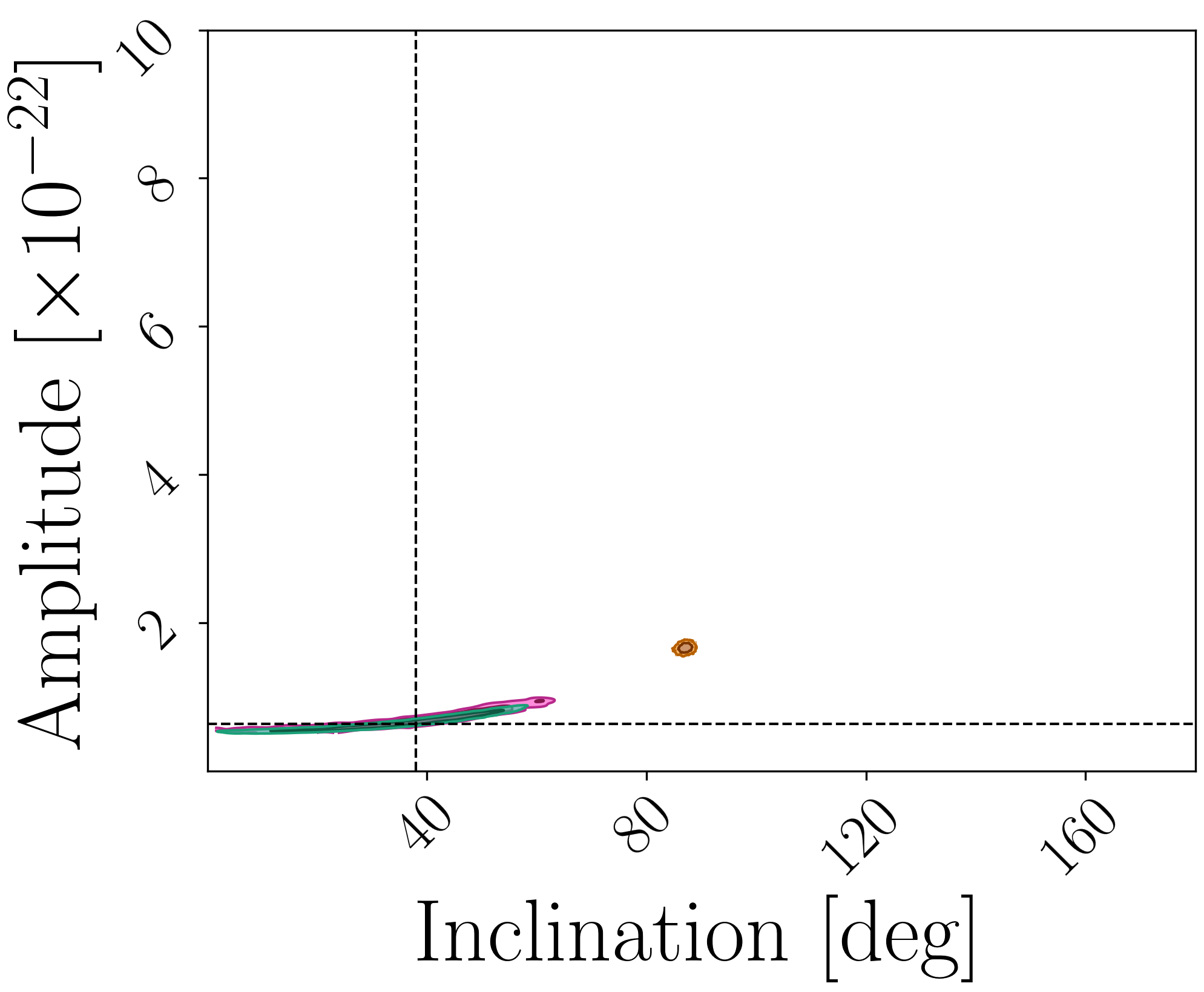}{0.15\textwidth}{{\hmcnc}}}
\gridline{\fig{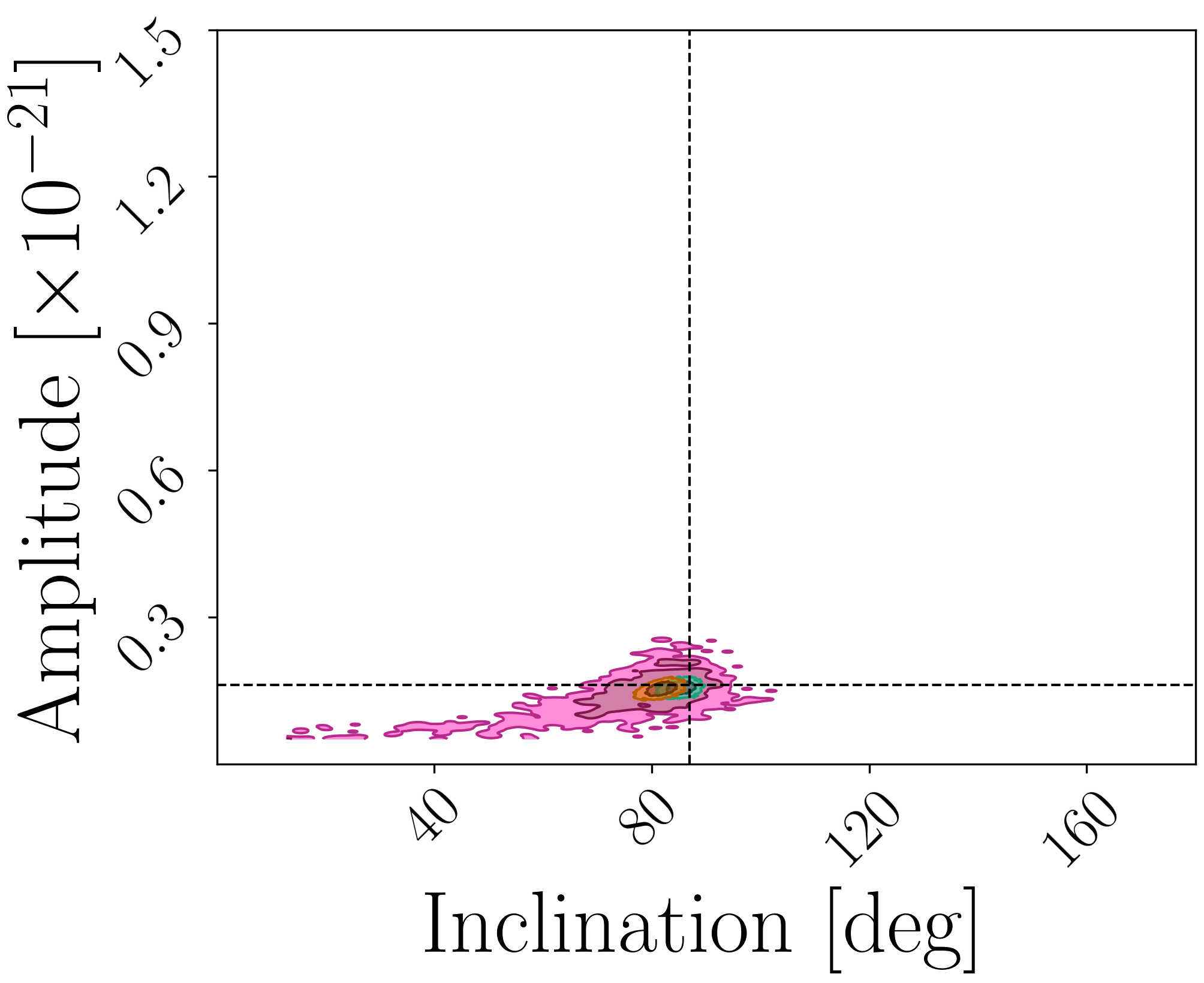}{0.15\textwidth}{{\sdss}}
          \fig{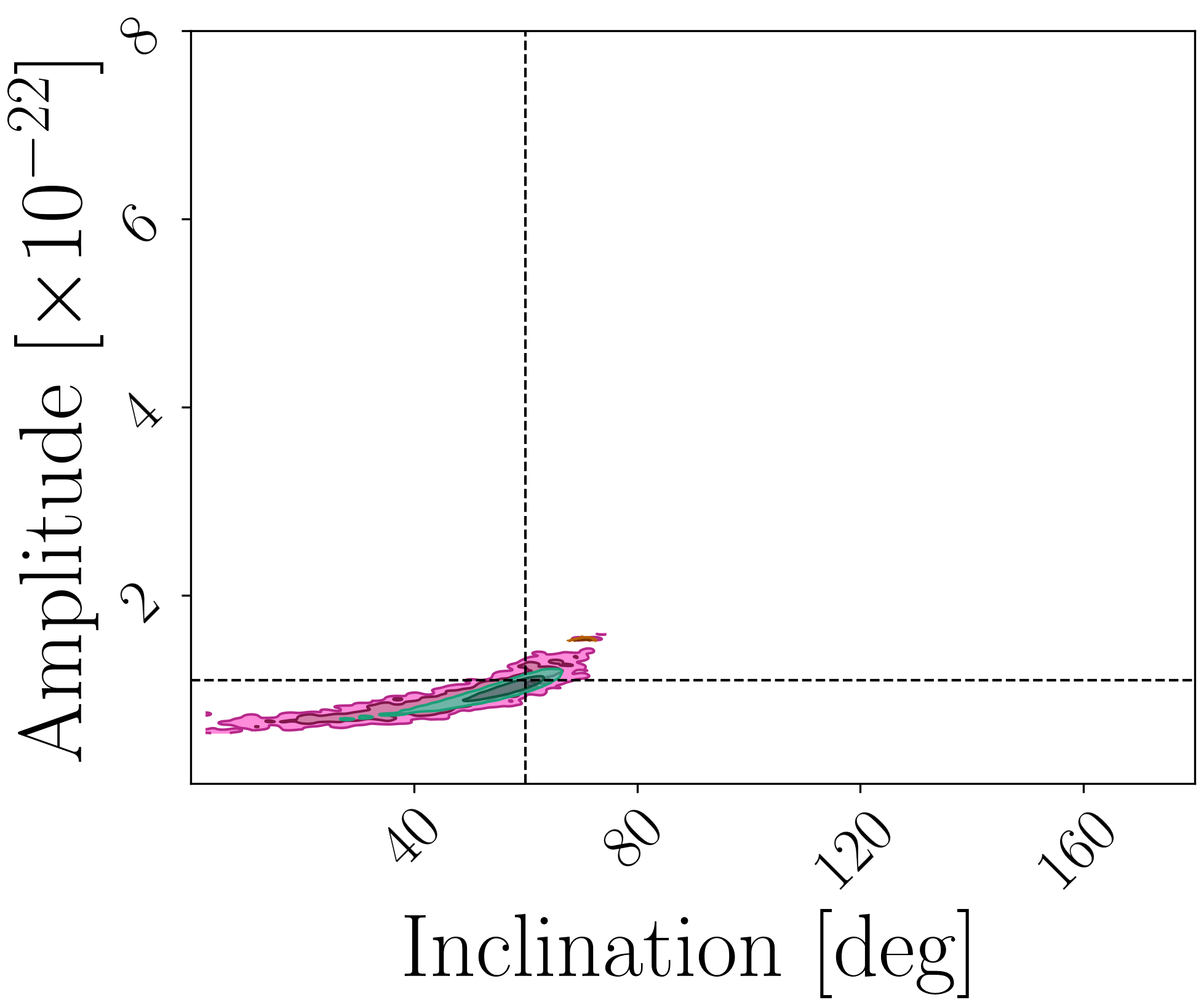}{0.15\textwidth}{{\vul}}
          \fig{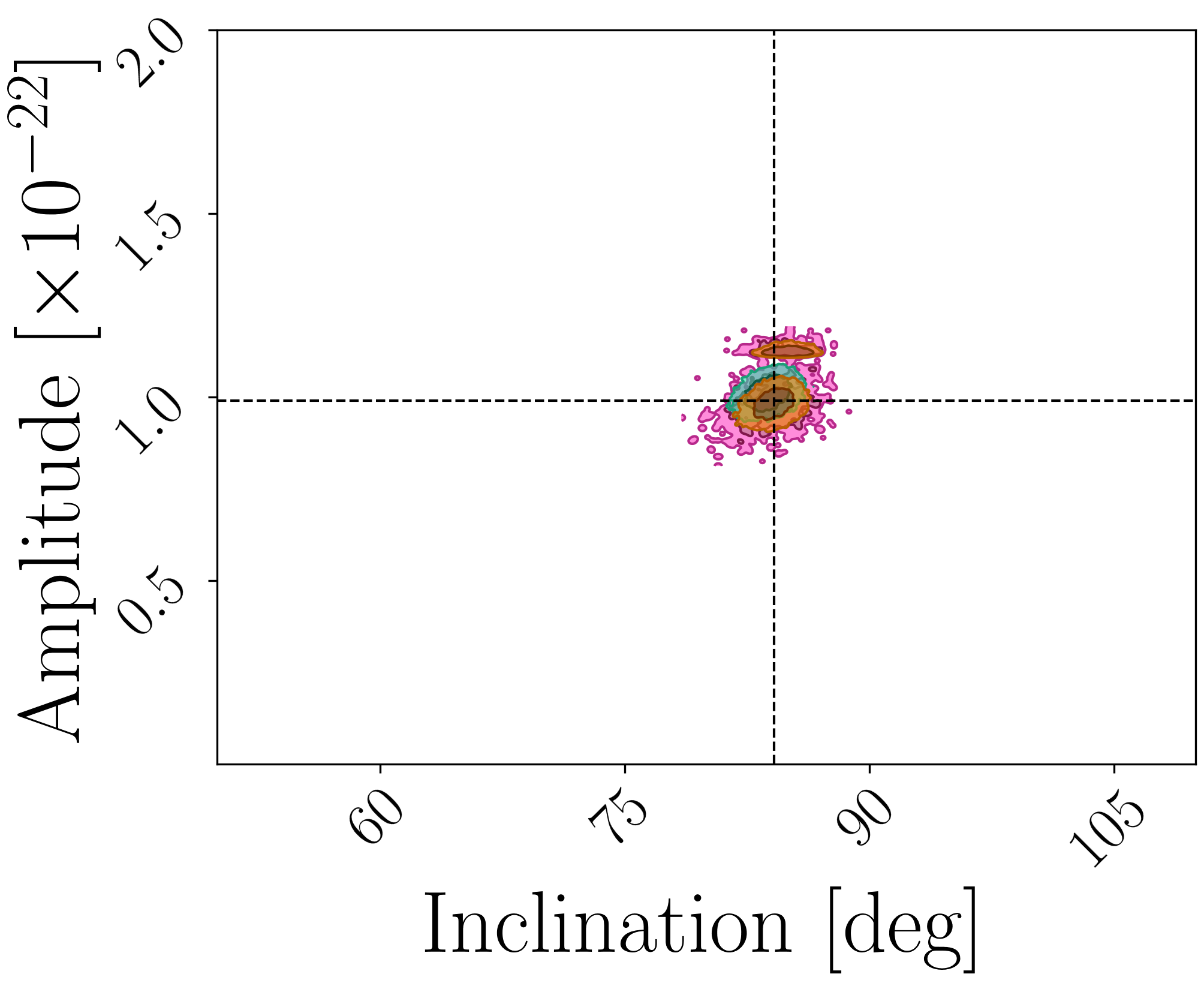}{0.15\textwidth}{{\ztf}}}
\caption{Comparison of amplitude parameters for 03 month observing times.  Magenta, green, and orange colors are for the {\vbglobal}, {\vbonly}, and {\vbgalaxy} analysis methods, respectively.  In several instances the systematic error, or bias, of the {\vbgalaxy} result is significant (i.e., larger than the statistical error) when compared to the baseline {\vbonly} method.  However, in all but the cases of {\hmcnc} and {\vul} the differences are obscured by the statistical error of the more-realistic {\vbglobal} analysis.  The biases in the case of {\hmcnc} and {\vul} are significant and problematic for the prospect of using known binaries as verification sources.}
\label{fig:03_amplitude}
\end{figure}

Figure~\ref{fig:03_amplitude} displays the amplitude parameters for a three month observing time.  
Two noticeable differences are the shifts between the {\vbonly} and {\vbgalaxy} distributions due to systematic errors introduced when including the rest of the galaxy in the data but ignoring it in the modeling, and the typically larger statistical errors in the {\vbglobal} analysis which is simultaneously fitting to the noise level which will be elevated at low frequency from the unresolved galactic foreground. 
The larger statistical error is most prevalent in the lower frequency binaries {\amcvn}, {\sdss}, and {\escet}.

The systematic errors for {\hmcnc} and {\vul} are particularly egregious in this example.  
In these two cases the bias found in the {\vbonly} analysis is larger than the statistical error from the {\vbglobal} analysis. 
This {\hmcnc} result is particularly disappointing for the prospects of having a straight-forward and simple method for using known binaries for data verification and validation. 
{\hmcnc} is the highest frequency and highest S/N currently known binary, above where the foreground noise is expected to dominate, as evidenced by the agreement between the {\vbglobal} and {\vbgalaxy} methods.  
Yet in the simulated galaxy other similarly high-S/N sources are sufficiently close to it in parameter space to cause the {\vbonly} method to fit to a combination of sources, despite the strong priors used in the model.
While all of the sources are individually resolvable by the global analysis (again citing the agreement between the green and megenta results) analyzing the data only for {\hmcnc} in isolation fails.

\begin{figure}
 \gridline{\fig{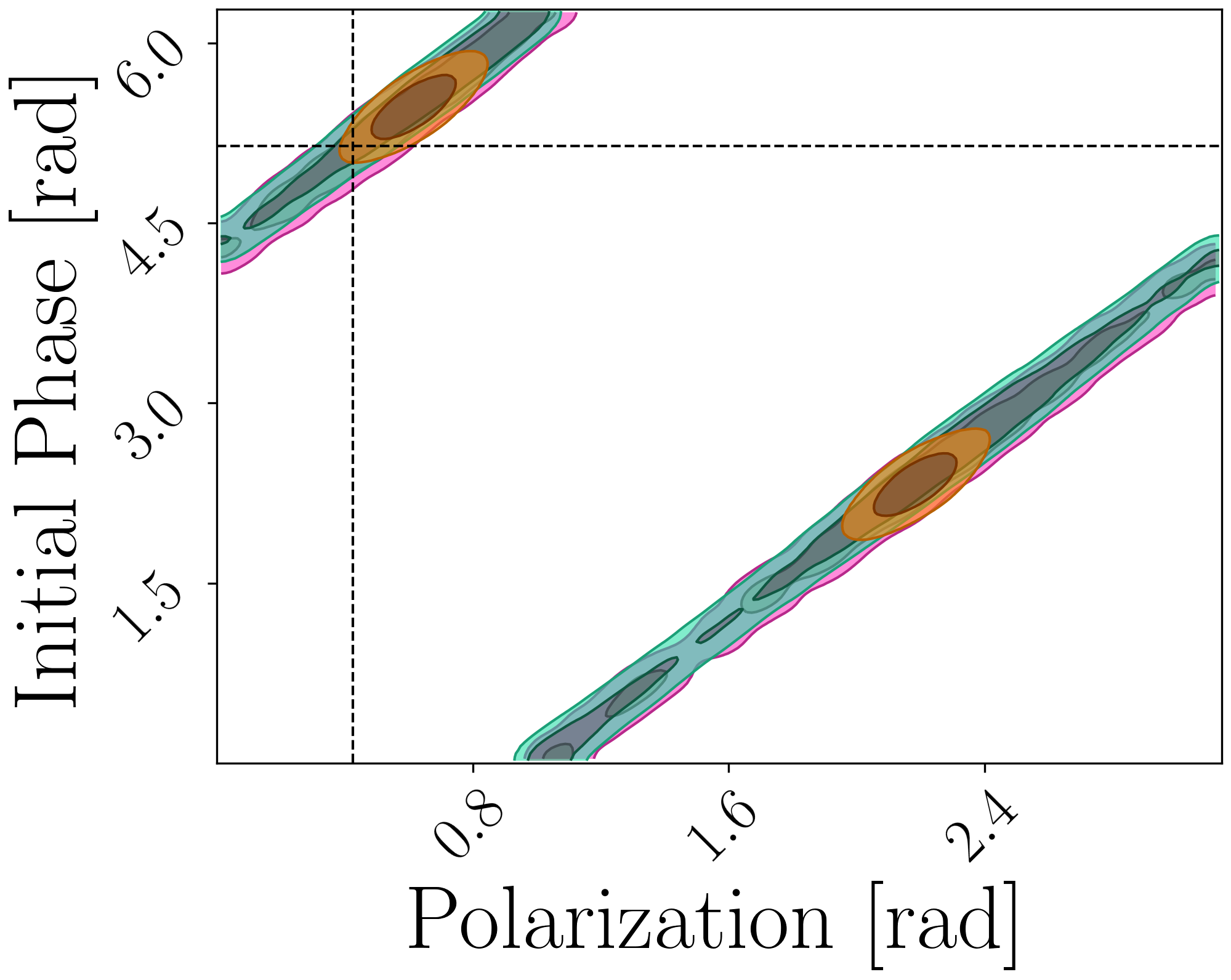}{0.15\textwidth}{{\amcvn}}
           \fig{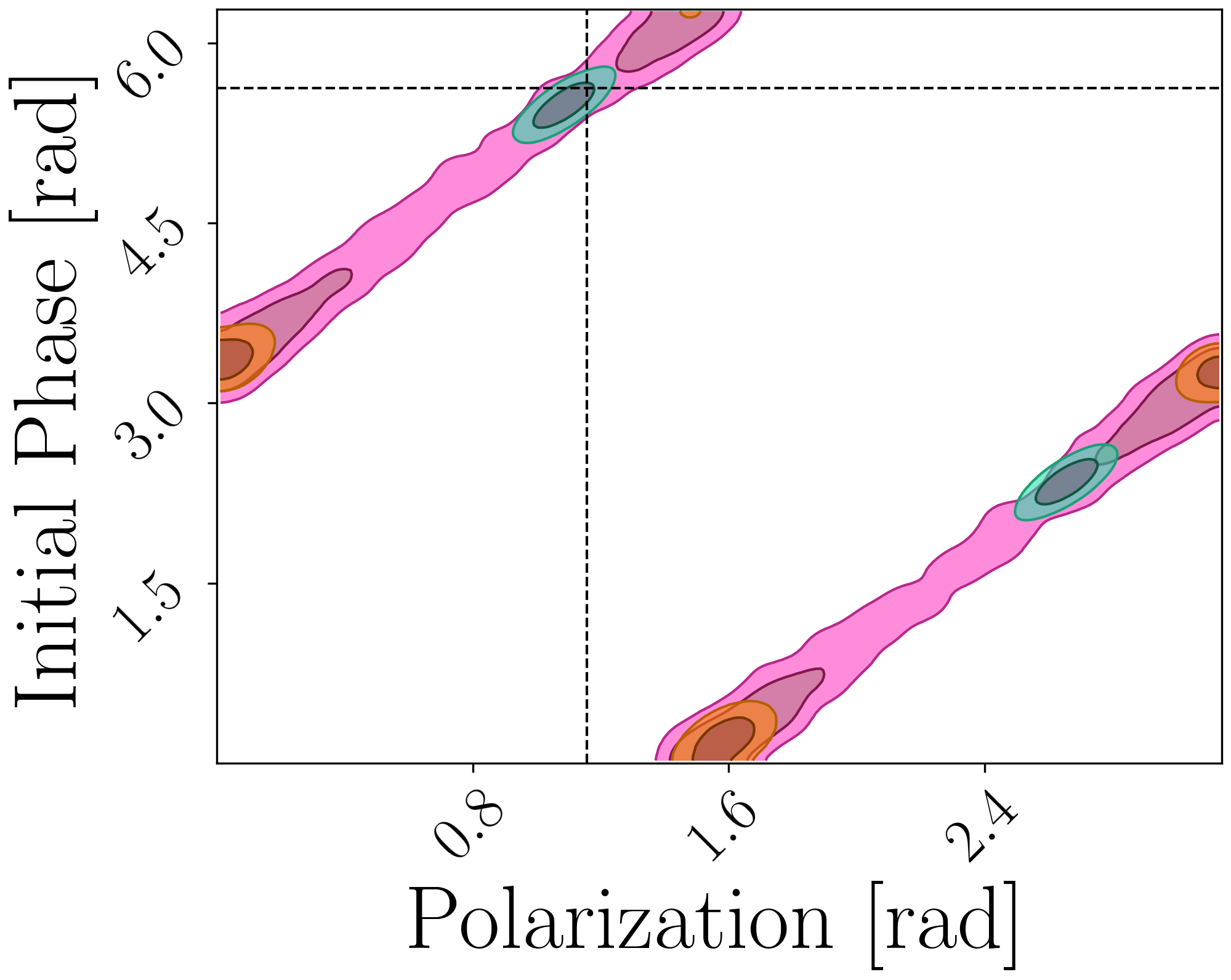}{0.15\textwidth}{{\escet}}
           \fig{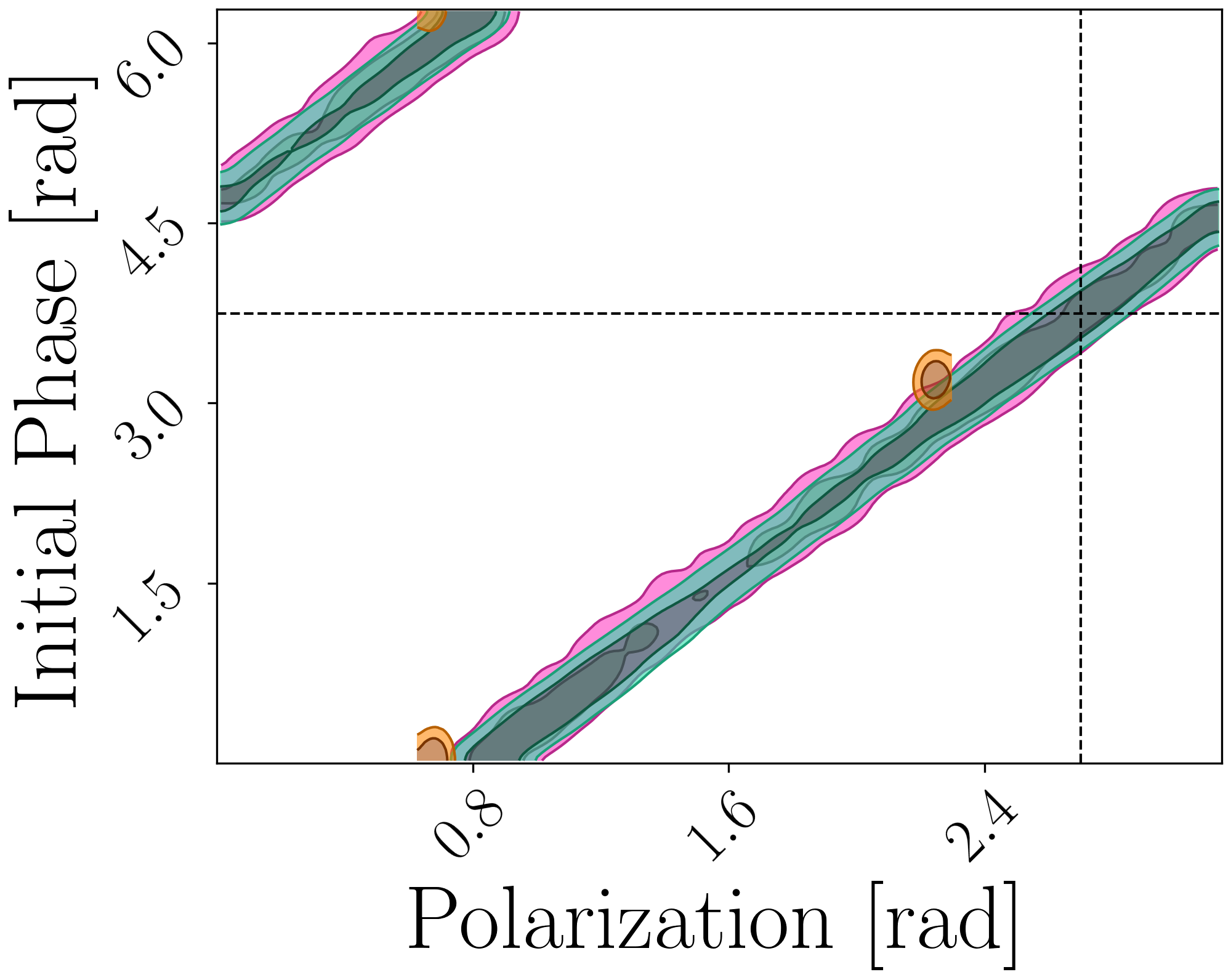}{0.15\textwidth}{{\hmcnc}}}
 \gridline{\fig{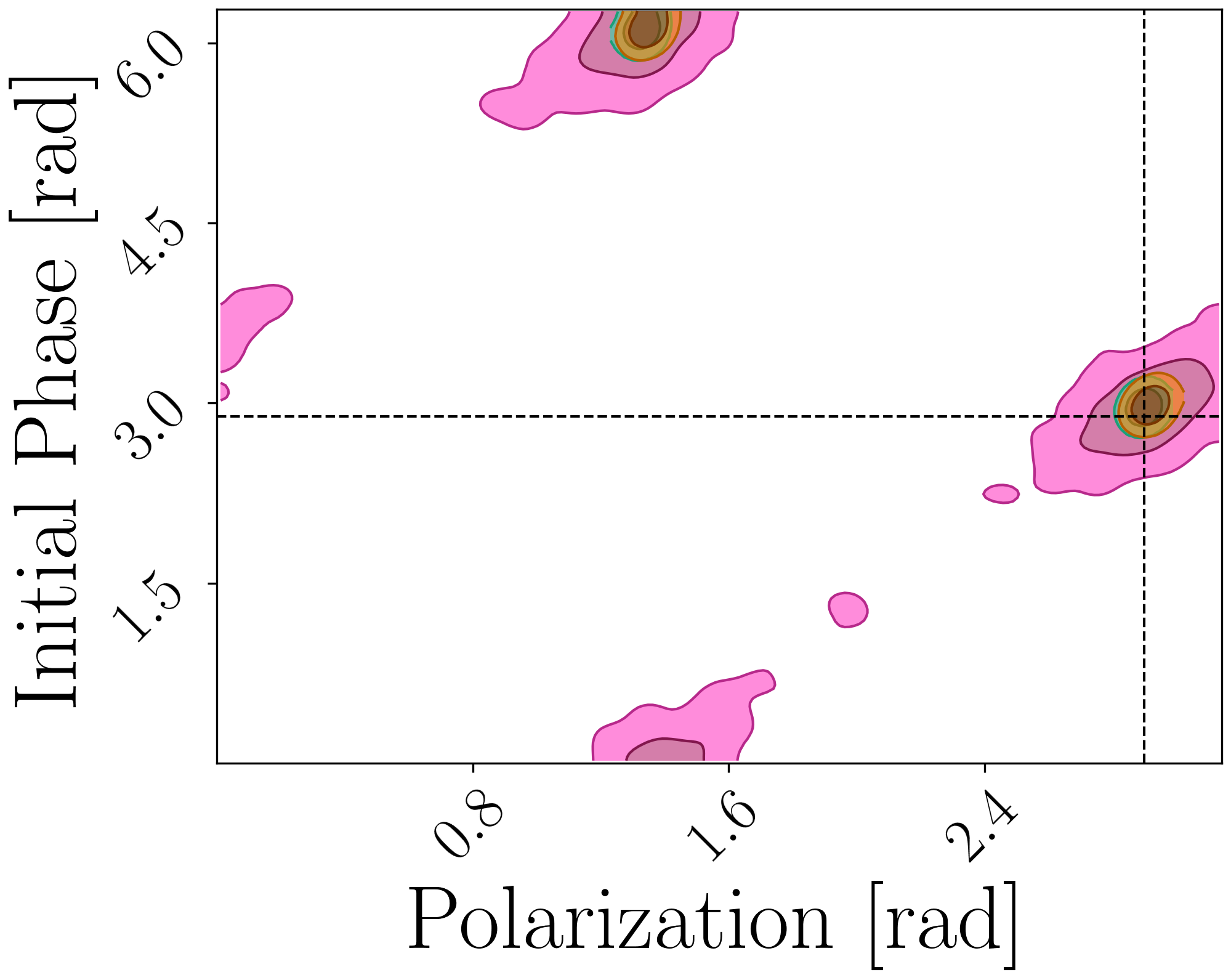}{0.15\textwidth}{{\sdss}}
           \fig{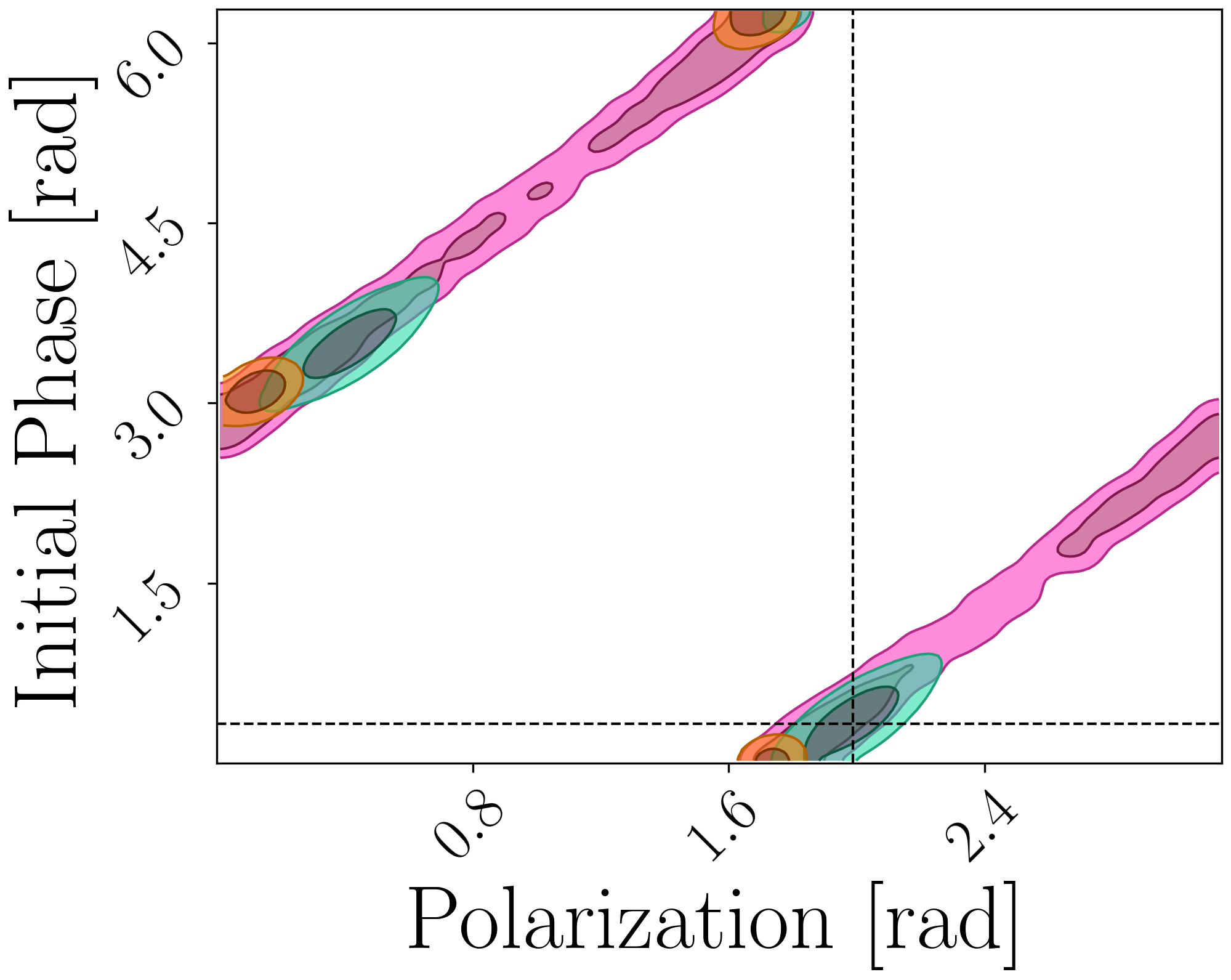}{0.15\textwidth}{{\vul}}
           \fig{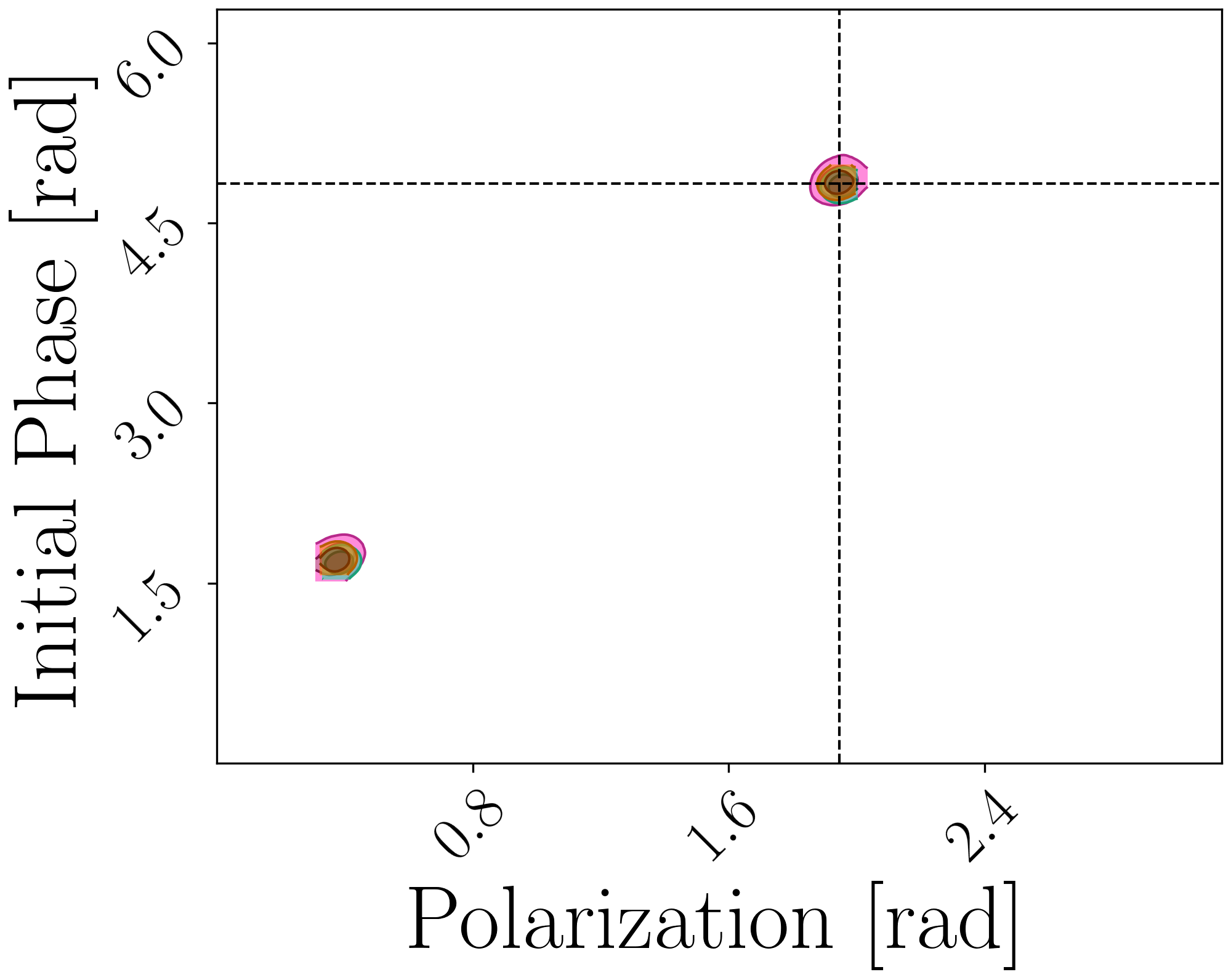}{0.15\textwidth}{{\ztf}}}
 \caption{Same as \ref{fig:03_amplitude} but comparing phase parameters.}
 \label{fig:03_phase}
 \end{figure}

Comparing the phase parameters in Fig.~\ref{fig:03_phase} leads to similar conclusions.  Systematic errors between the {\vbonly} and {\vbgalaxy} analyses are occasionally larger than the statistical error ({\escet} being particularly noteworthy here), but are masked by the larger statistical error of the {\vbglobal} analysis except for the case of {\hmcnc}.

\subsection{Scenario 2}

One of the more exciting, and perhaps under-appreciated, prospects of LISA's galactic binary observing capabilities is that many sources will be resolvable within much shorter observing times.  
The six systems we are focusing on in this work are expected to reach detectable S/Ns by a targeted search within a week of LISA observing. 
This gives us the opportunity to explore a scenario where the sources are monitored during the early phases of mission operations to check for consistency of the inferred source properties as the integrated signal strength grows.

\begin{figure}
\gridline{\fig{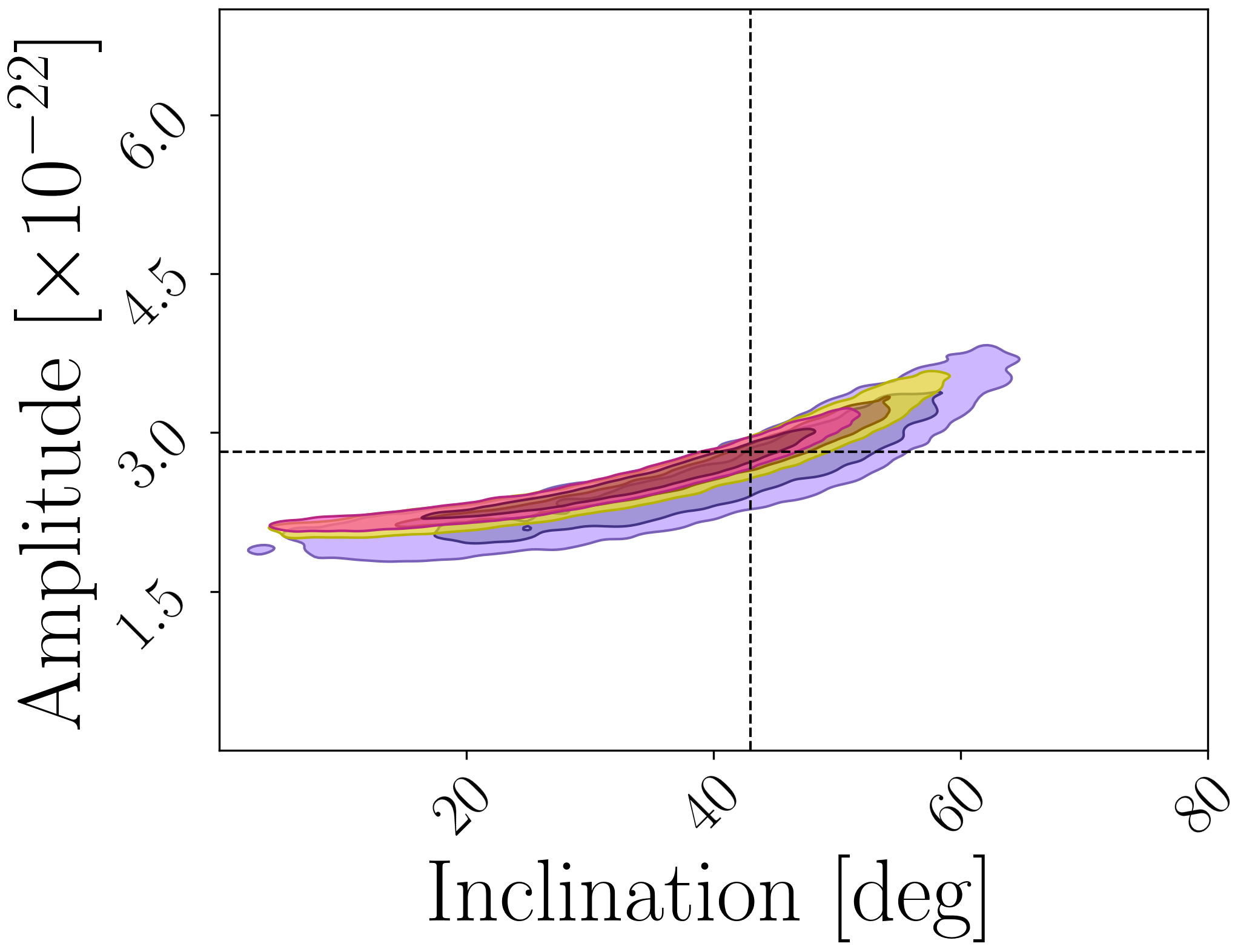}{0.15\textwidth}{{\amcvn}}
          \fig{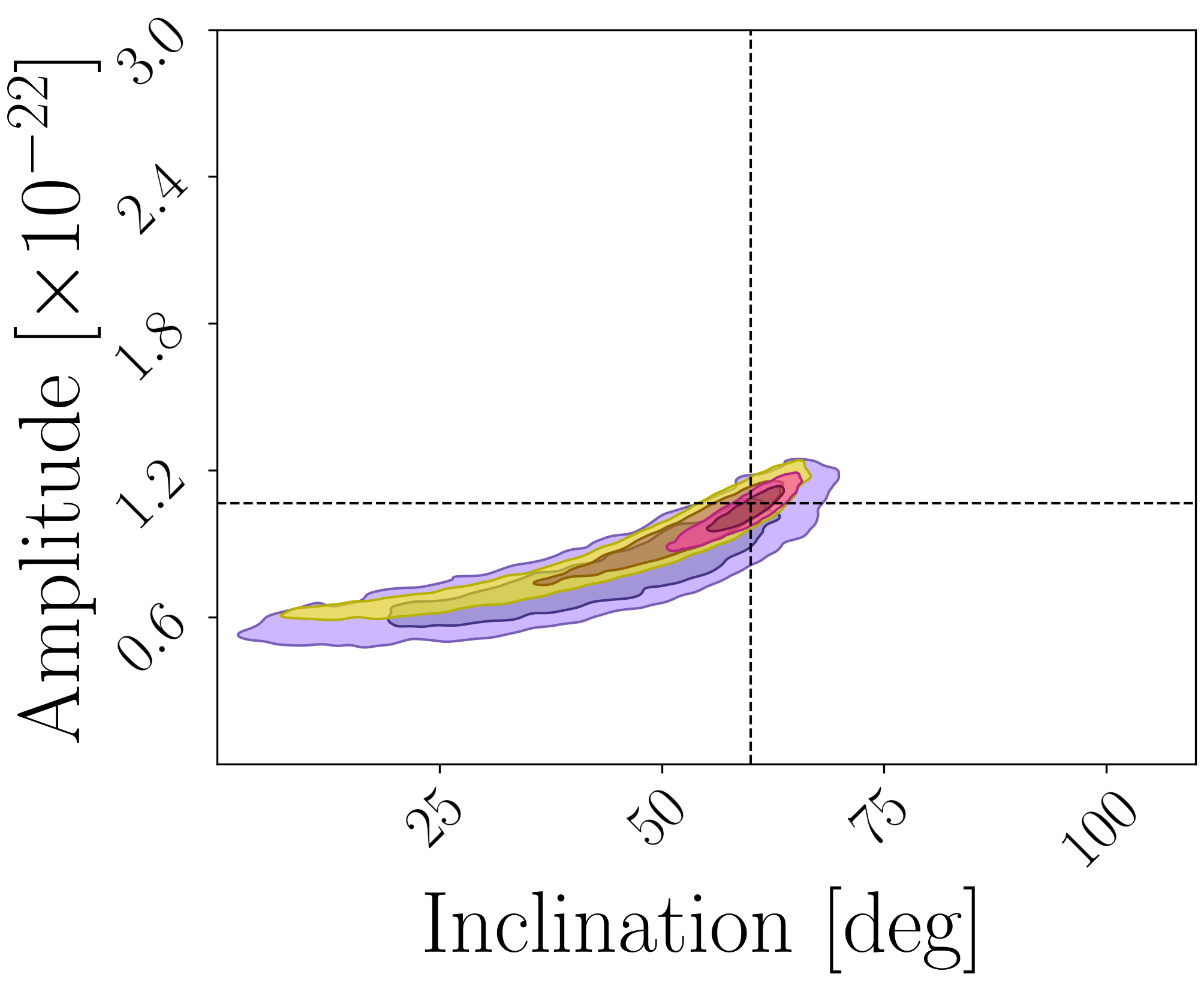}{0.15\textwidth}{{\escet}}
          \fig{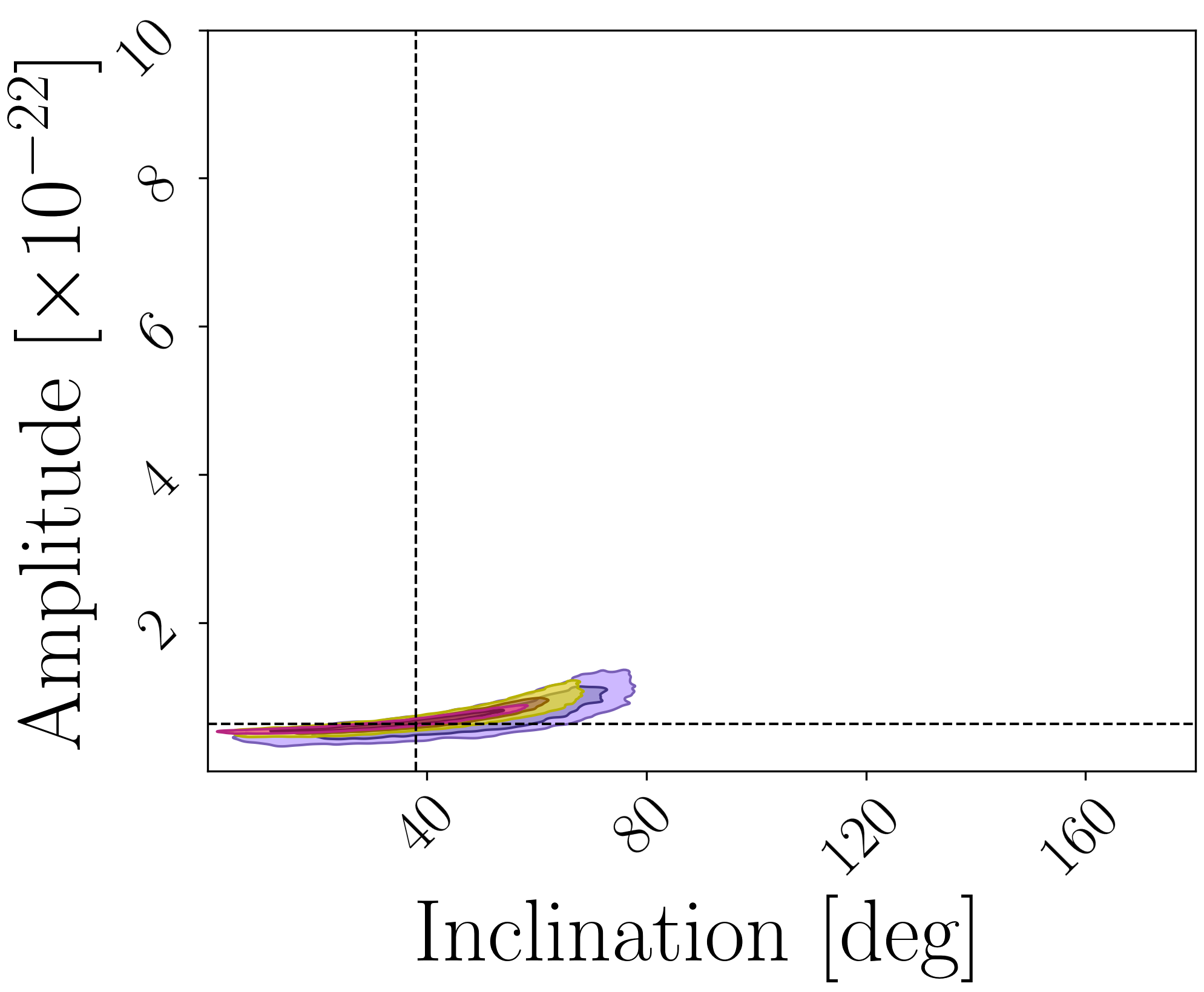}{0.15\textwidth}{{\hmcnc}}}
\gridline{\fig{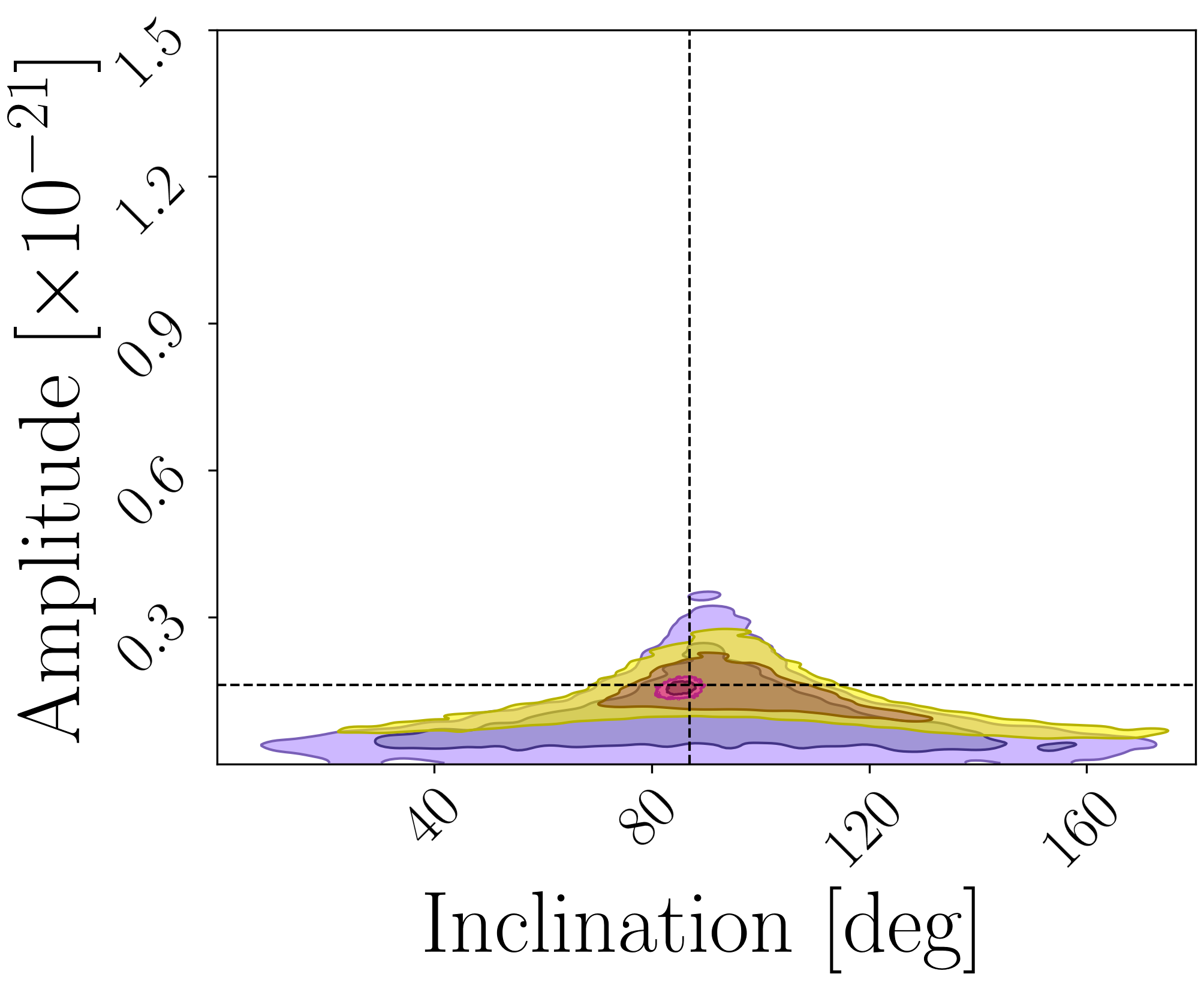}{0.15\textwidth}{{\sdss}}
          \fig{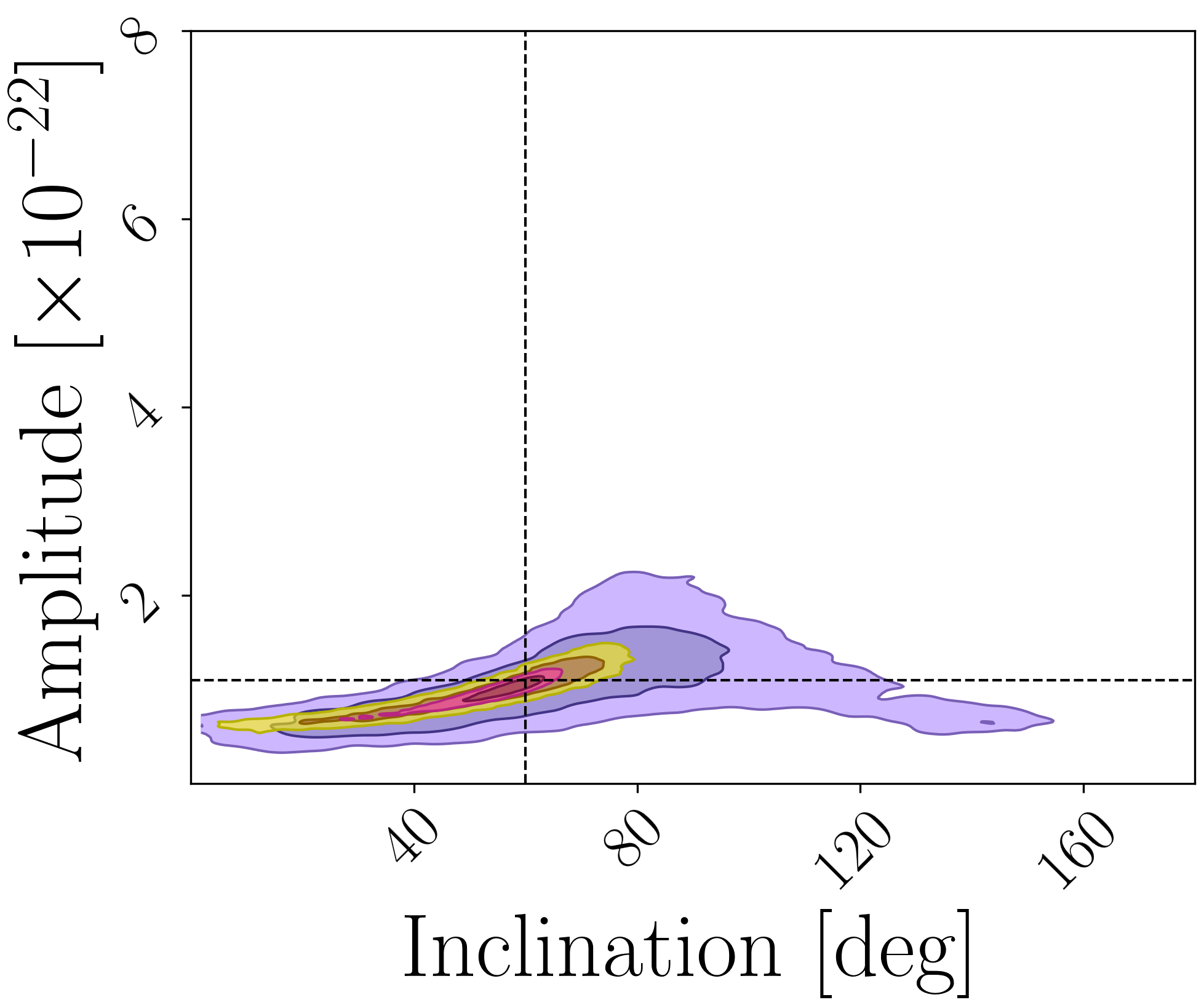}{0.15\textwidth}{{\vul}}
          \fig{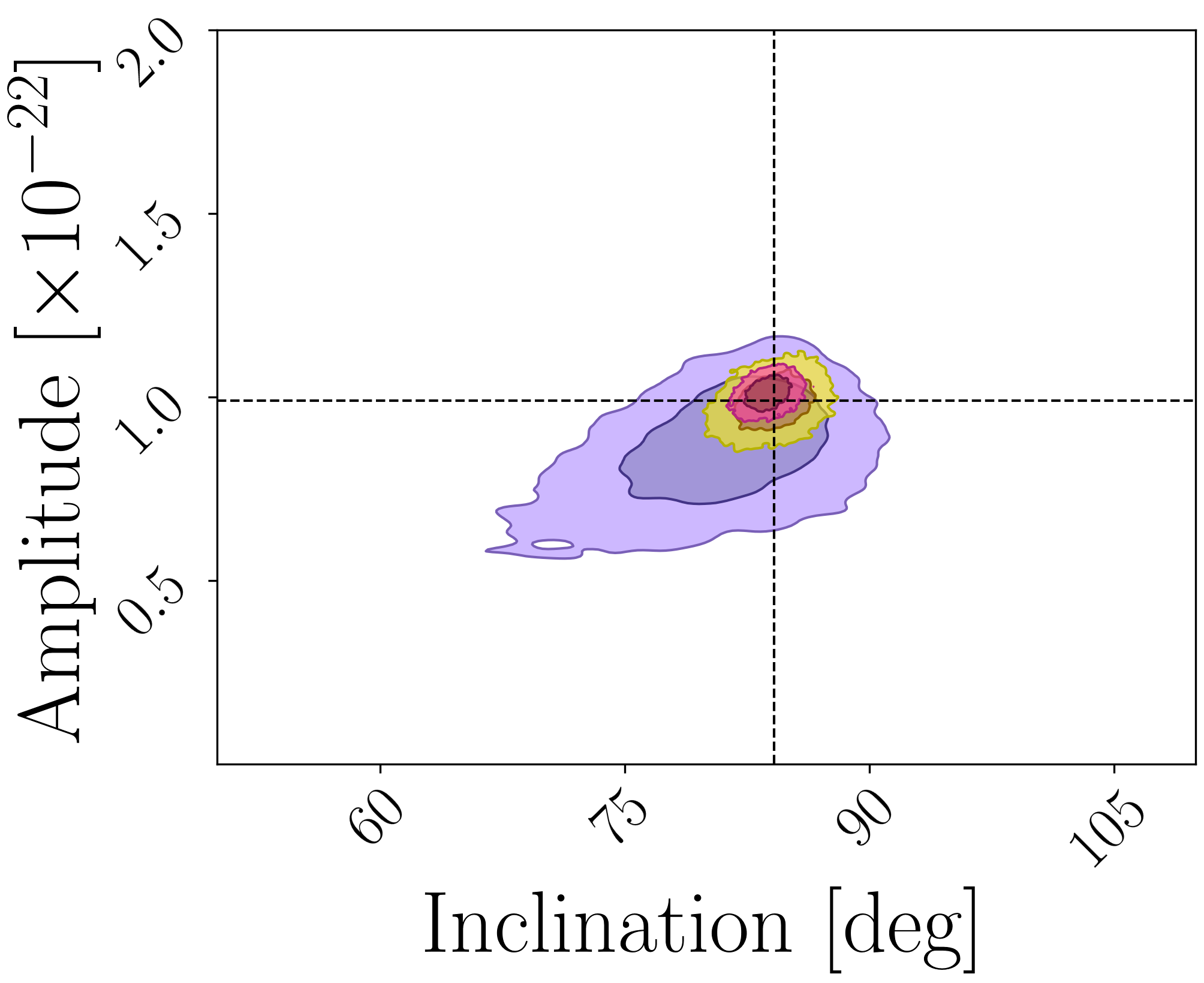}{0.15\textwidth}{{\ztf}}}
\caption{Time-evolving amplitude parameters using {\vbonly} model. Purple, yellow, and magenta contours are for 1 week, 1 month, and 3 month observation times, respectively.  The contours denote the 1 and 2$\sigma$ credible intervals. Dashed lines mark the true parameters values for the simulated signal.}
\label{fig:vgbonly_time_amplitude}
\end{figure}

\begin{figure}
\gridline{\fig{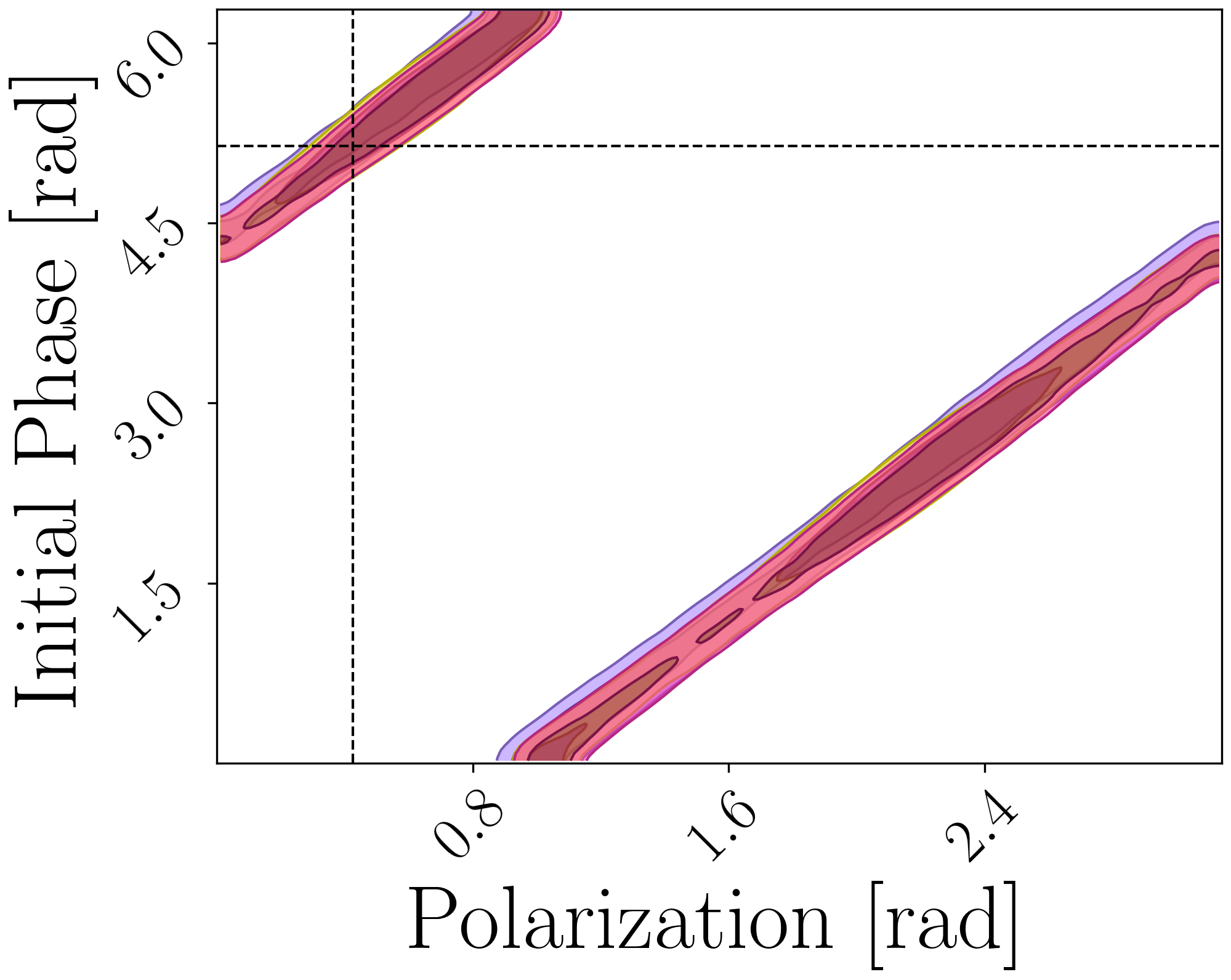}{0.15\textwidth}{{\amcvn}}
          \fig{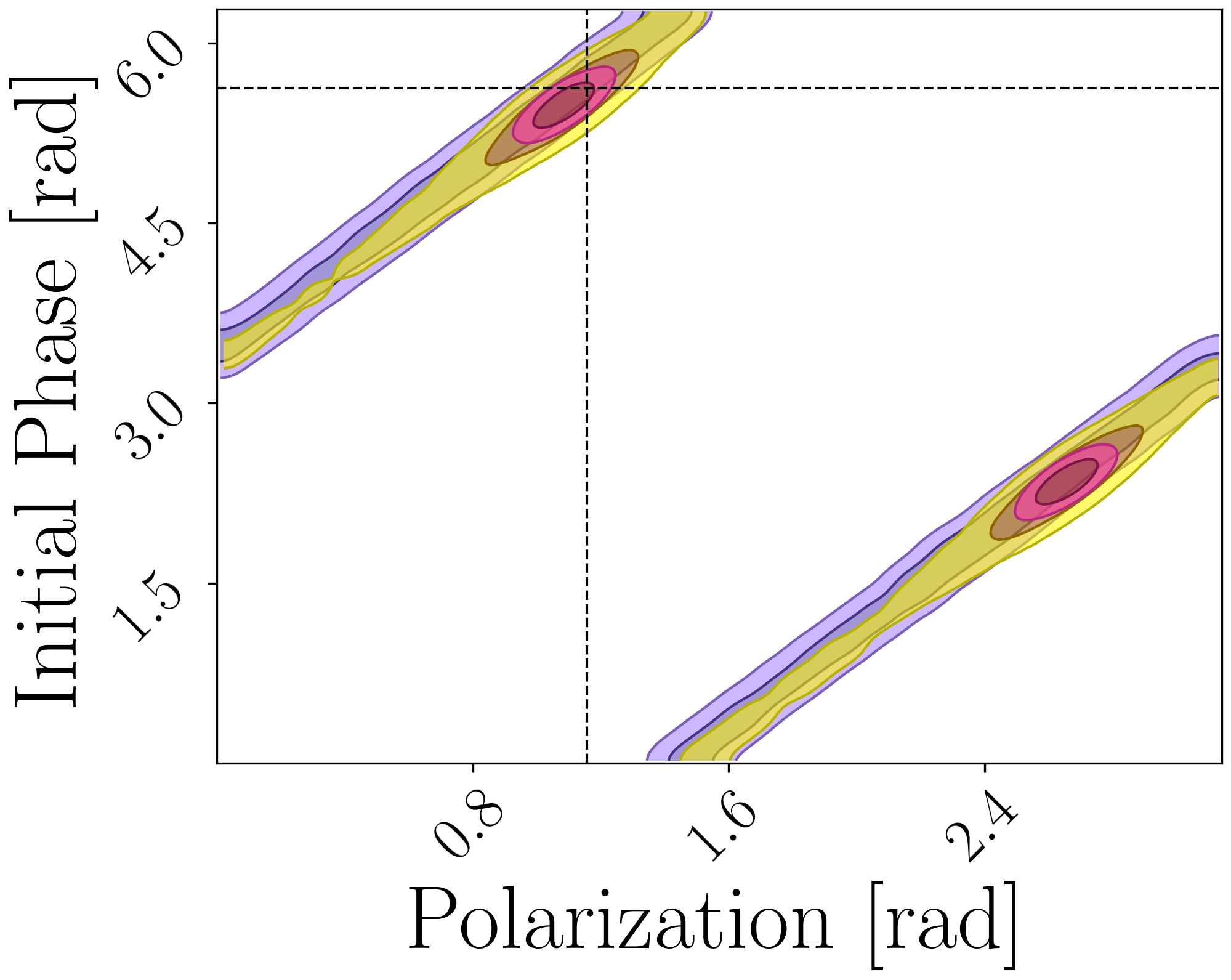}{0.15\textwidth}{{\escet}}
          \fig{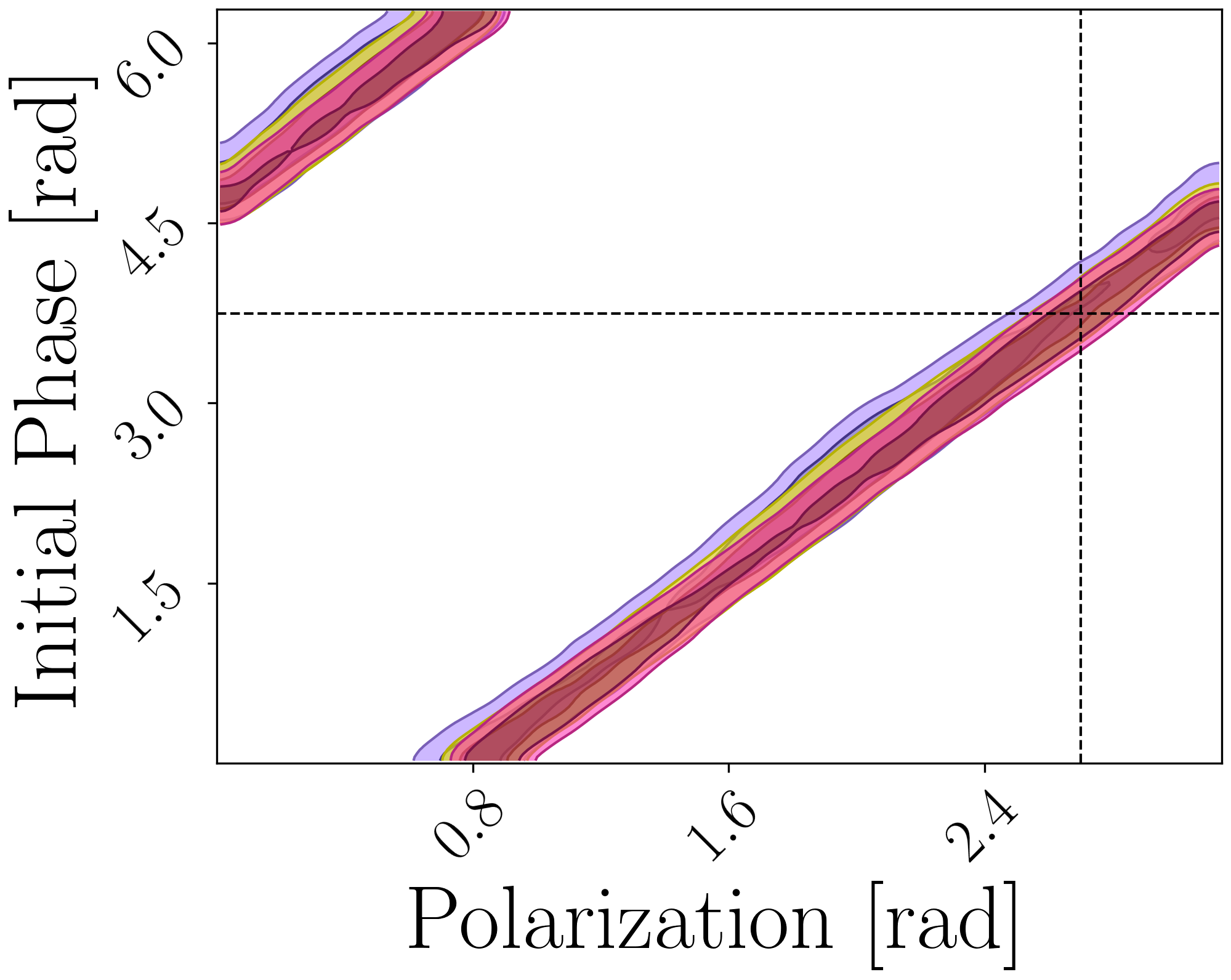}{0.15\textwidth}{{\hmcnc}}}
\gridline{\fig{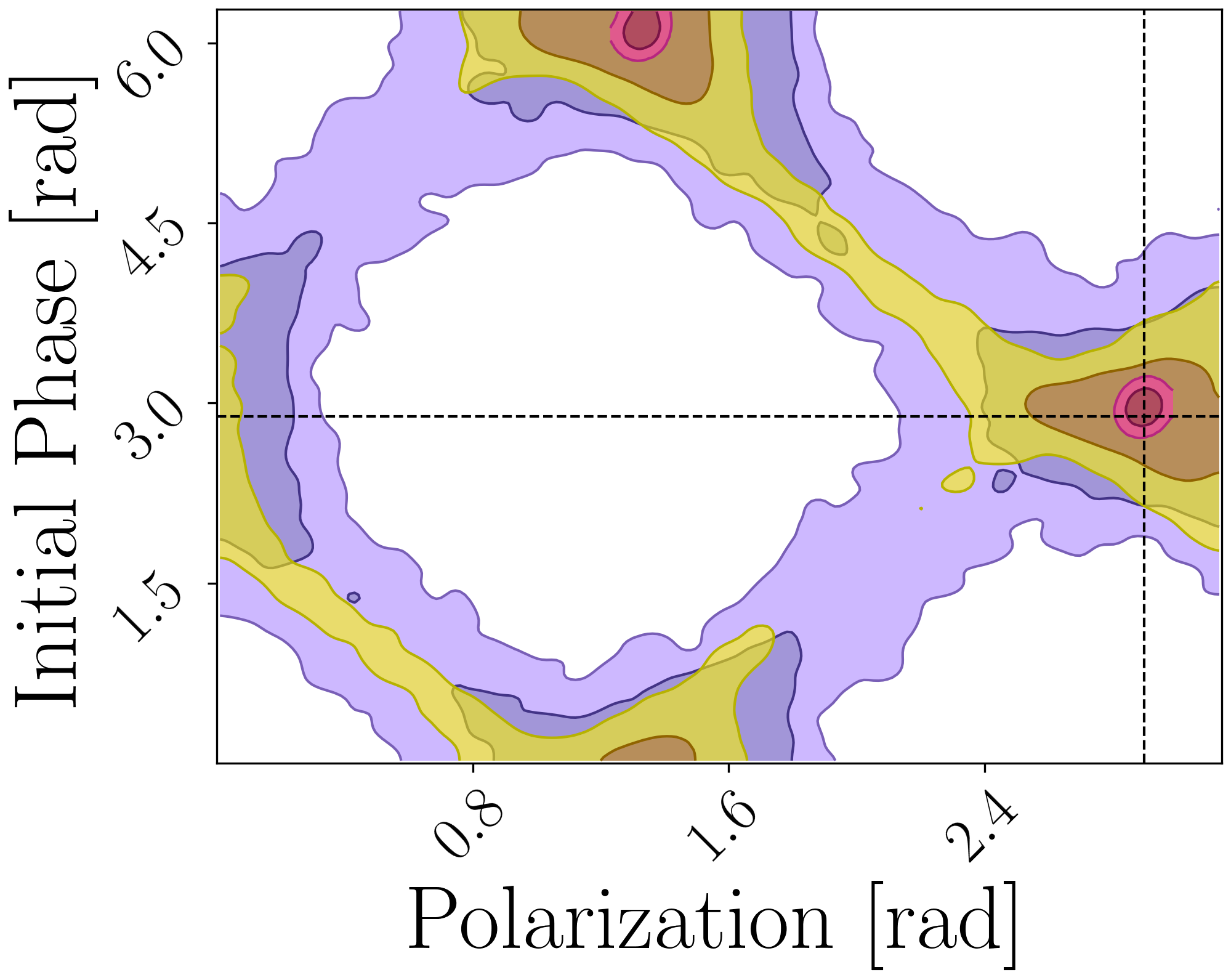}{0.15\textwidth}{{\sdss}}
          \fig{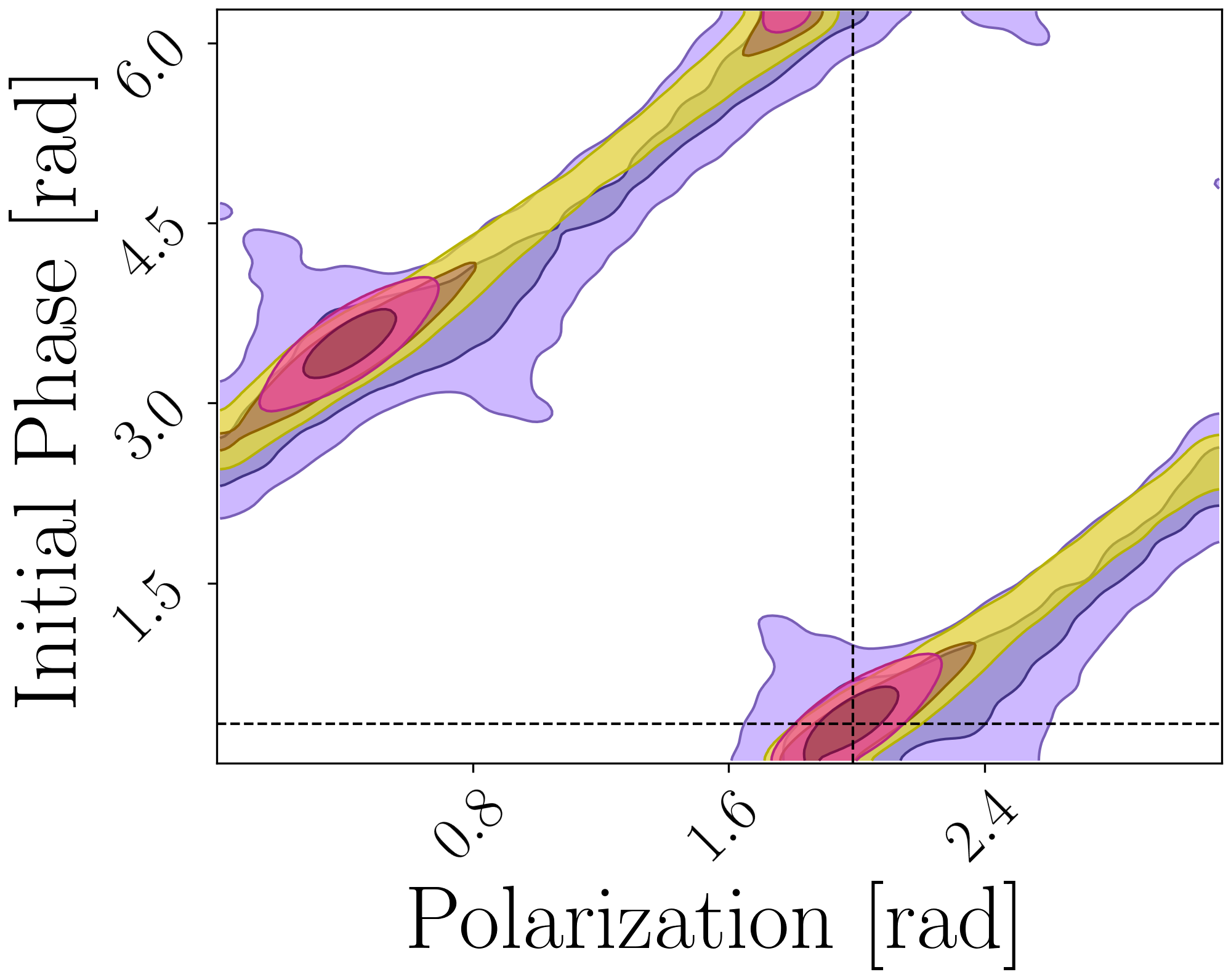}{0.15\textwidth}{{\vul}}
          \fig{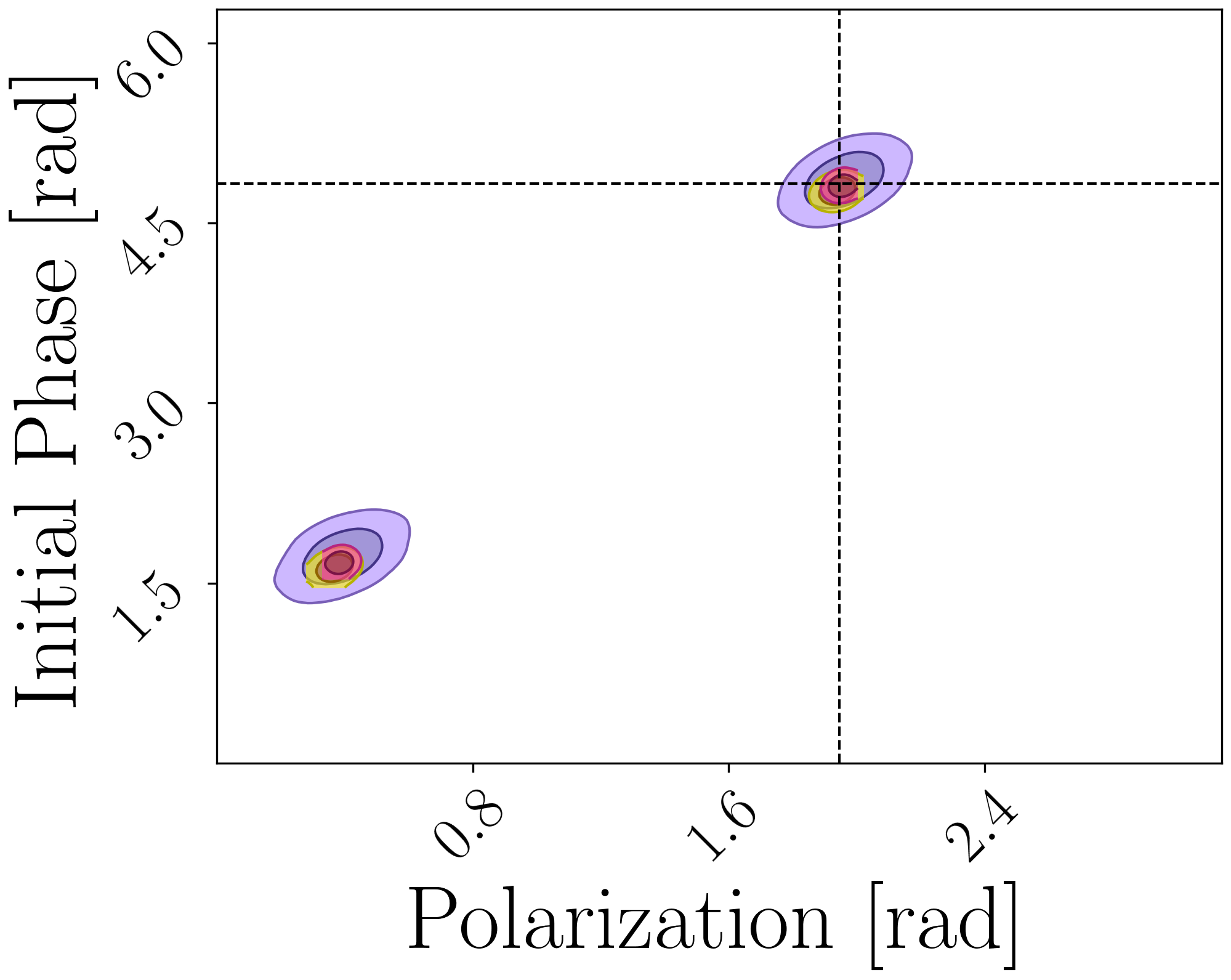}{0.15\textwidth}{{\ztf}}}
\caption{Same as Fig.~\ref{fig:vgbonly_time_amplitude} but for phase parameters.}
\label{fig:vgbonly_time_phase}
\end{figure}

To that end Figures~\ref{fig:vgbonly_time_amplitude} and \ref{fig:vgbonly_time_phase} show how that exploitation of the data would proceed if the known binaries were completely isolated and could be analyzed independently using the {\vbonly} paradigm. 
In these figures the colors correspond to different observing times, with purple, yellow, and magenta corresponding to 1 week, 1 month, and 3 months of LISA data. 
As before the contours are for 1 and 2$\sigma$ and the dashed lines mark the true values. 

In each example here the expected behavior is evident:  The true parameters are recovered accurately for each observing time (the truth is contained within the contours), while the precision increases (the contours shrink) with increased observing time (and therefore increased S/N). 

\begin{figure}
\gridline{\fig{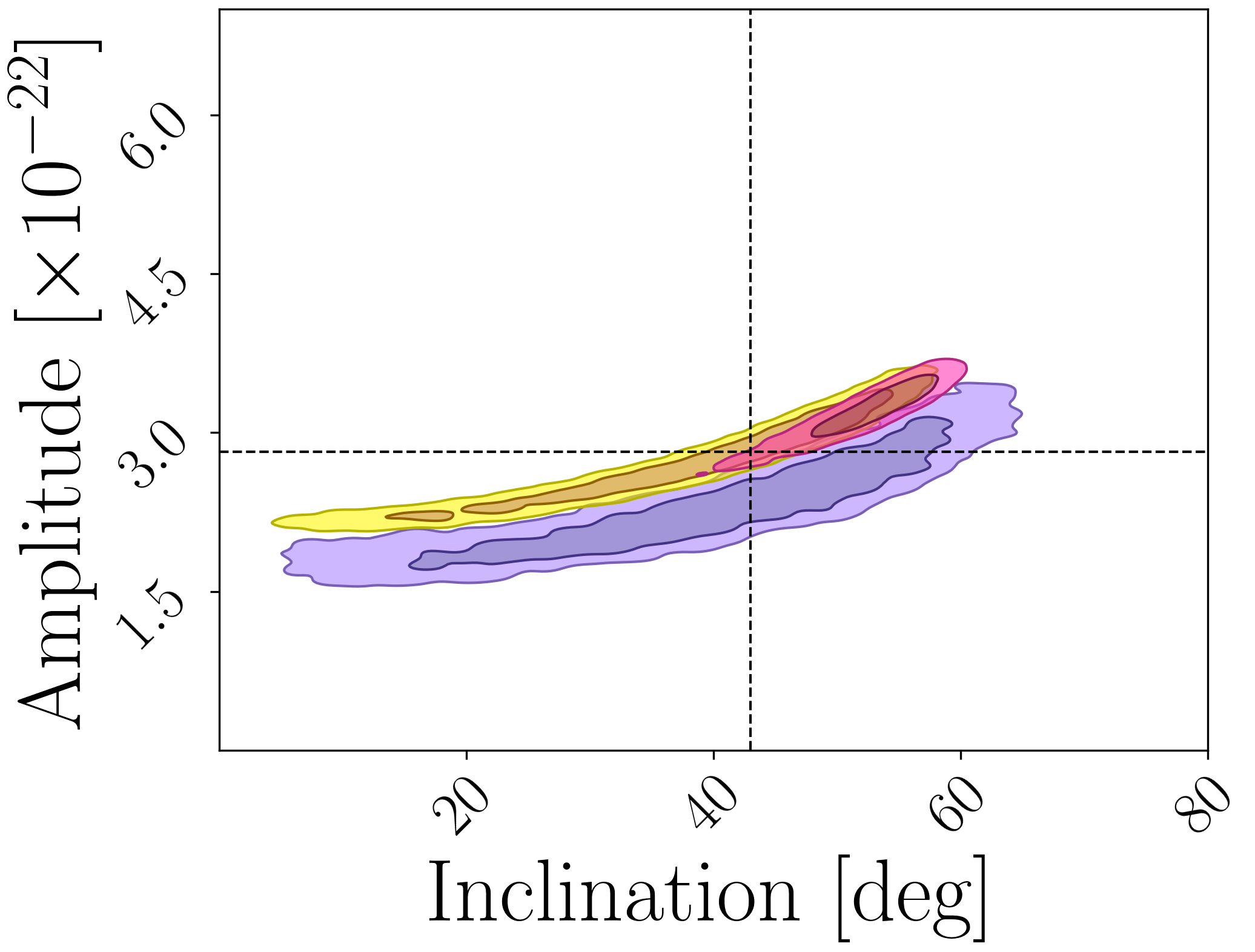}{0.15\textwidth}{{\amcvn}}
          \fig{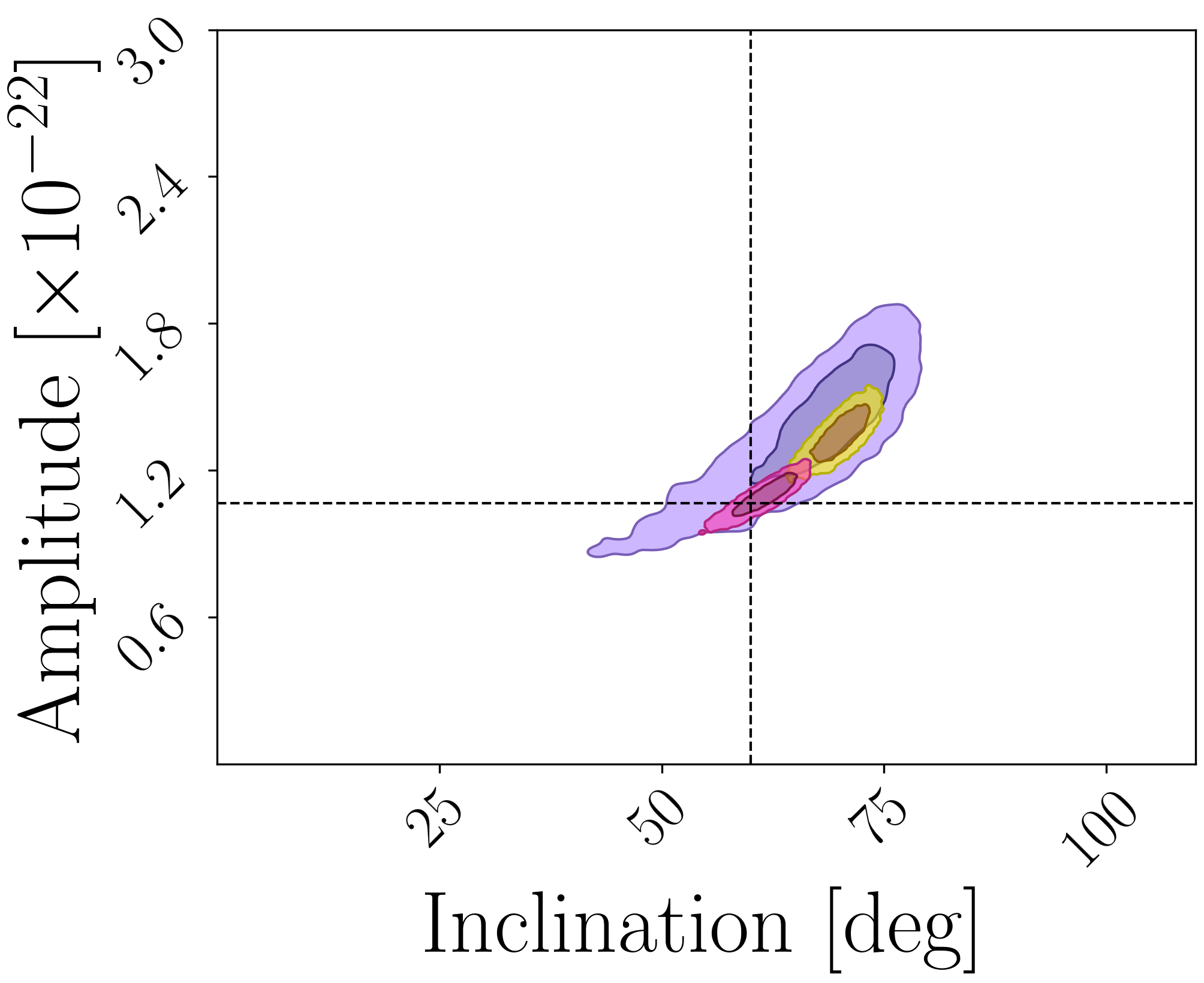}{0.15\textwidth}{{\escet}}
          \fig{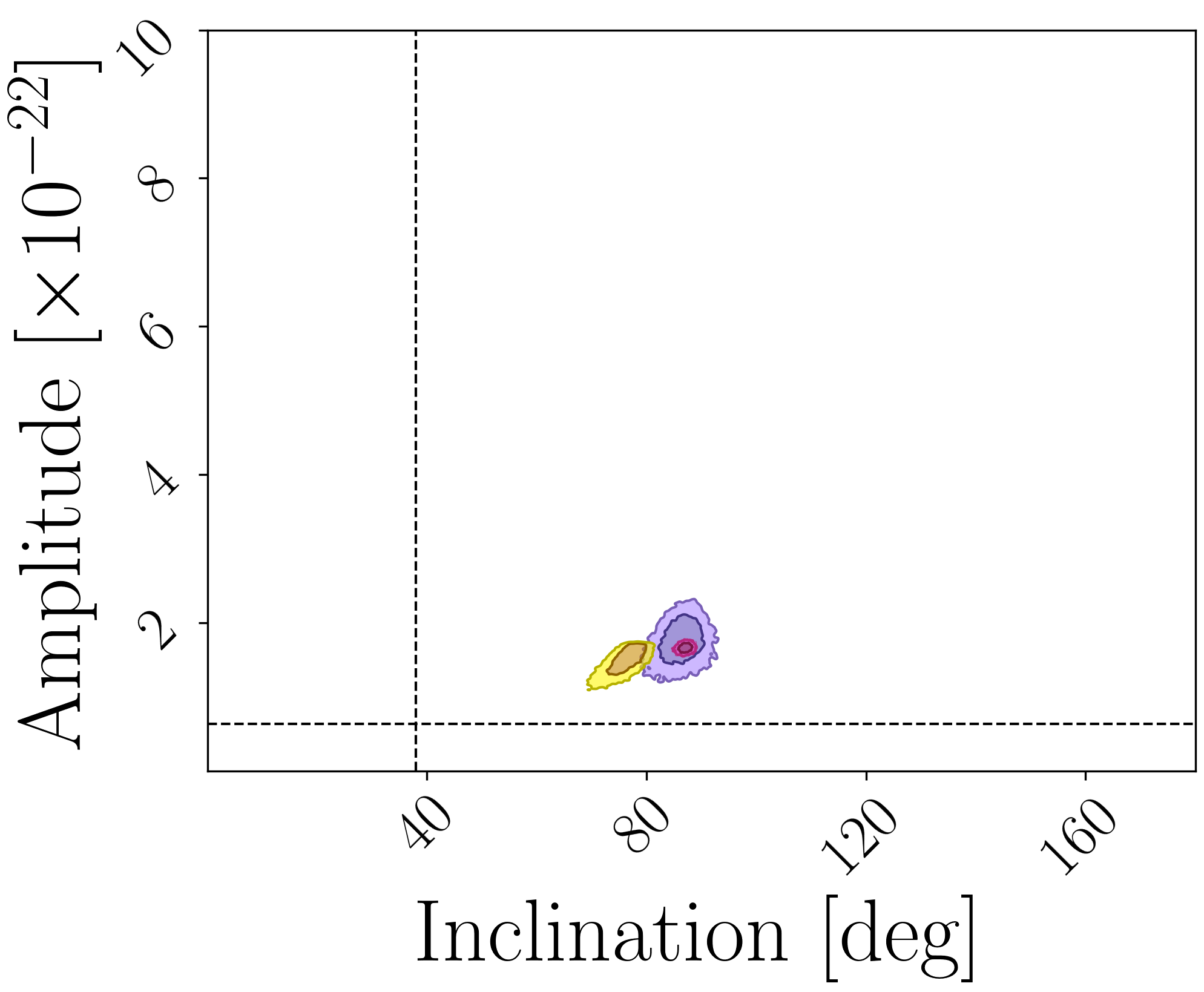}{0.15\textwidth}{{\hmcnc}}}
\gridline{\fig{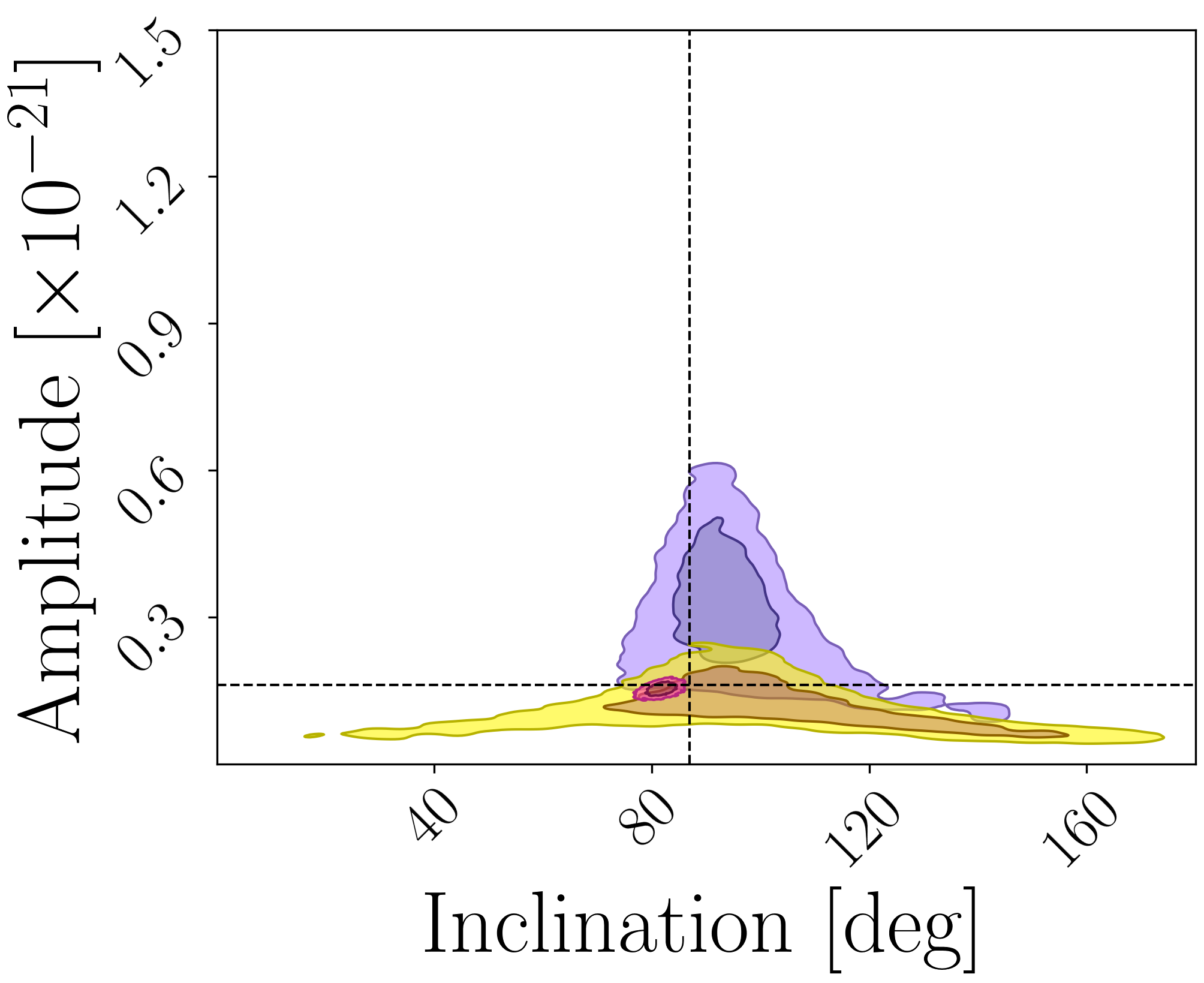}{0.15\textwidth}{{\sdss}}
          \fig{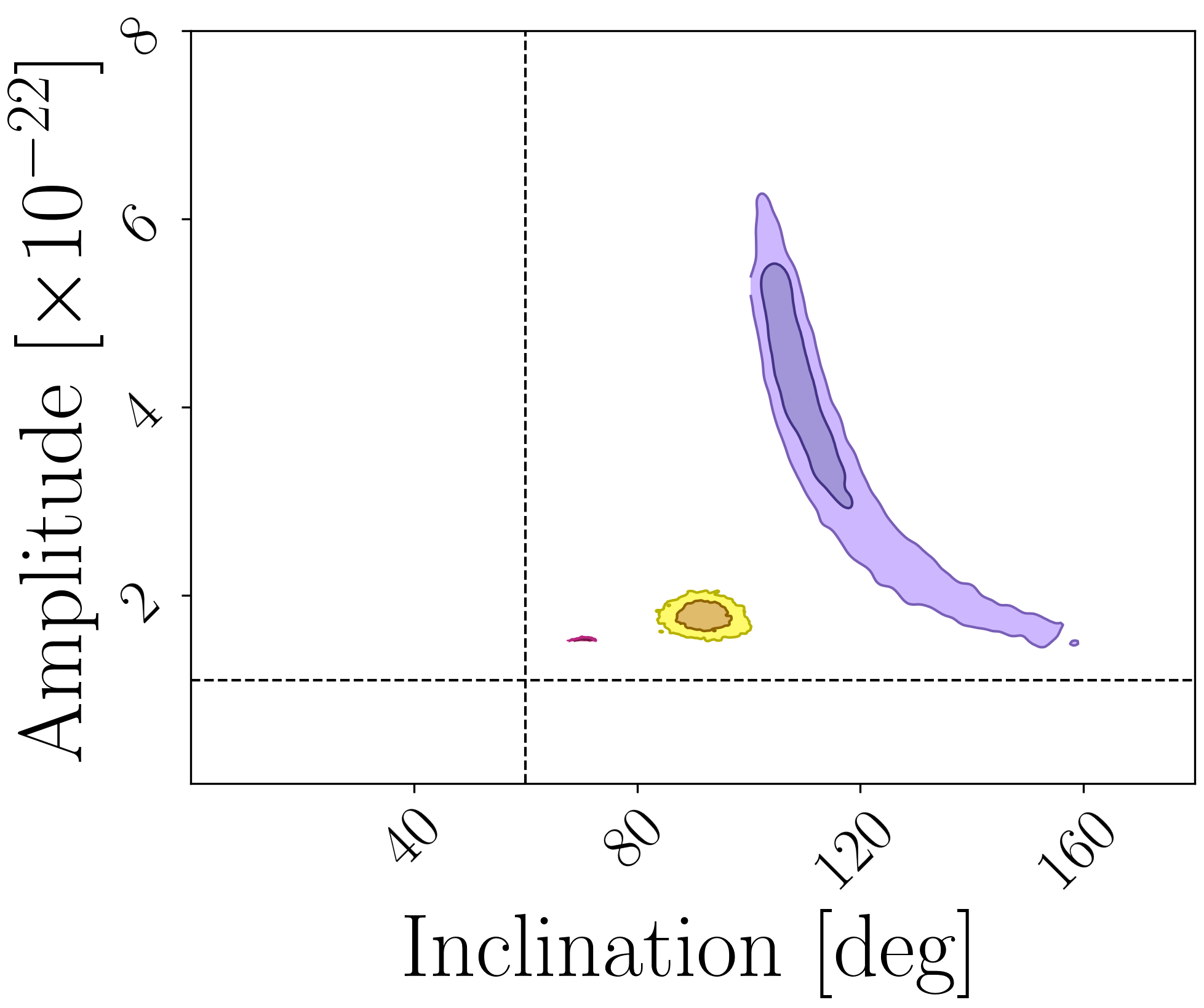}{0.15\textwidth}{{\vul}}
          \fig{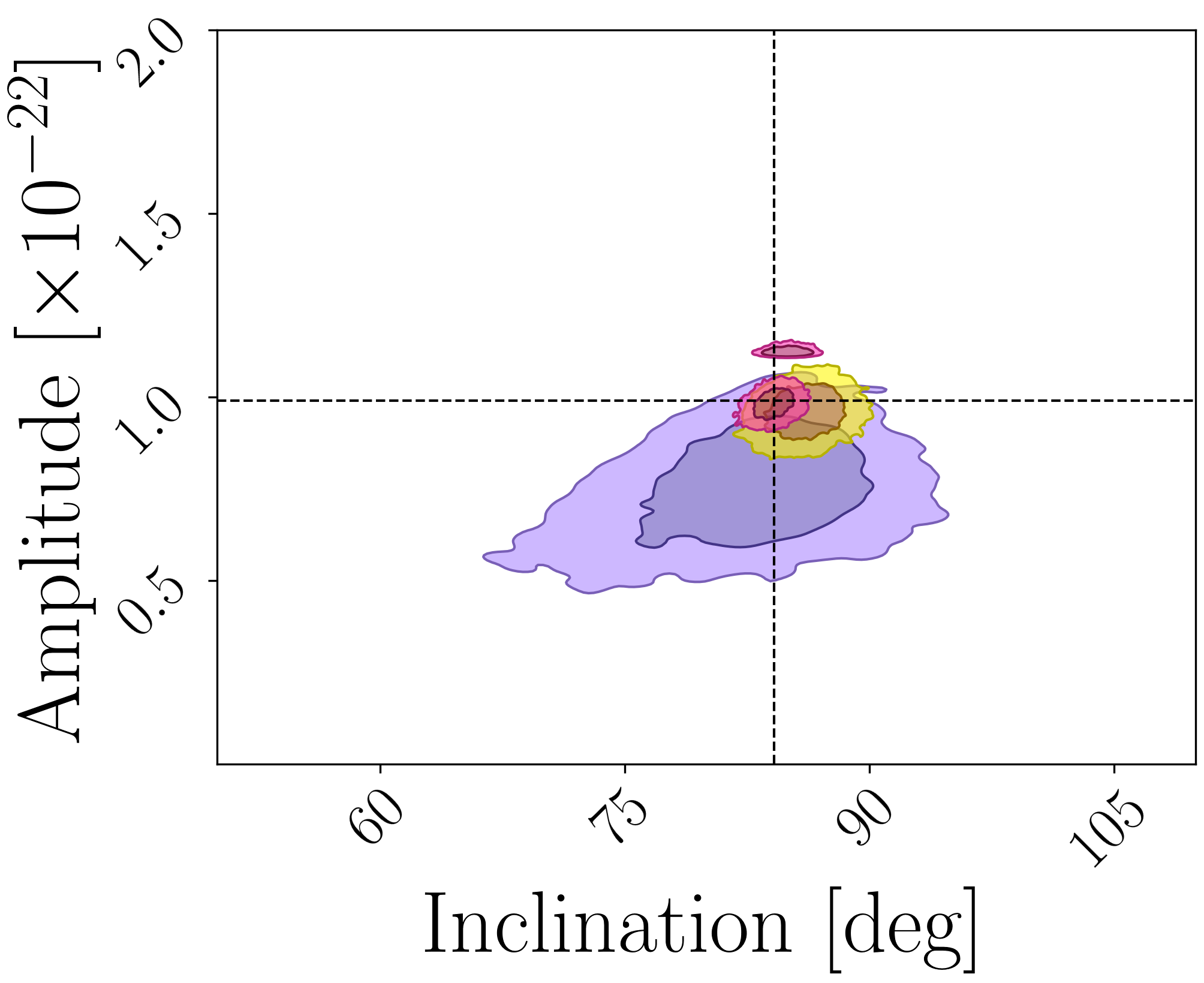}{0.15\textwidth}{{\ztf}}}
\caption{Same as Fig~\ref{fig:vgbonly_time_amplitude} but for {\vbgalaxy} model.}
\label{fig:time_amplitude}
\end{figure}

When the same analysis is performed on data that includes the rest of the galaxy ({\vbgalaxy}) problems arise. 
Figure~\ref{fig:time_amplitude} shows the amplitude parameters for the {\vbgalaxy} analysis.  
Here we find significant inconsistencies over time for {\hmcnc}, {\vul}, and {\escet}, while {\amcvn} is a borderline case.
Even without the benefit of knowing the true source parameters these results could be misinterpreted as inconsistency in the quality of the science data.
In fact the differences are due to mis-modeling of the data resulting in contamination of the recovered signals by other, \emph{a priori} undiscovered, galactic sources.

\begin{figure}
\gridline{\fig{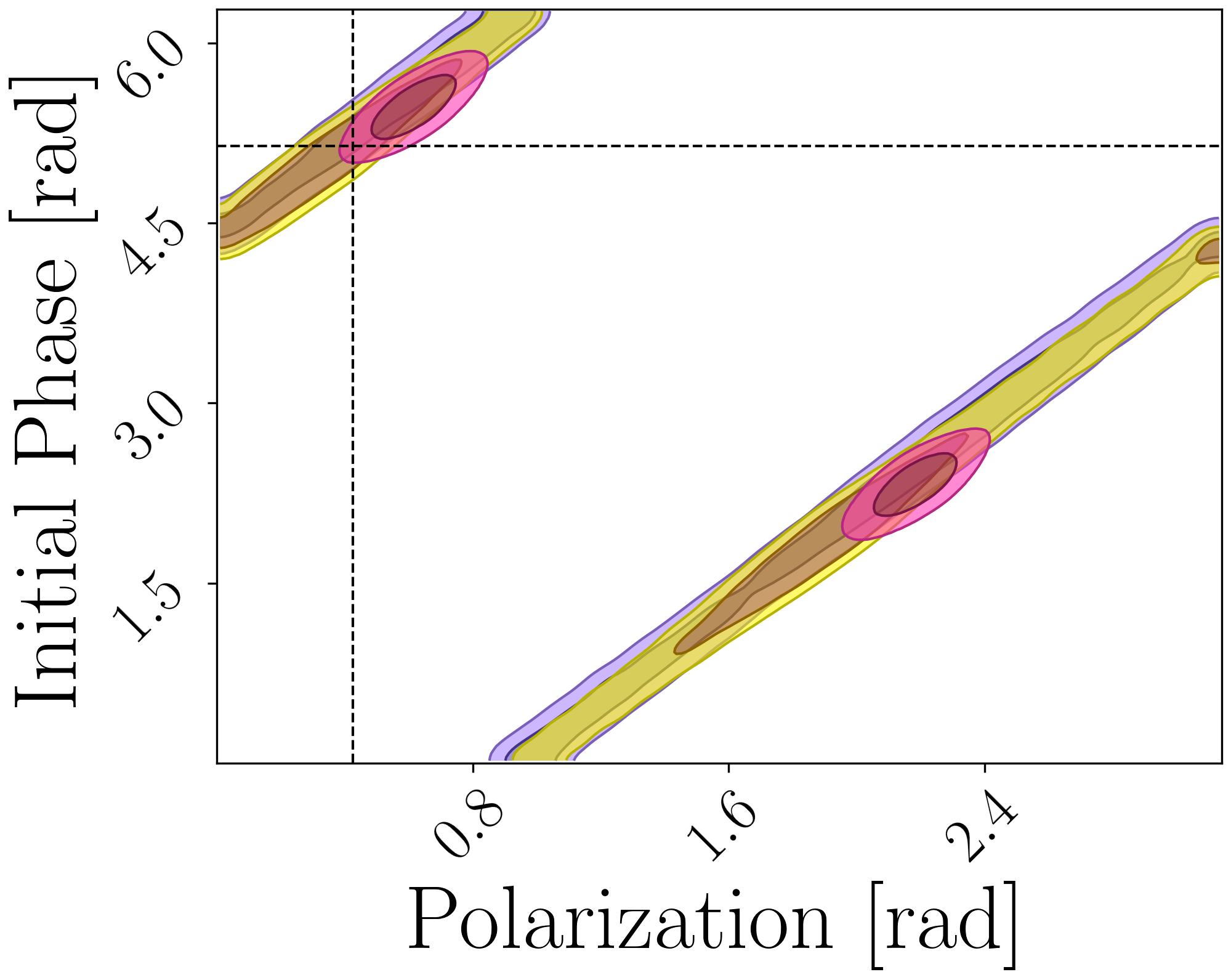}{0.15\textwidth}{{\amcvn}}
          \fig{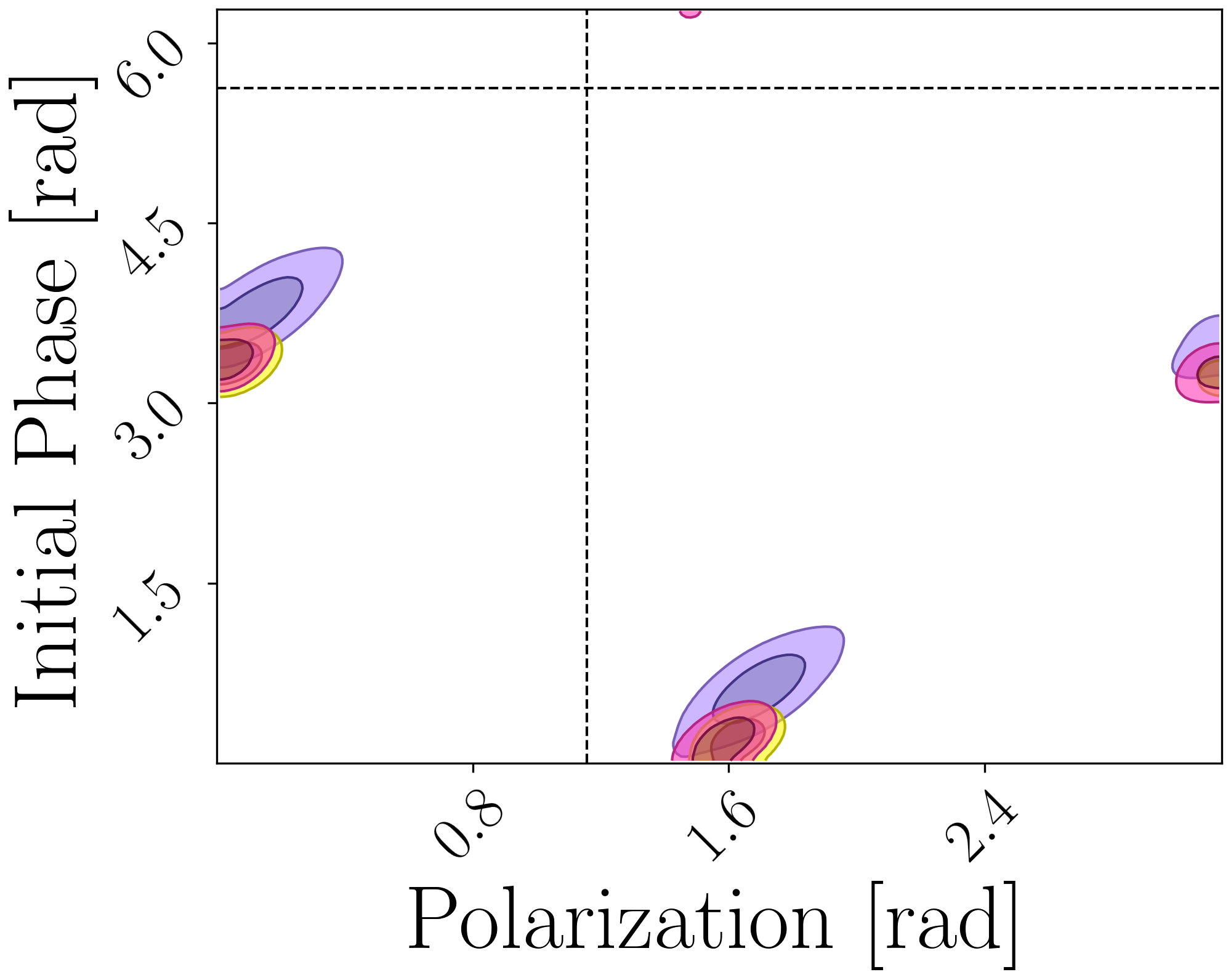}{0.15\textwidth}{{\escet}}
          \fig{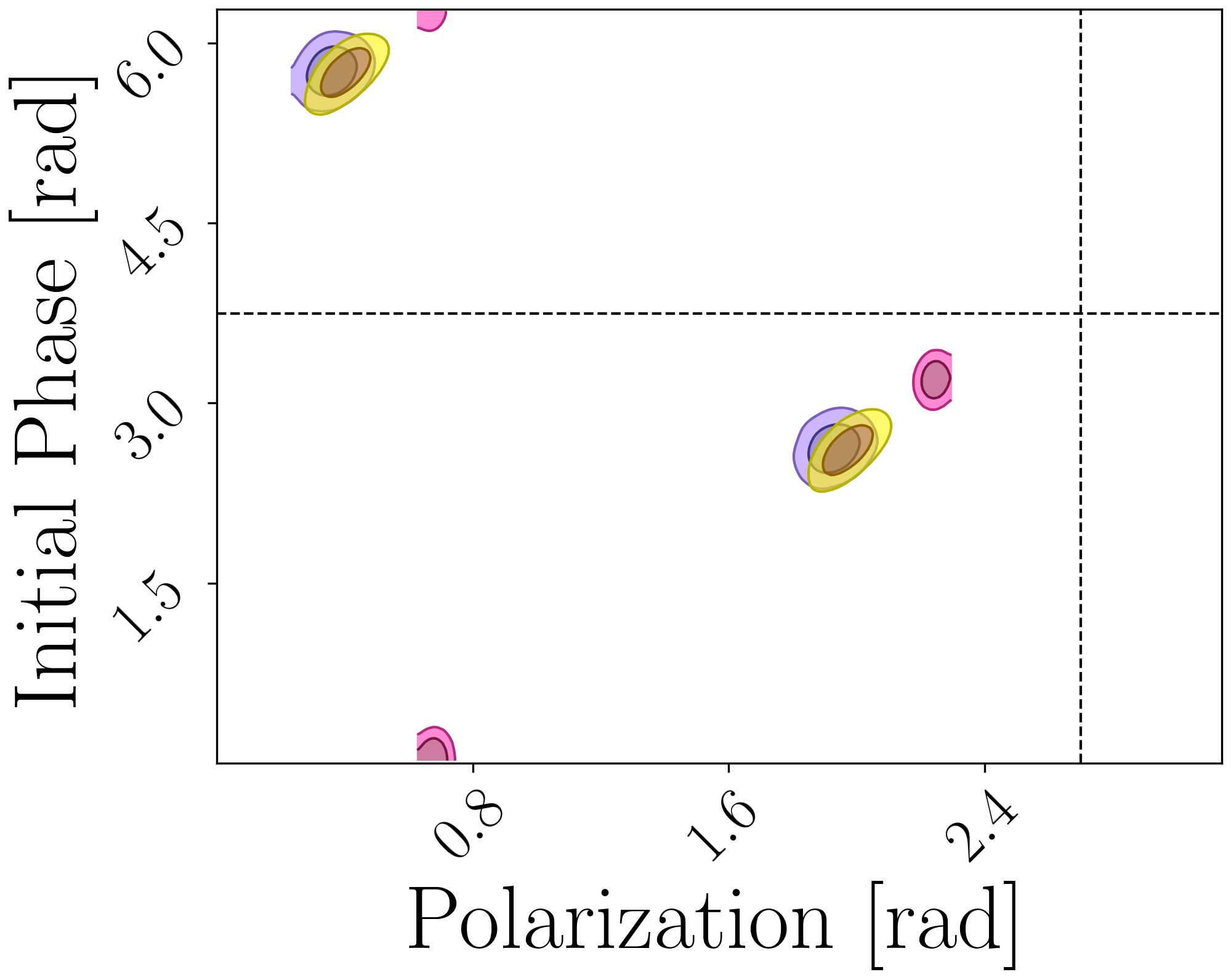}{0.15\textwidth}{{\hmcnc}}}
\gridline{\fig{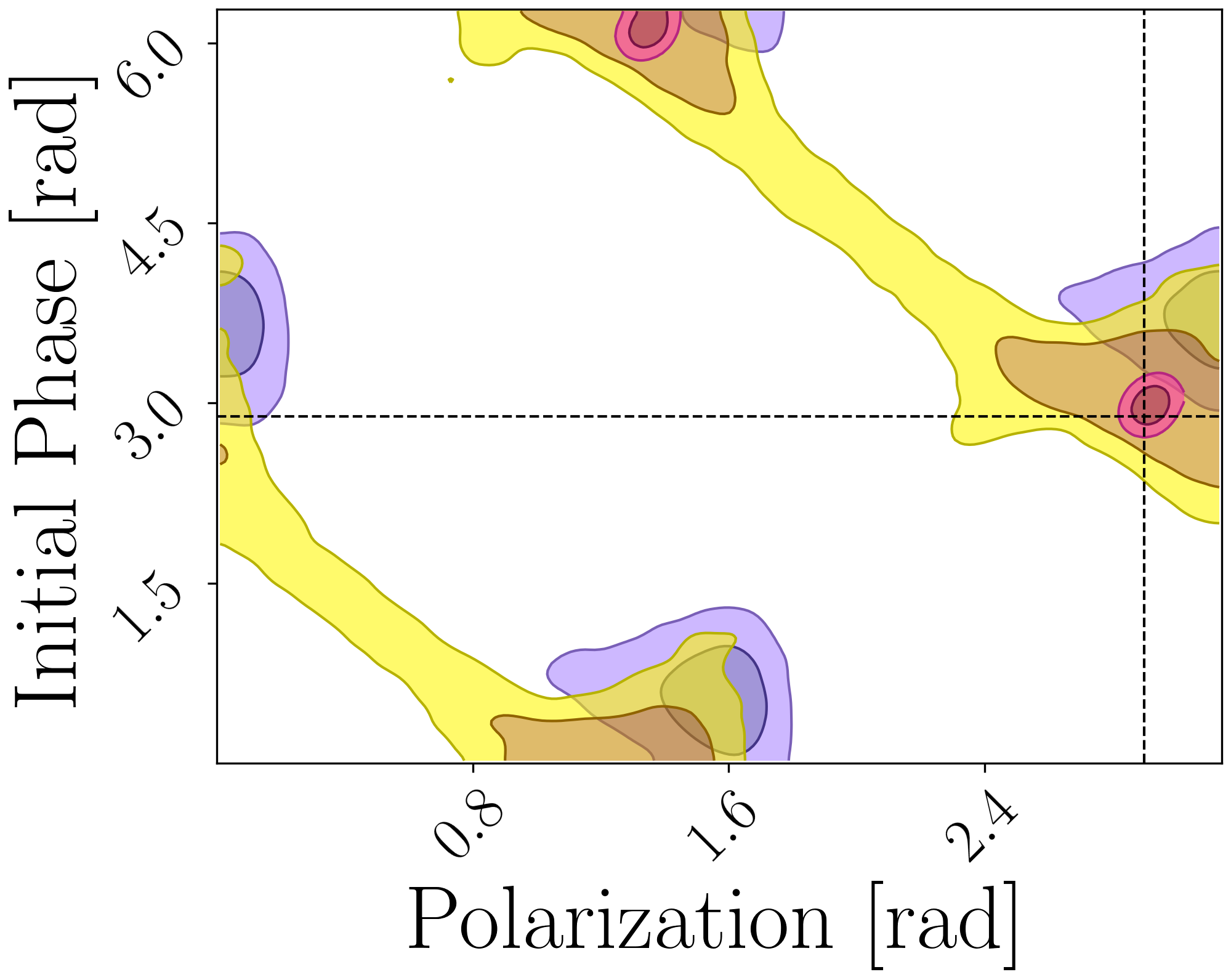}{0.15\textwidth}{{\sdss}}
          \fig{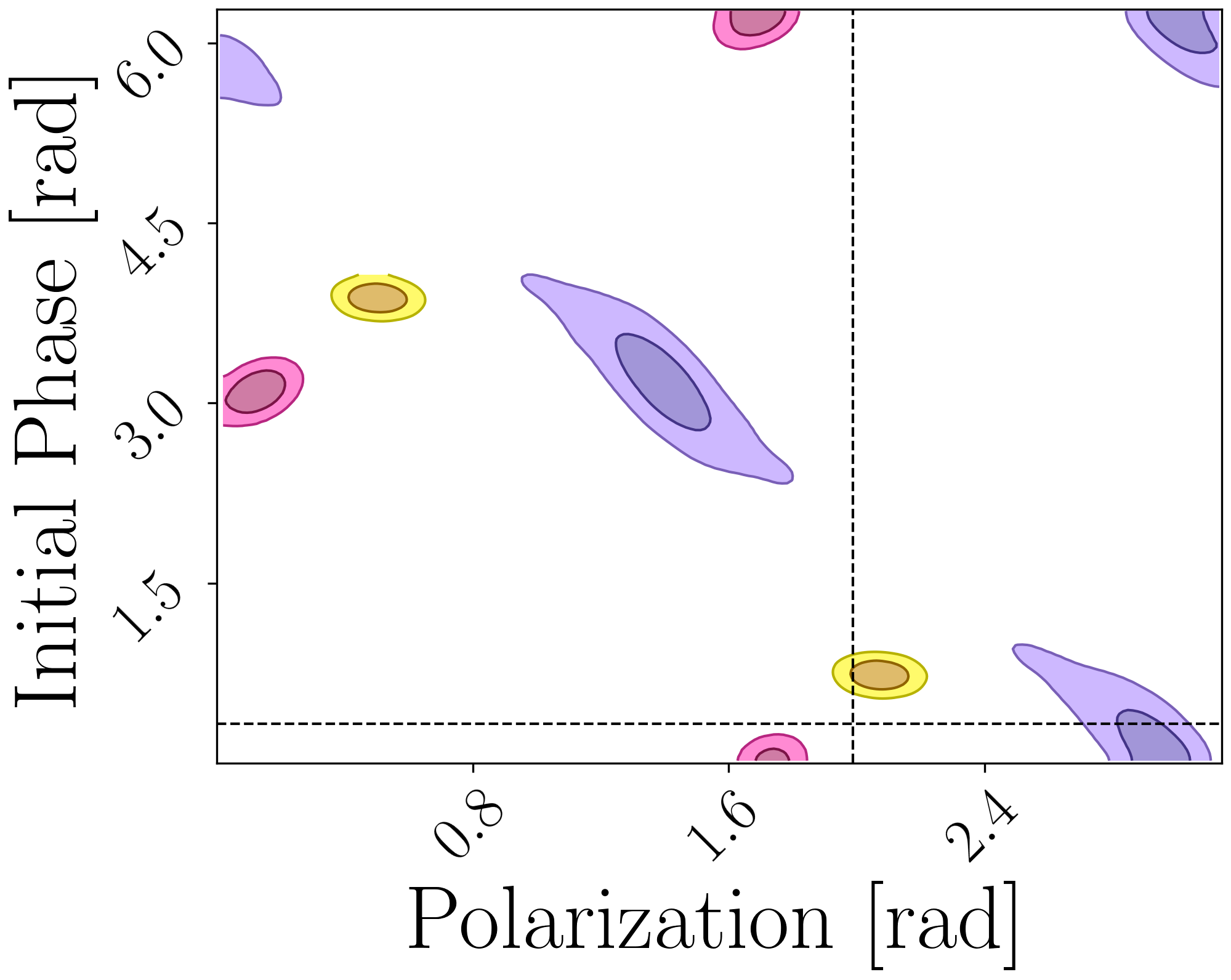}{0.15\textwidth}{{\vul}}
          \fig{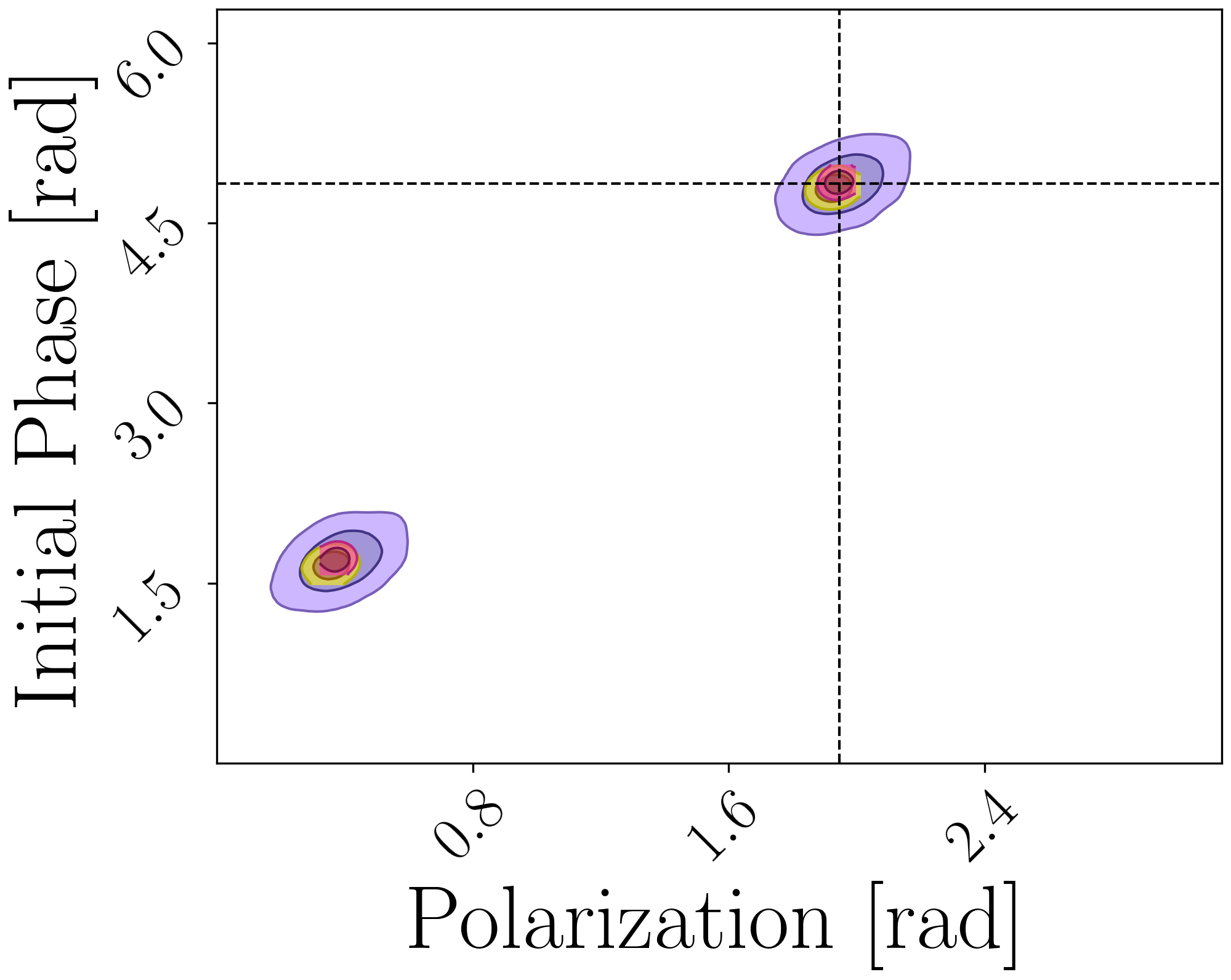}{0.15\textwidth}{{\ztf}}}
\caption{Same as Fig~\ref{fig:vgbonly_time_phase} but for {\vbgalaxy} model.}
\label{fig:time_phase}
\end{figure}

\subsection{Scenario 3}

The final observing scenario we consider exploits the short integration times needed for detection of these six known binaries, by dividing 1 month of data into four week-long segments that are analyzed independently and compared for consistency.
Results from each week-long analysis are color coded in order of week as purple, yellow, magenta, and blue.
Here we do not expect the results to be identical even in the ideal case. 
Each week-long segment will have a different realization of the instrument/foreground noise so the location of the maximum likelihood point (i.e. the centroid of the distribution) will shift by ${\mathcal O}(1\sigma)$ from week to week.
Furthermore, the S/N in each week will not be constant due to the changing orientation of the LISA antenna pattern relative to the source location as the detector moves through its orbit.

\begin{figure}
\gridline{\fig{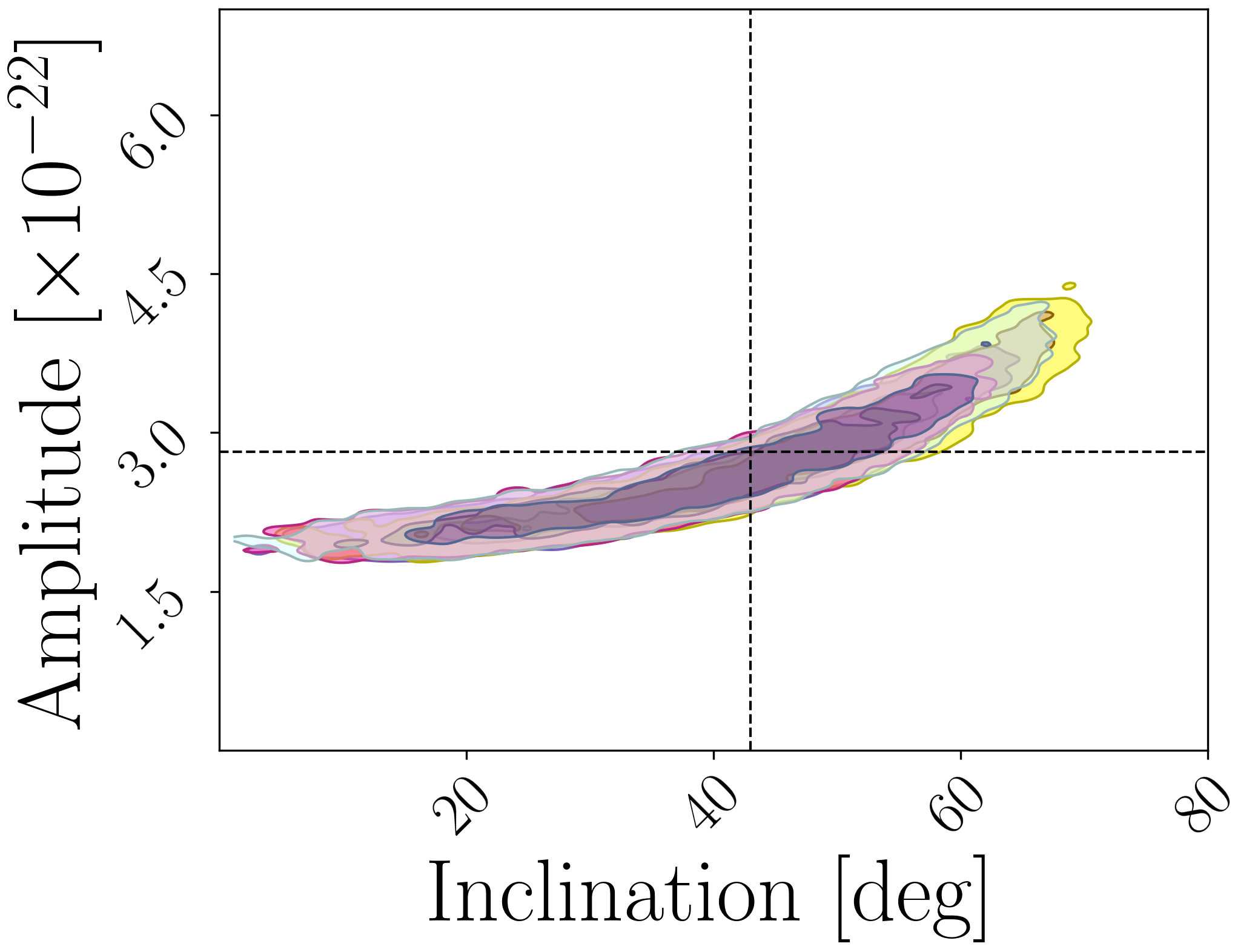}{0.15\textwidth}{{\amcvn}}
          \fig{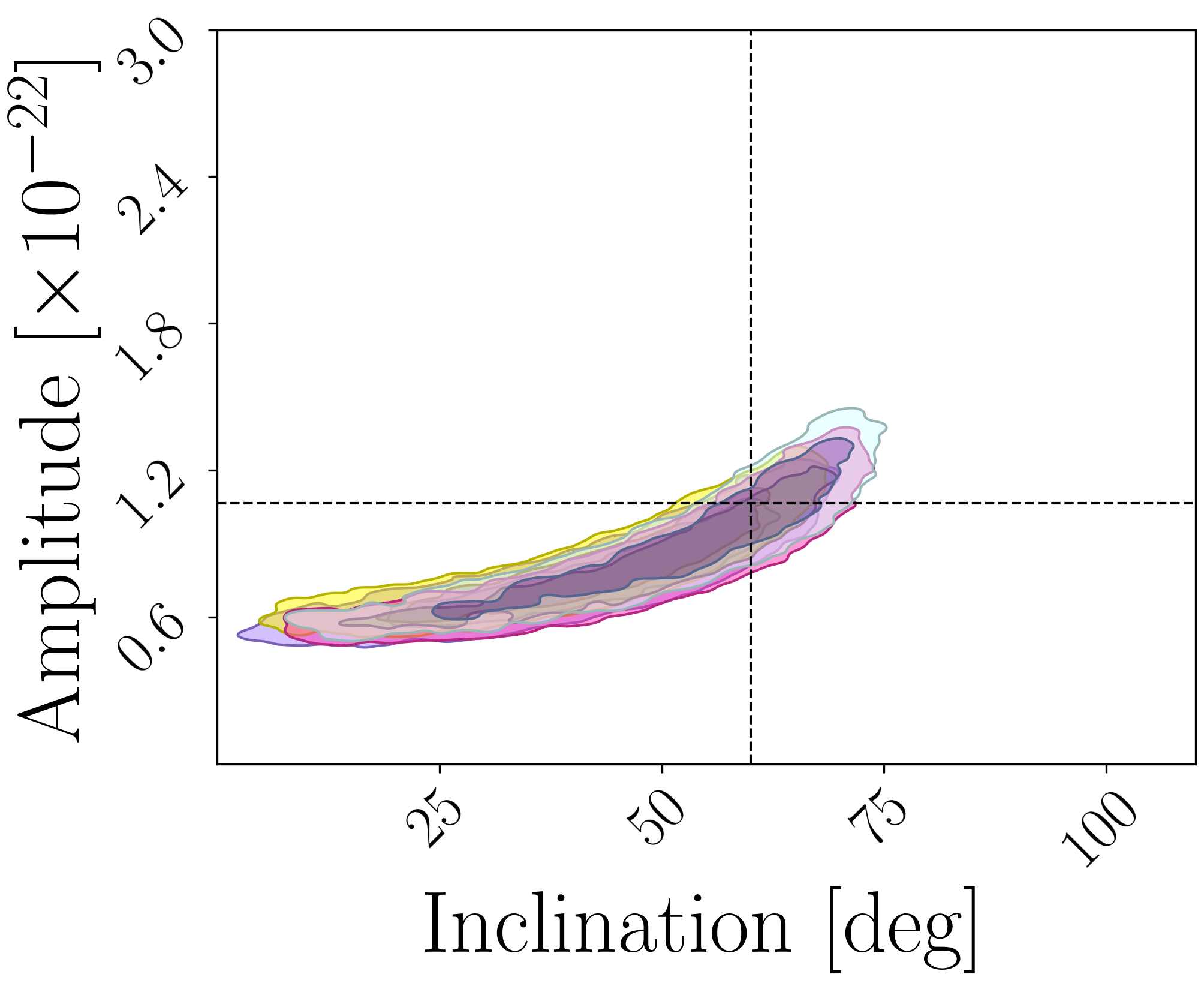}{0.15\textwidth}{{\escet}}
          \fig{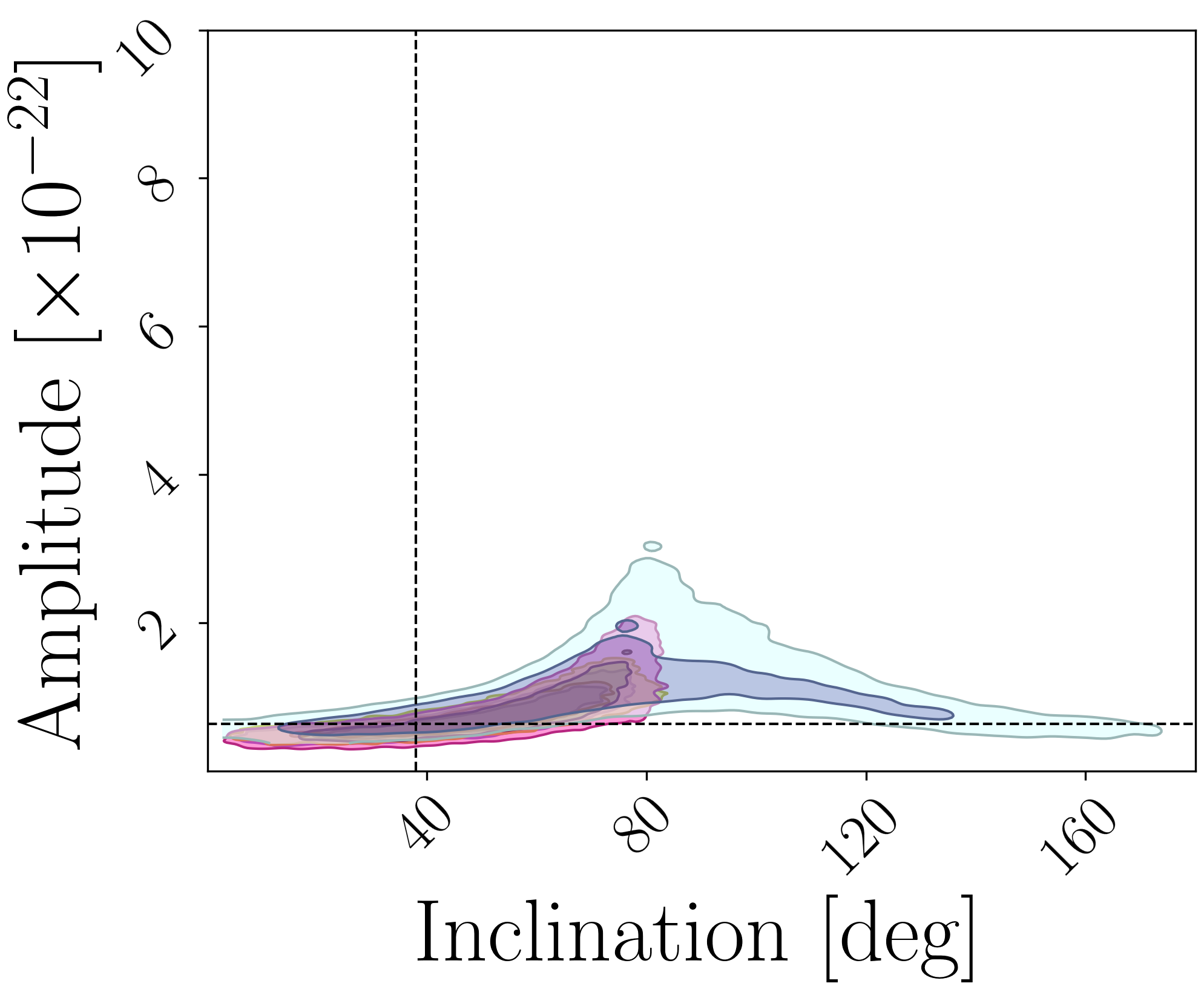}{0.15\textwidth}{{\hmcnc}}}
\gridline{\fig{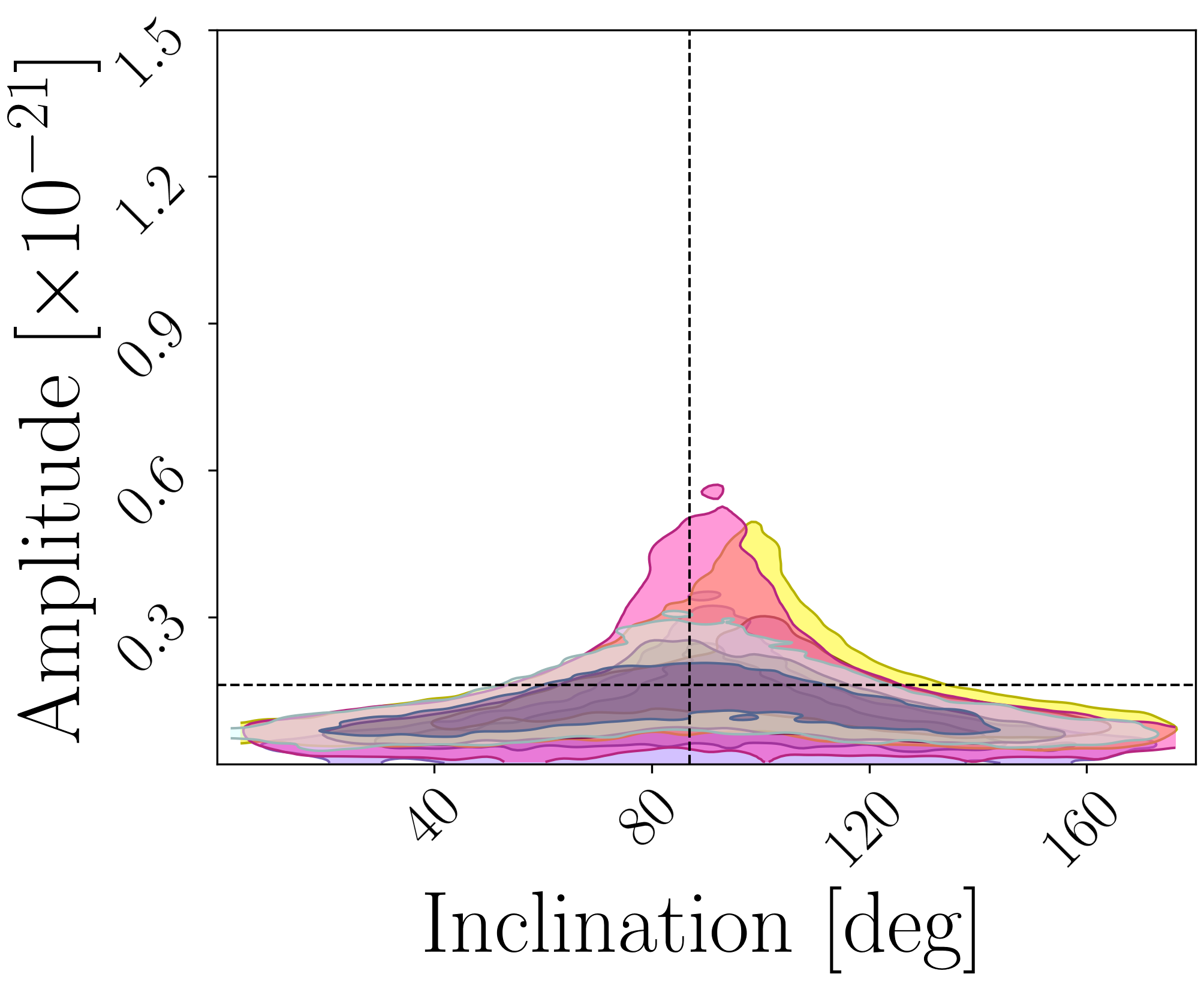}{0.15\textwidth}{{\sdss}}
          \fig{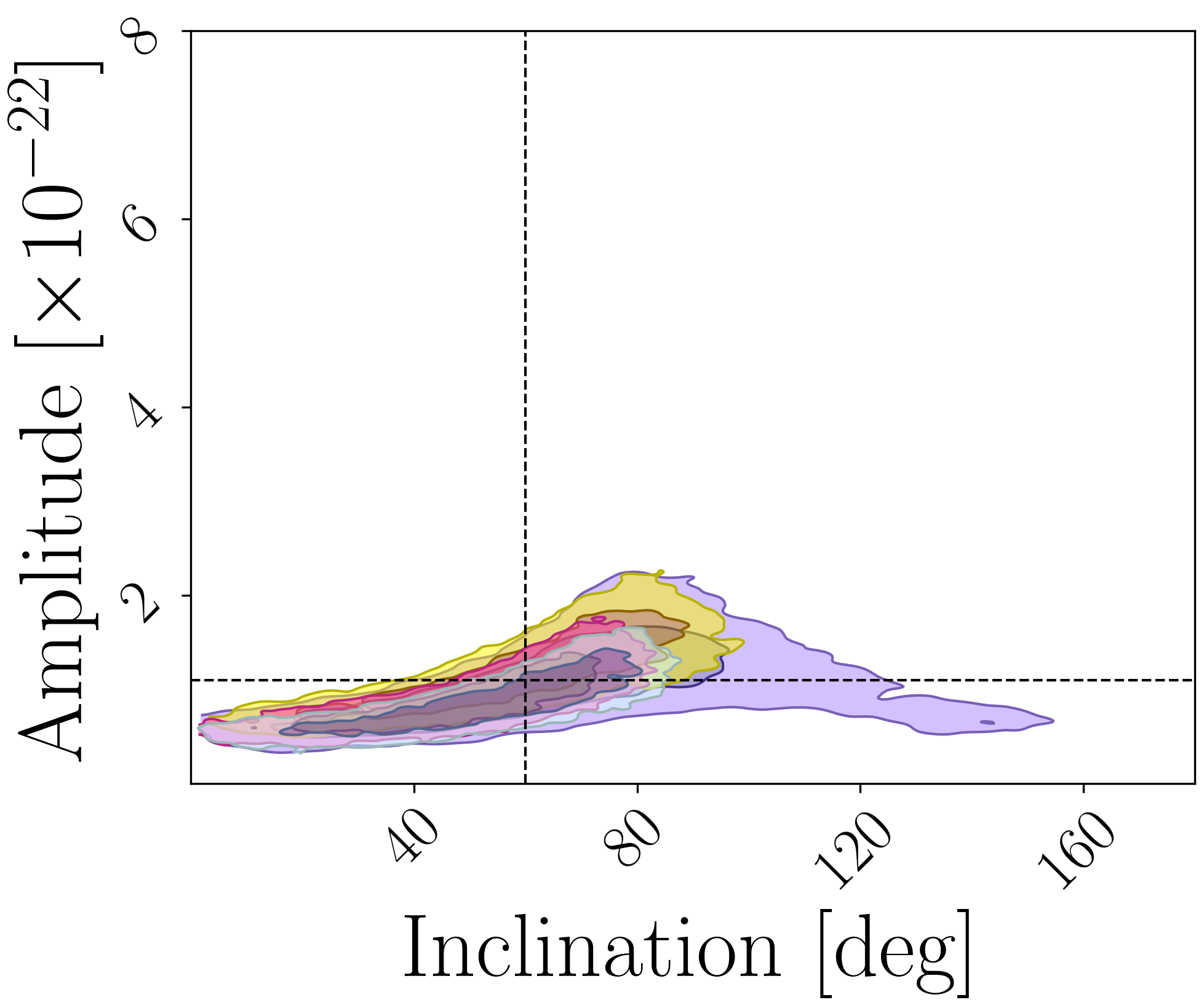}{0.15\textwidth}{{\vul}}
          \fig{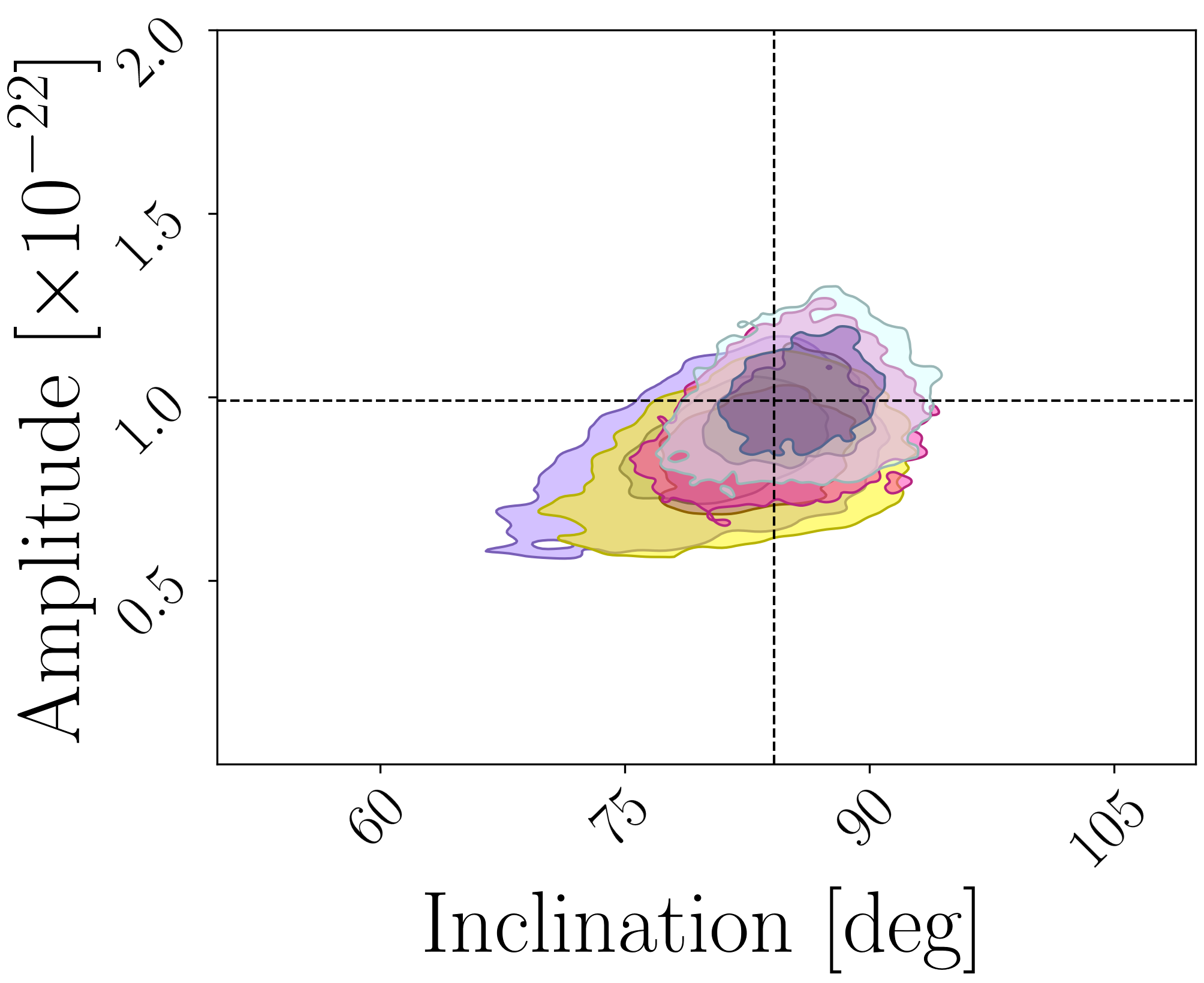}{0.15\textwidth}{{\ztf}}}
\caption{Jackknife tests of amplitude parameters using {\vbonly} model. Different colors represent independently analyzed week-long segments from a 1 month observation time.  Contours demarcate the 1 and 2$\sigma$ credible intervals. The dashed lines indicate the true parameter values for the simulated signal.}
\label{fig:vgbonly_by_week_amplitude}
\end{figure}

\begin{figure}
\gridline{\fig{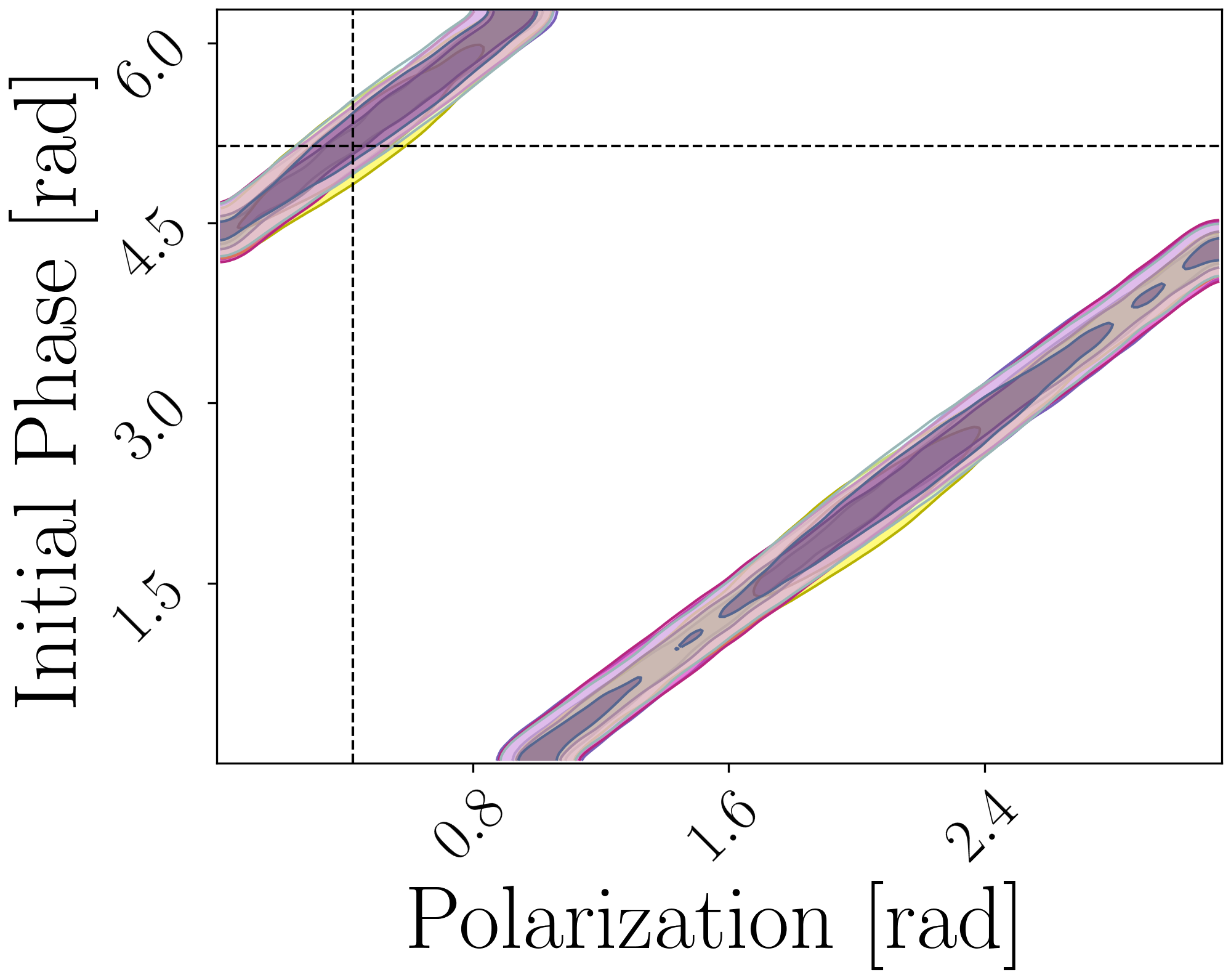}{0.15\textwidth}{{\amcvn}}
          \fig{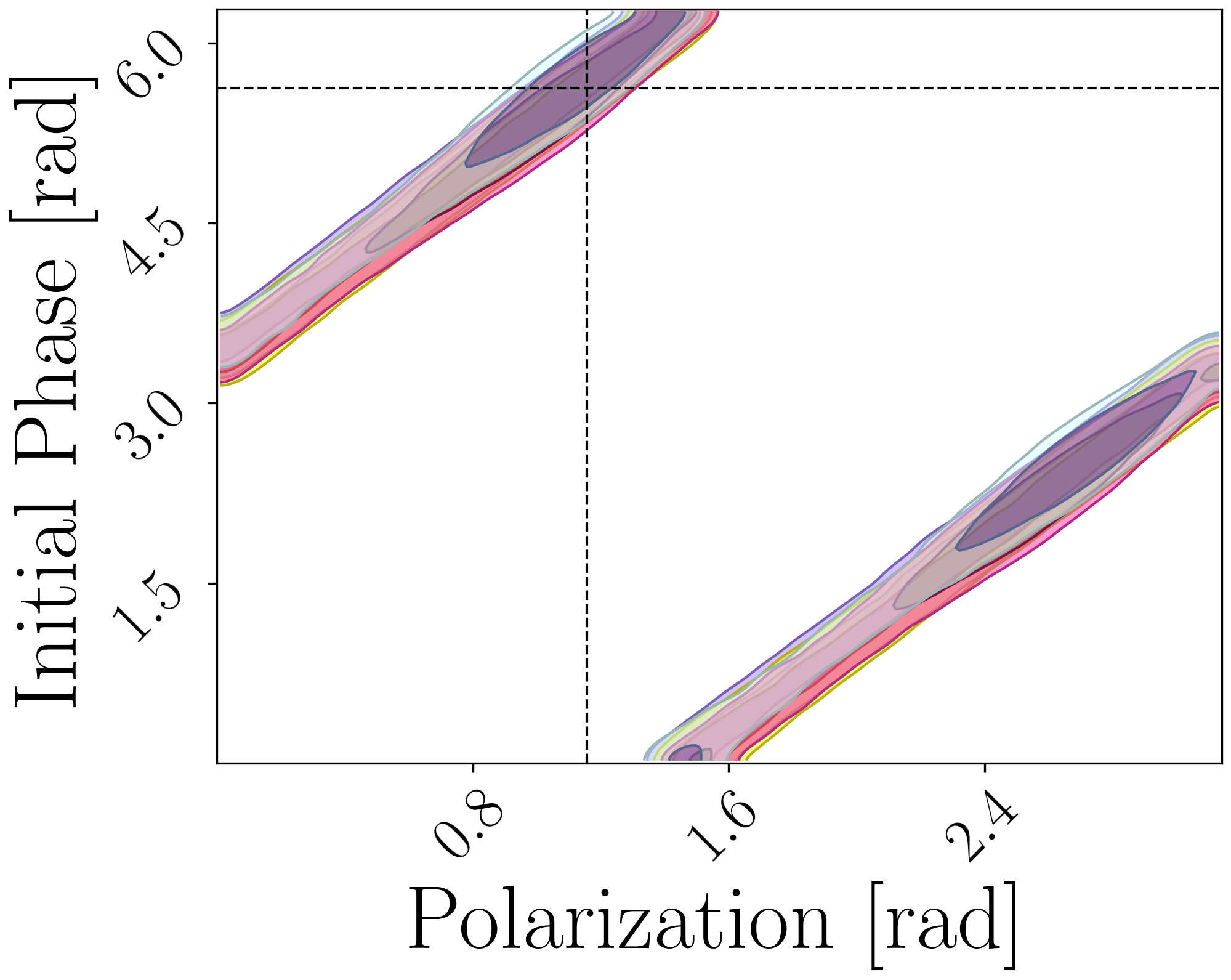}{0.15\textwidth}{{\escet}}
          \fig{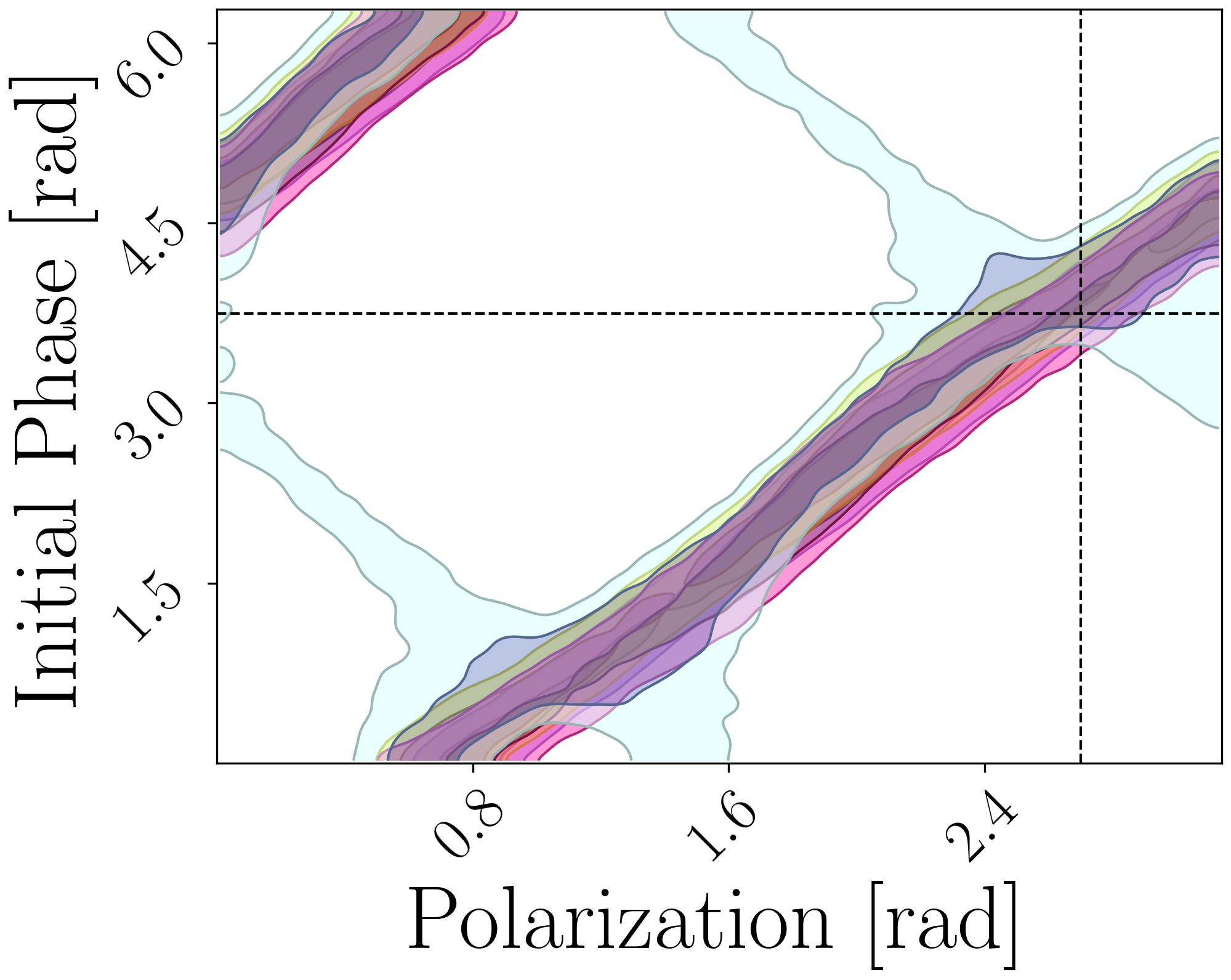}{0.15\textwidth}{{\hmcnc}}}
\gridline{\fig{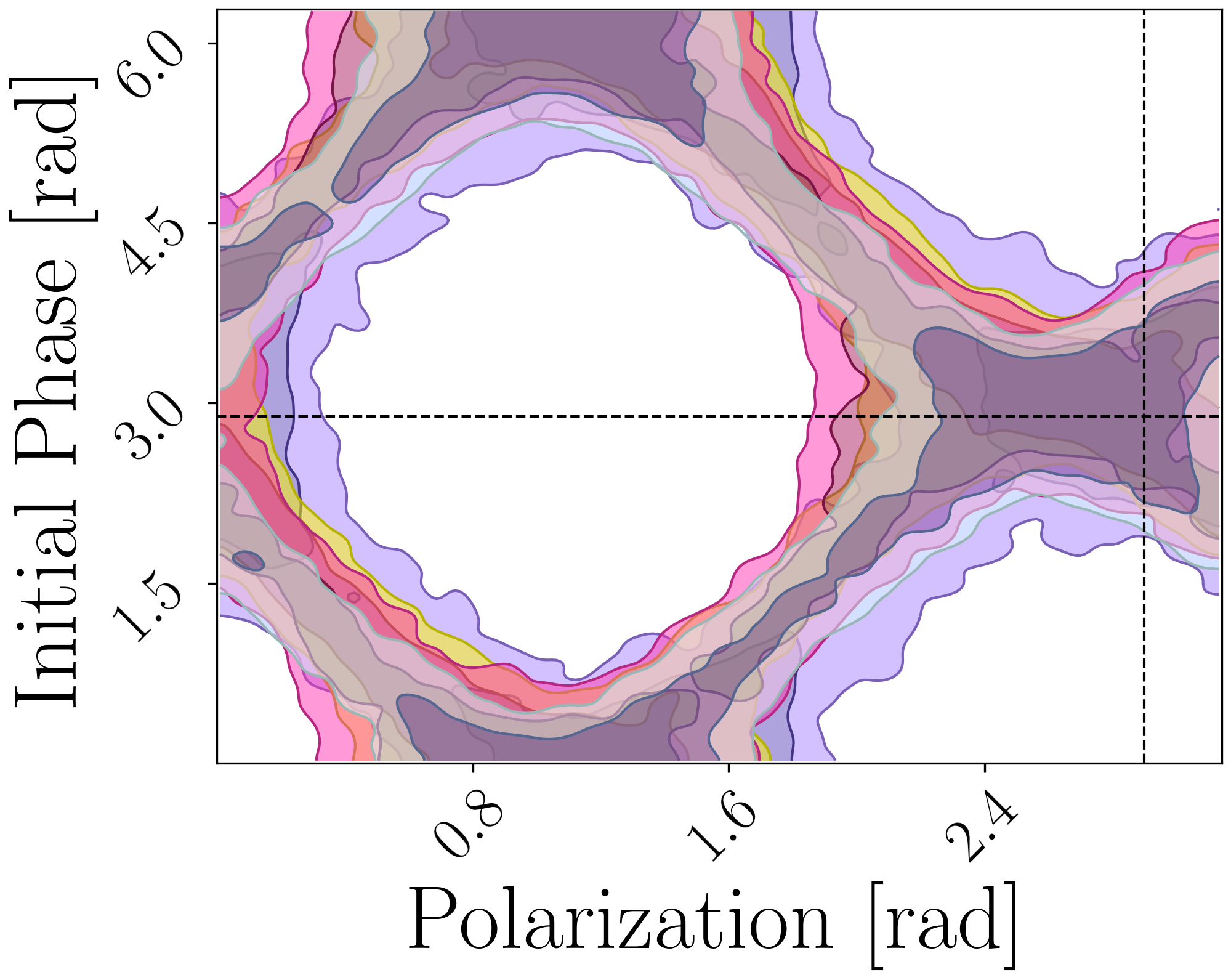}{0.15\textwidth}{{\sdss}}
          \fig{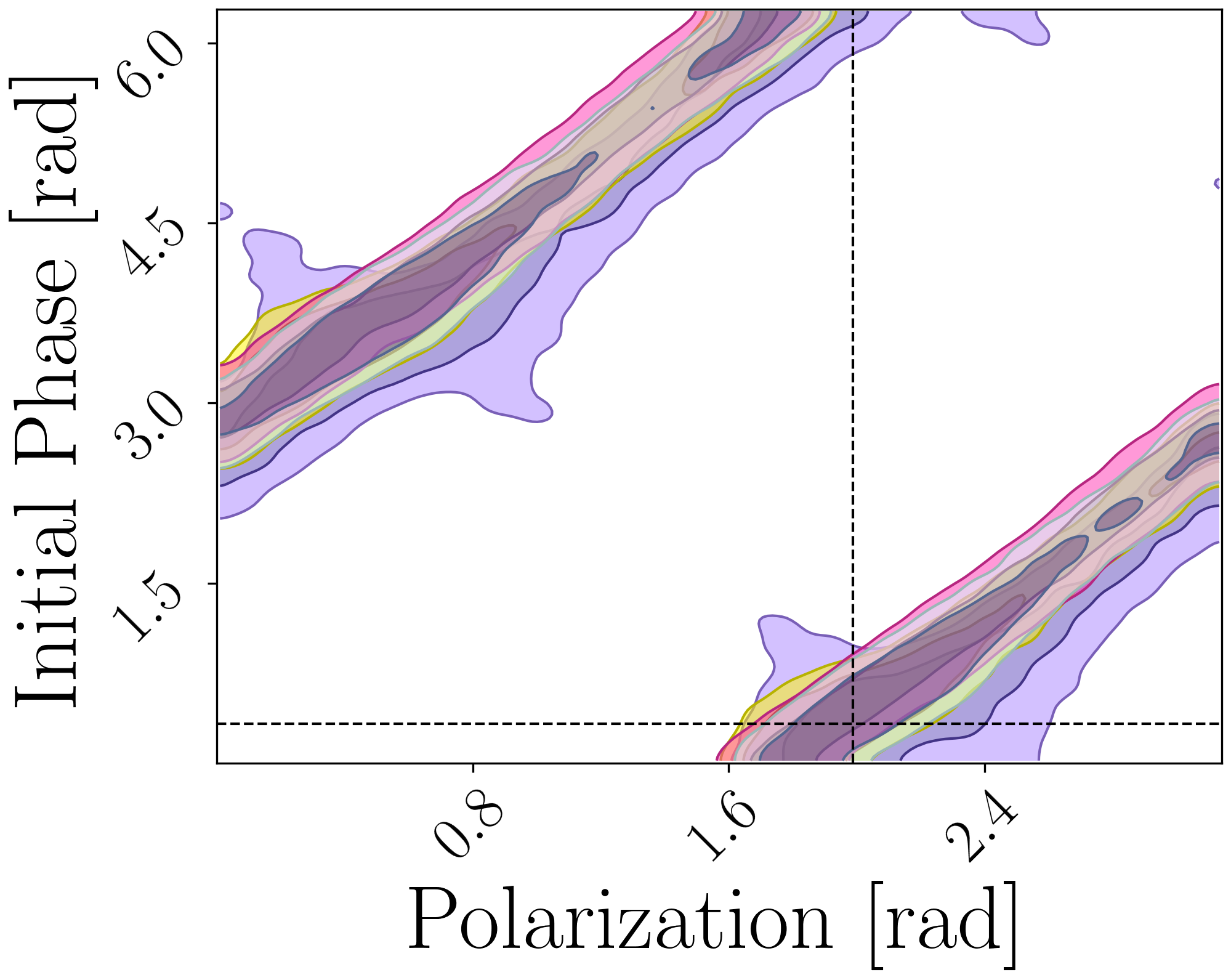}{0.15\textwidth}{{\vul}}
          \fig{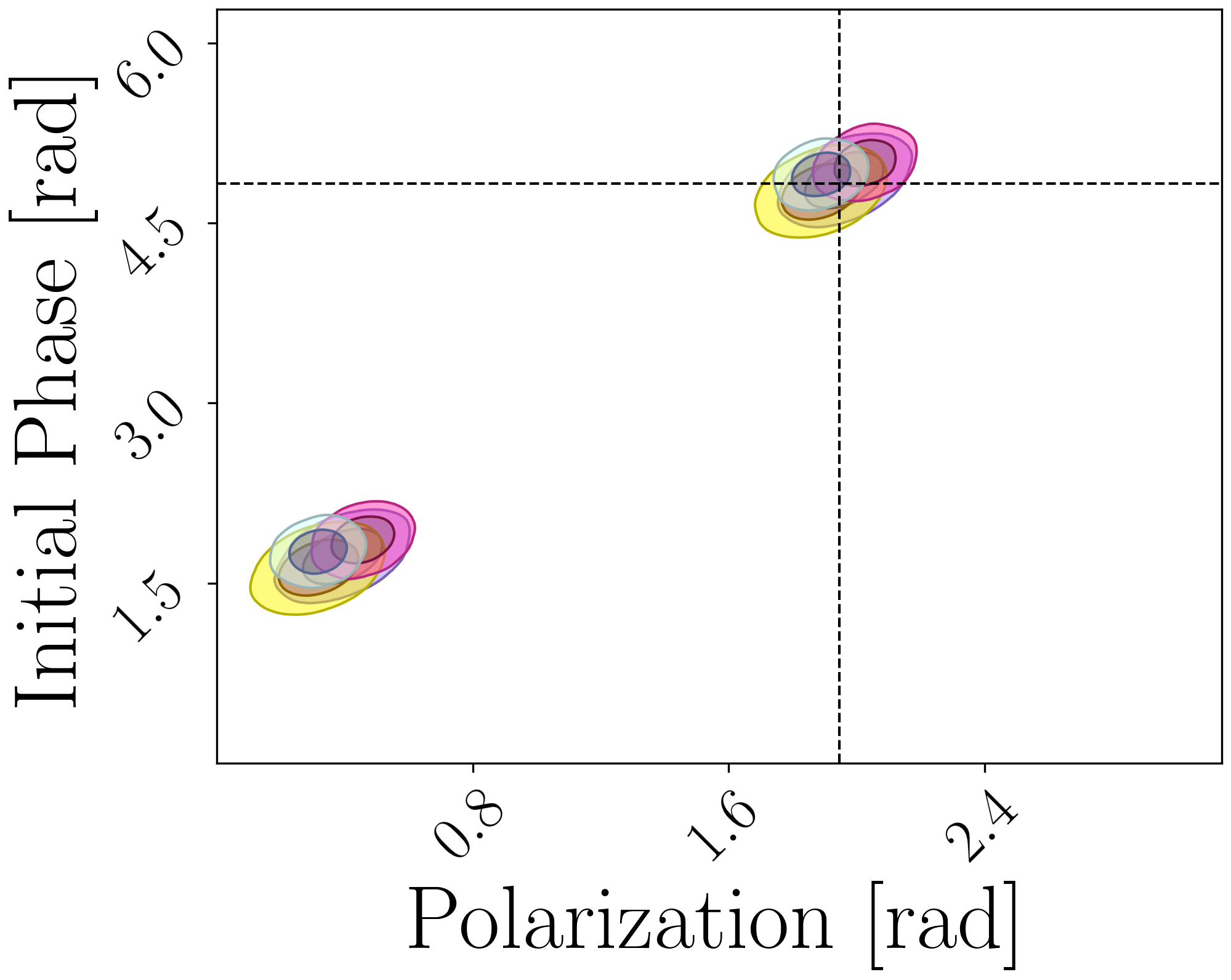}{0.15\textwidth}{{\ztf}}}
\caption{Same as Fig.~\ref{fig:vgbonly_by_week_amplitude} but for phase parameters.}
\label{fig:vgbonly_by_week_phase}
\end{figure}

As before we begin with the {\vbonly} analysis that only includes the known binaries in the data simulation to establish a baseline expectation for what we would consider a successful test outcome. 
As predicted we see shifts in the locations of the contours by less than their typical size, and slight differences in their sizes from week to week. 
The amplitude parameter inferences of {\hmcnc} and {\vul} in Fig.~\ref{fig:vgbonly_by_week_amplitude} are good examples of this latter effect, while the the phase parameter inferences for {\ztf} in Fig.~\ref{fig:vgbonly_by_week_phase} are a particularly clean example of the slight shifts in the location of the posteriors due to the different noise realization.

\begin{figure}
\gridline{\fig{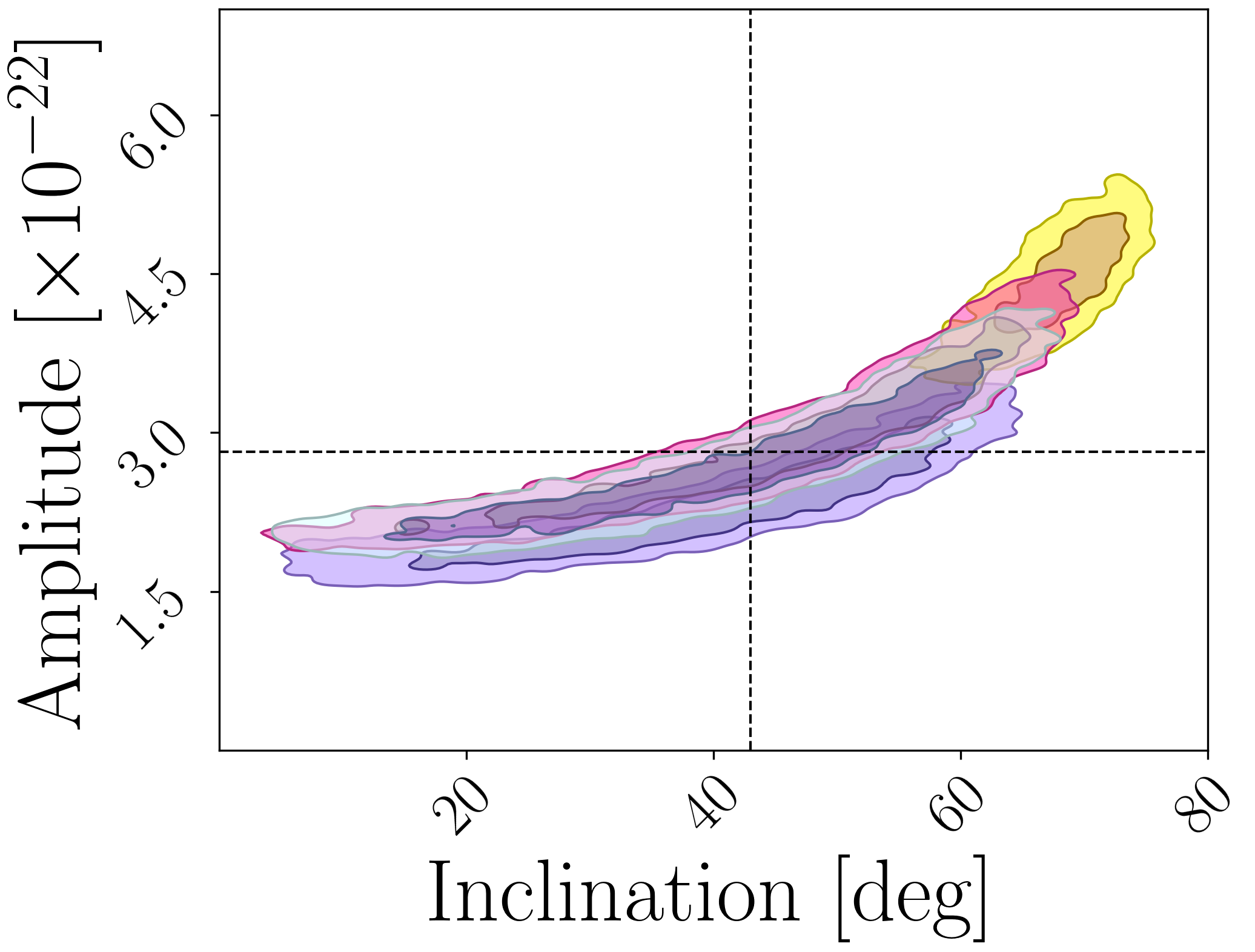}{0.15\textwidth}{{\amcvn}}
          \fig{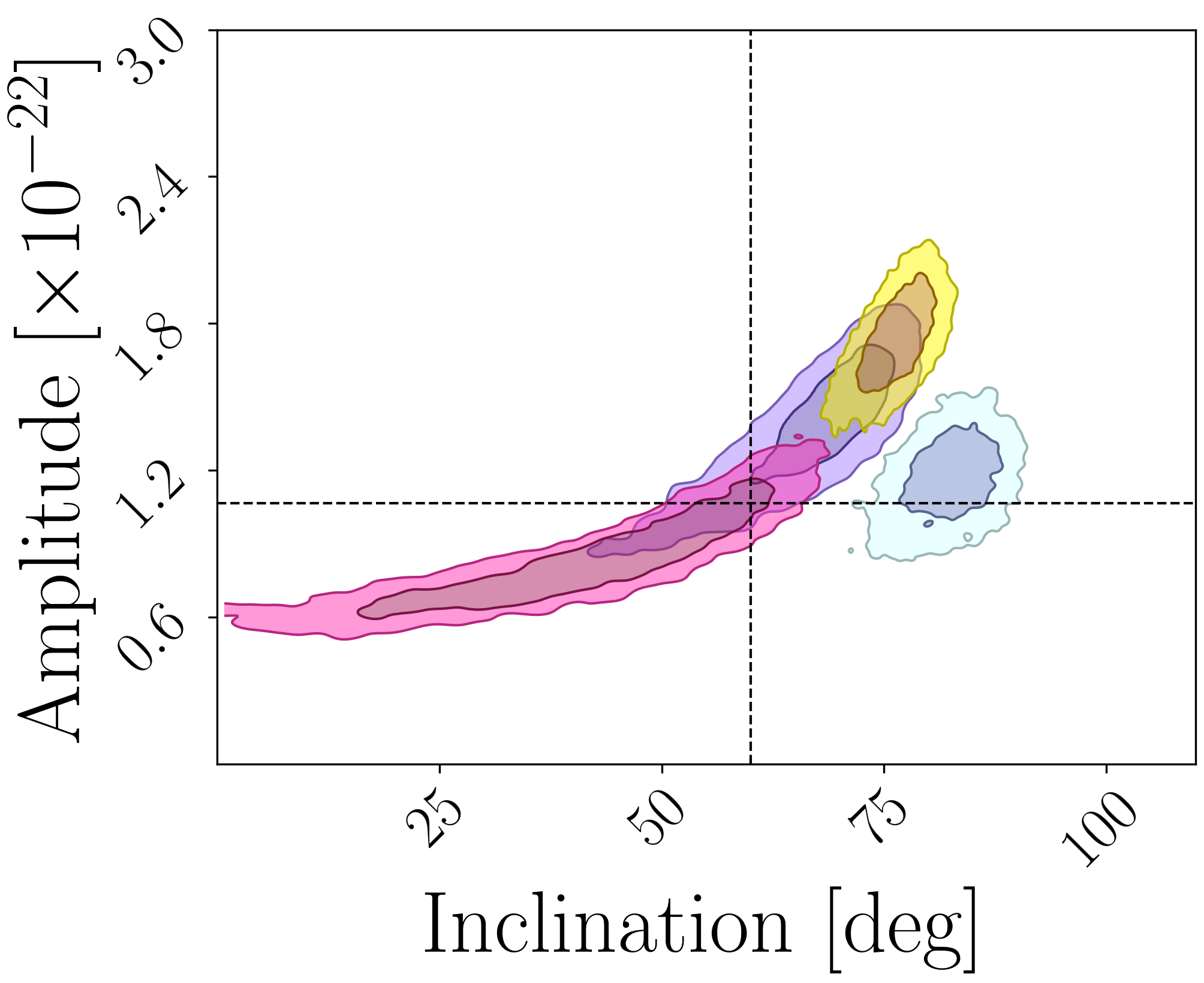}{0.15\textwidth}{{\escet}}
          \fig{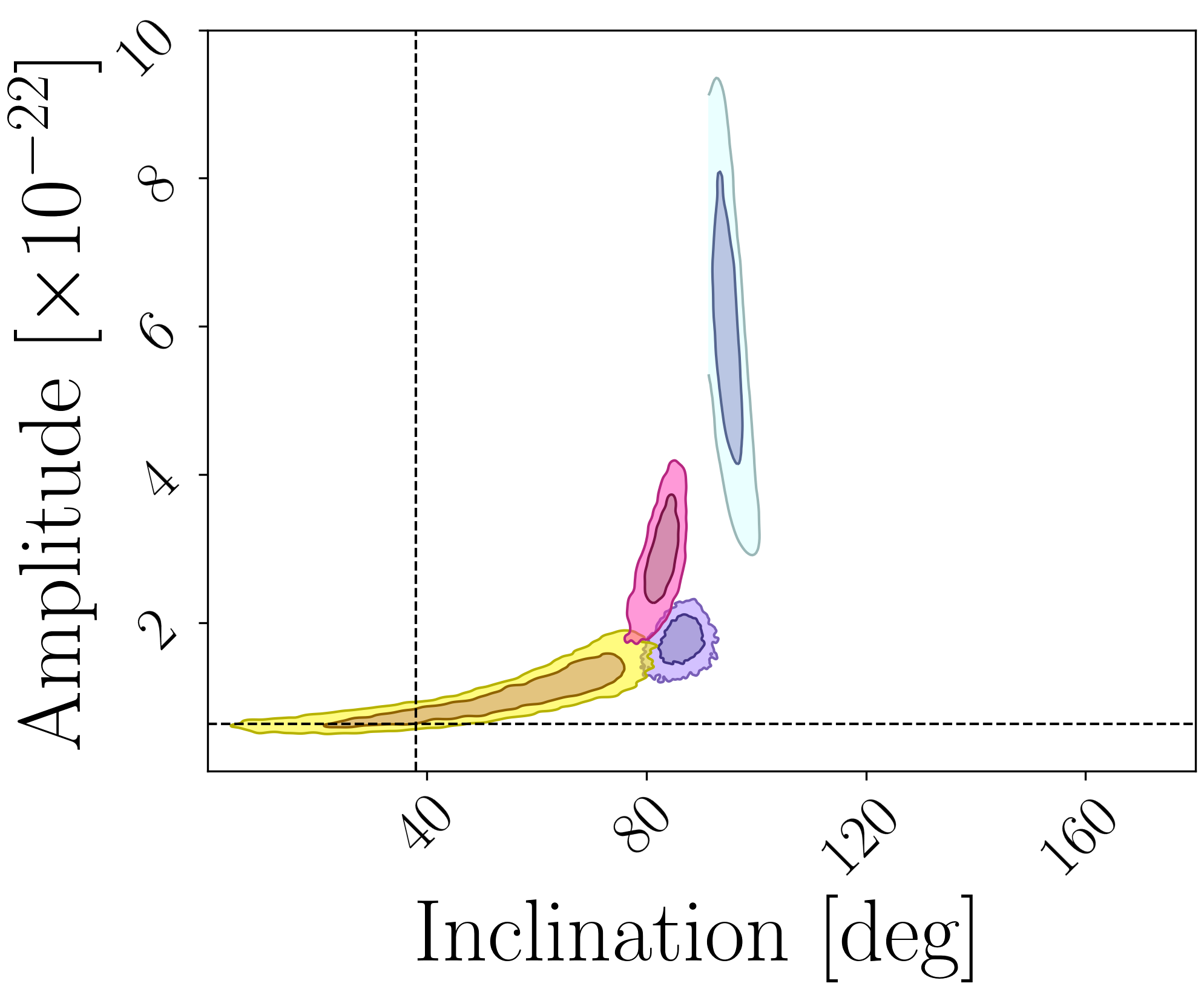}{0.15\textwidth}{{\hmcnc}}}
\gridline{\fig{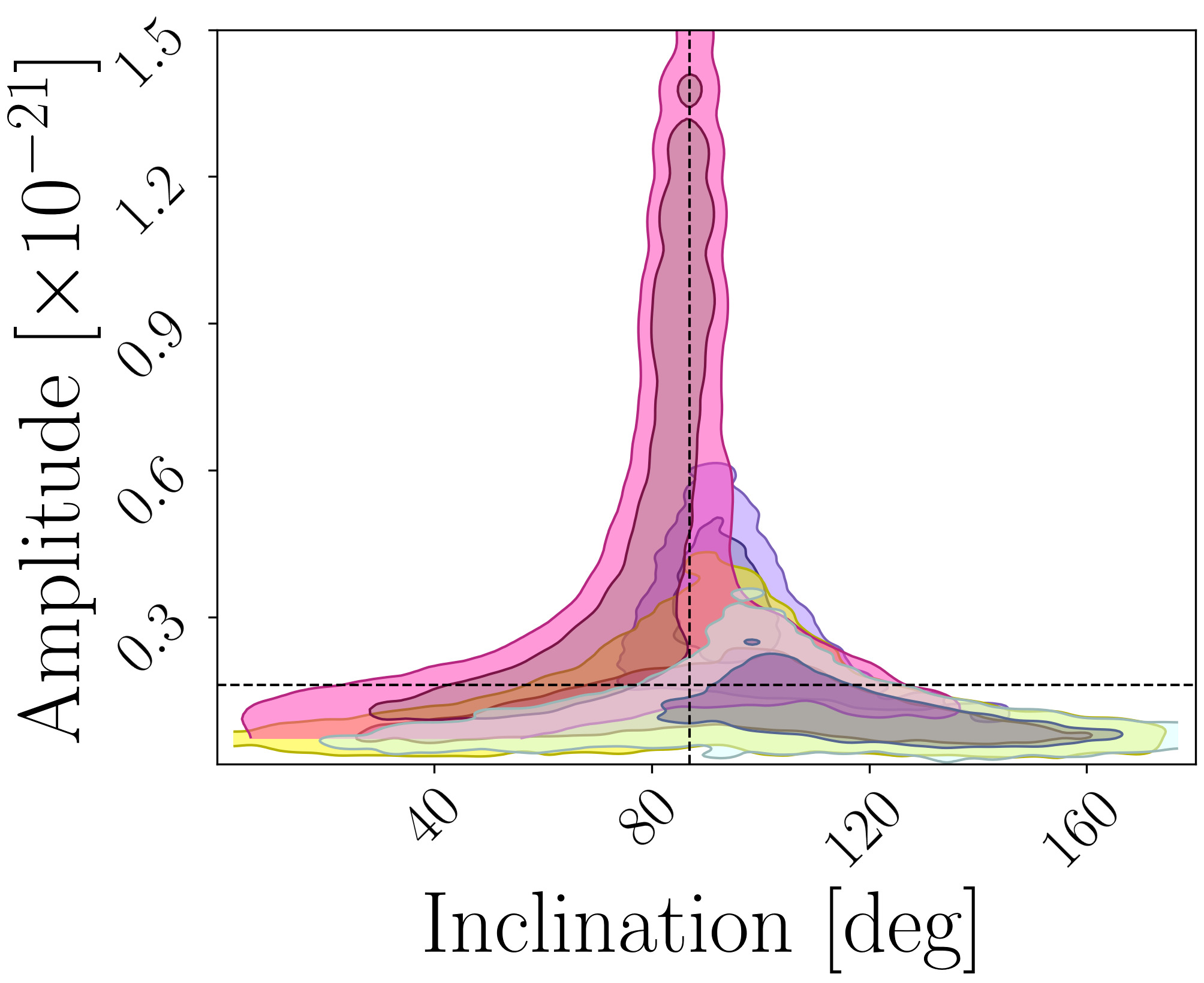}{0.15\textwidth}{{\sdss}}
          \fig{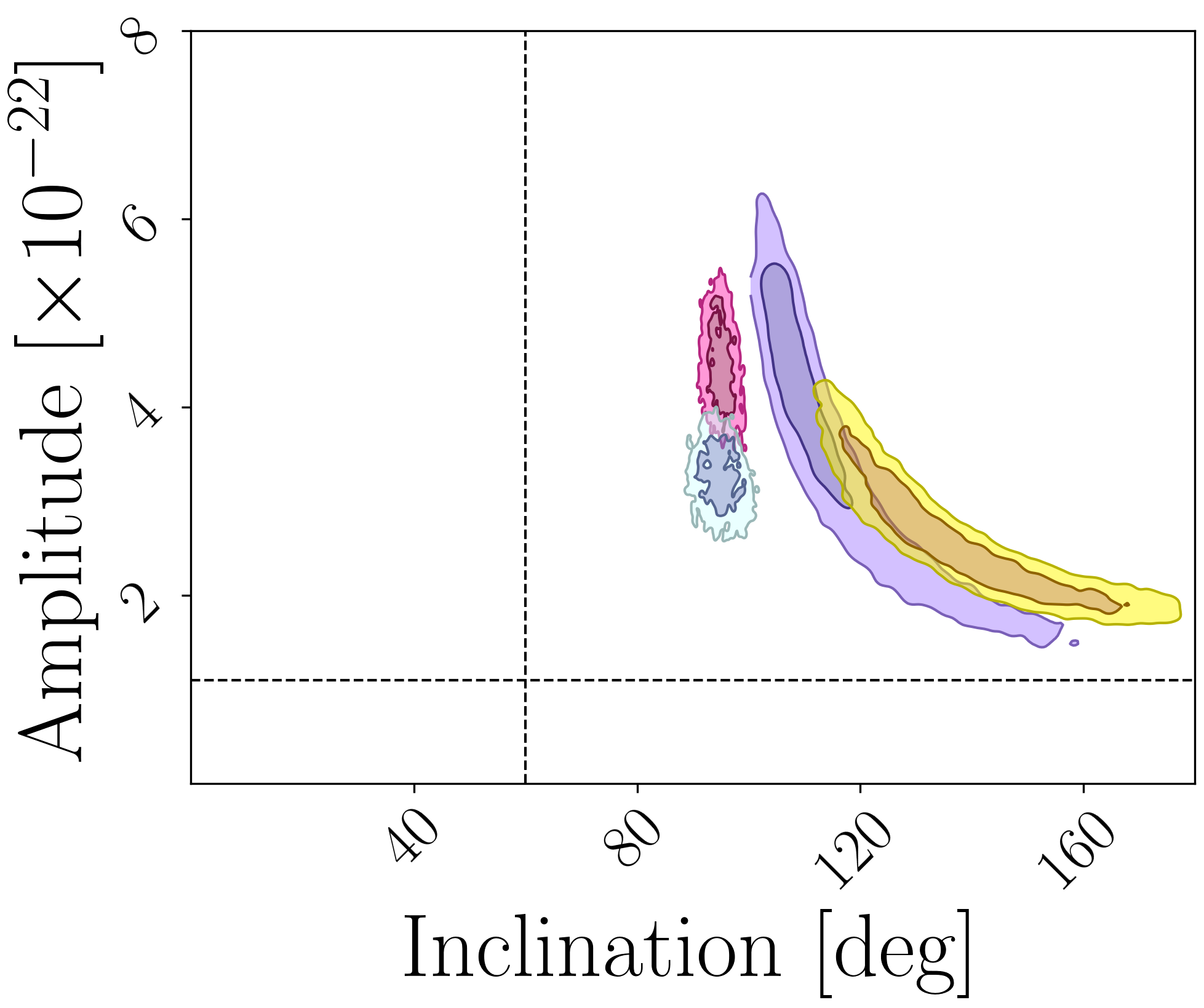}{0.15\textwidth}{{\vul}}
          \fig{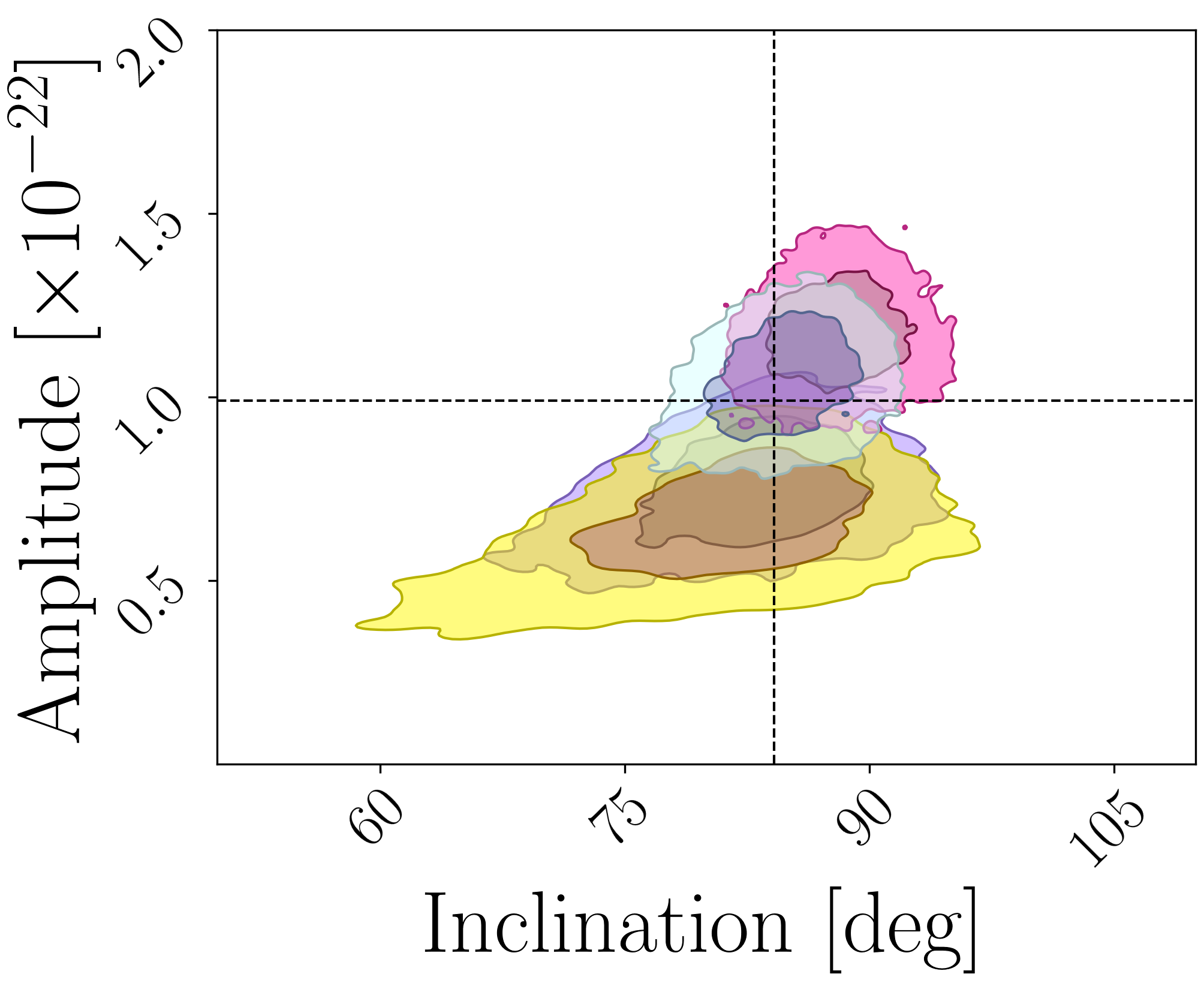}{0.15\textwidth}{{\ztf}}}
\caption{Same as Fig~\ref{fig:vgbonly_by_week_amplitude} but for {\vbgalaxy} model.}
\label{fig:by_week_amplitude}
\end{figure}

\begin{figure}
\gridline{\fig{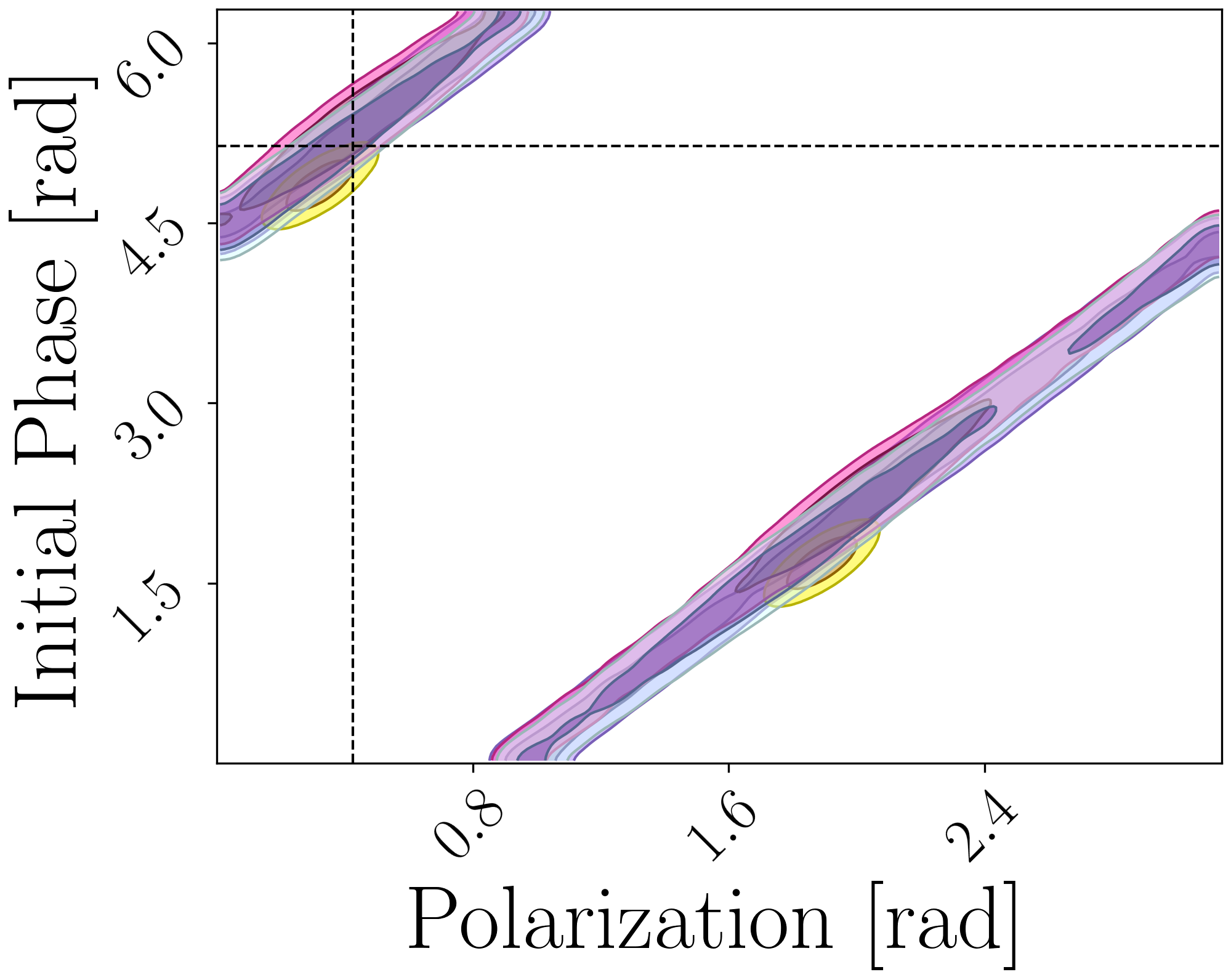}{0.15\textwidth}{{\amcvn}}
          \fig{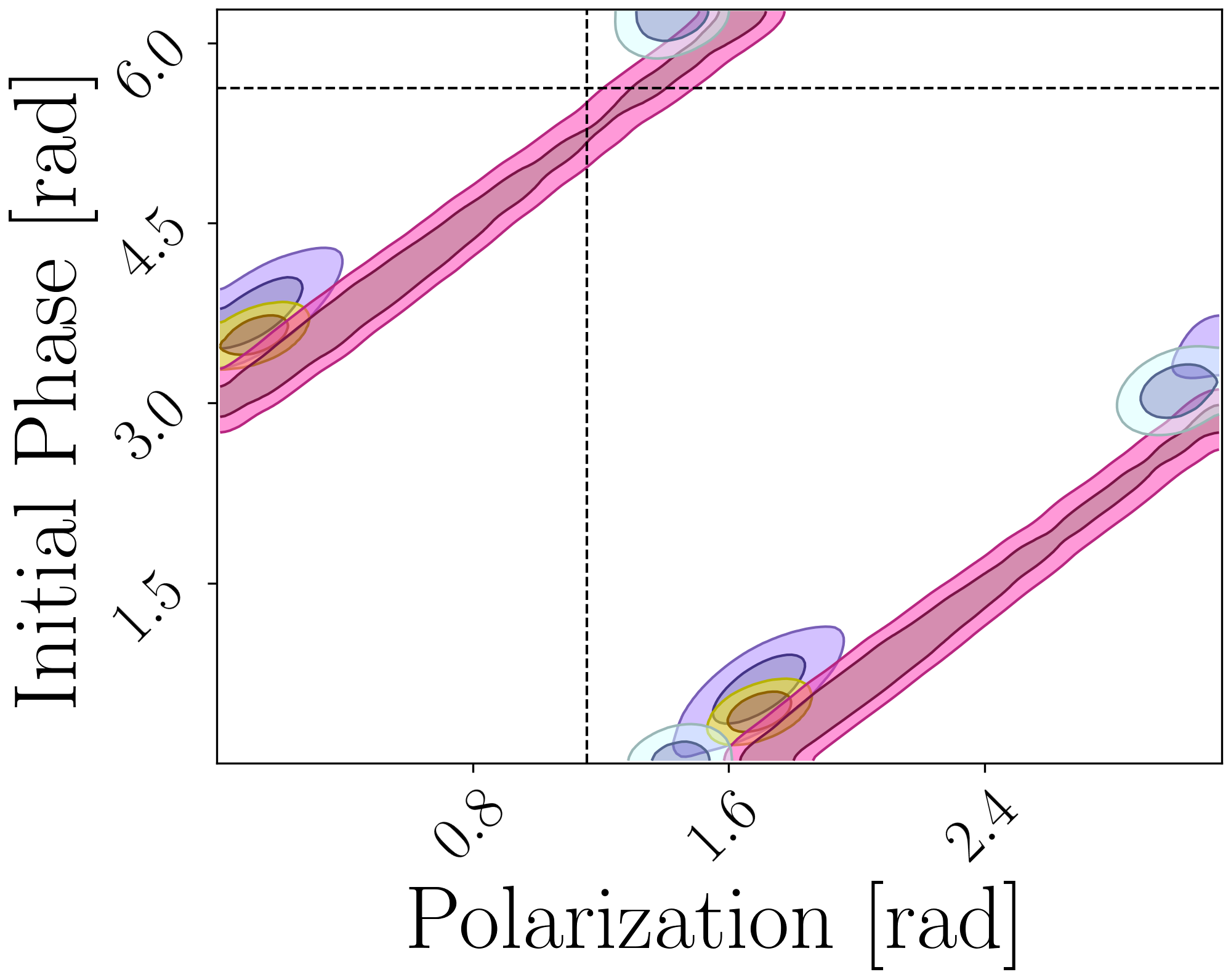}{0.15\textwidth}{{\escet}}
          \fig{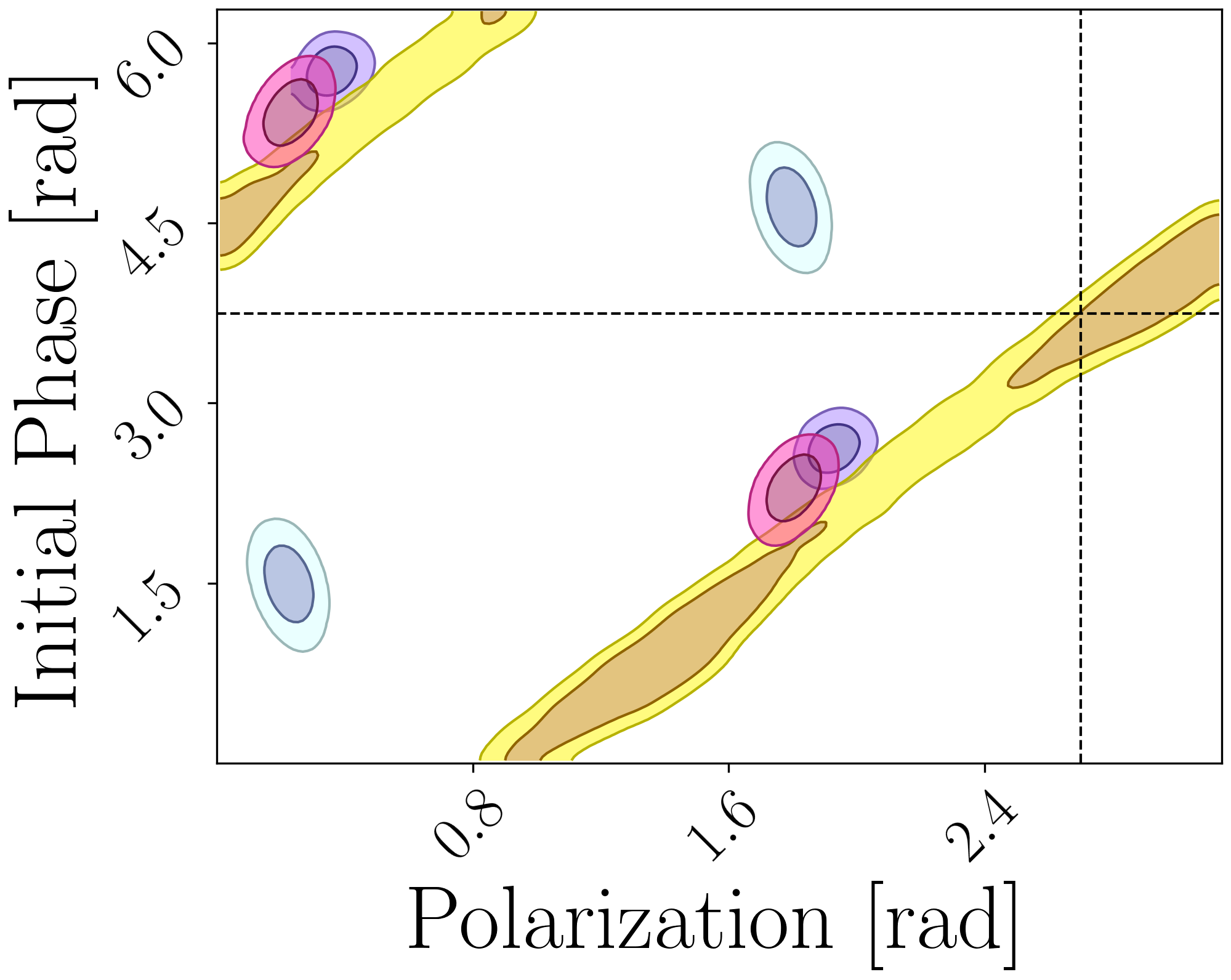}{0.15\textwidth}{{\hmcnc}}}
\gridline{\fig{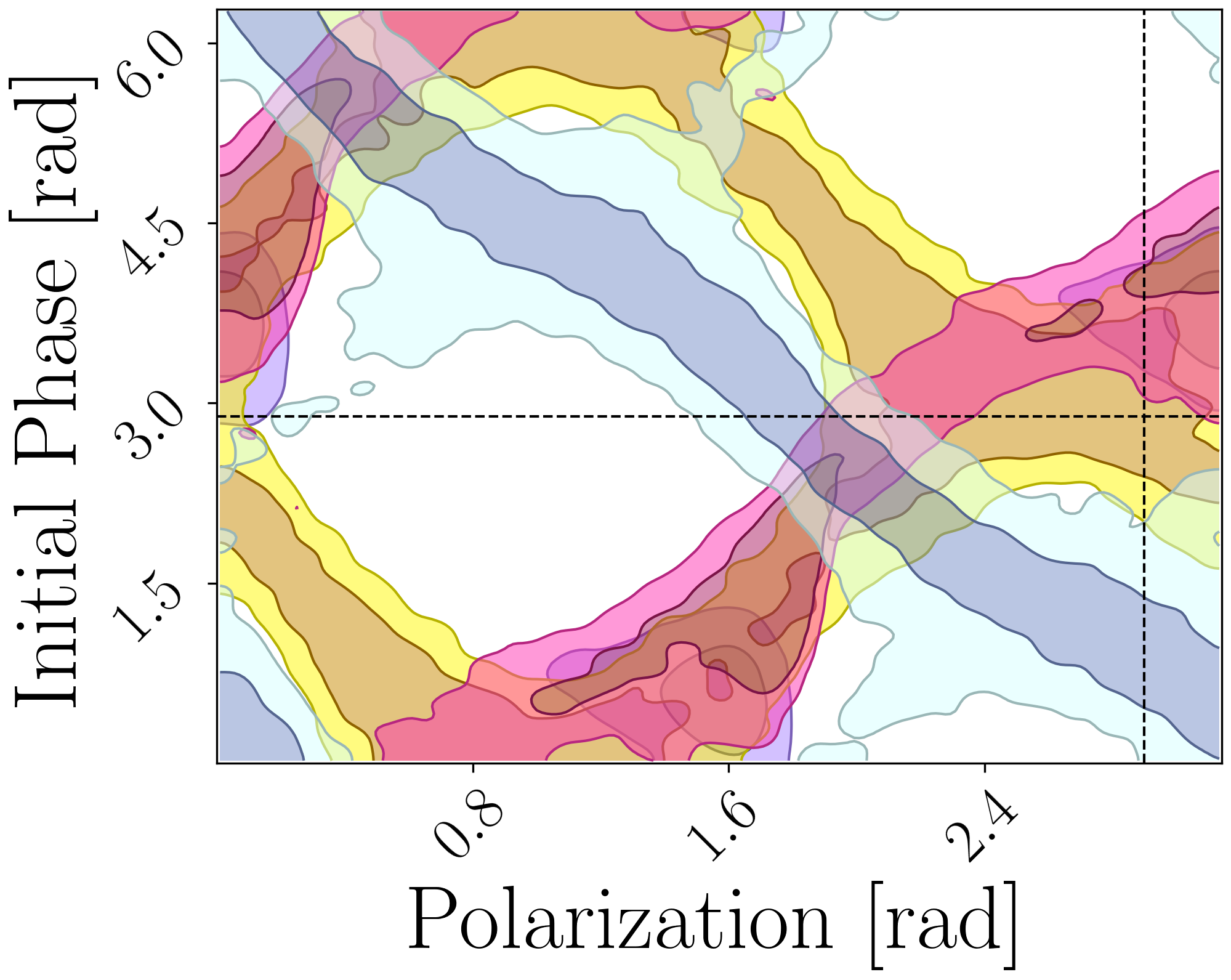}{0.15\textwidth}{{\sdss}}
          \fig{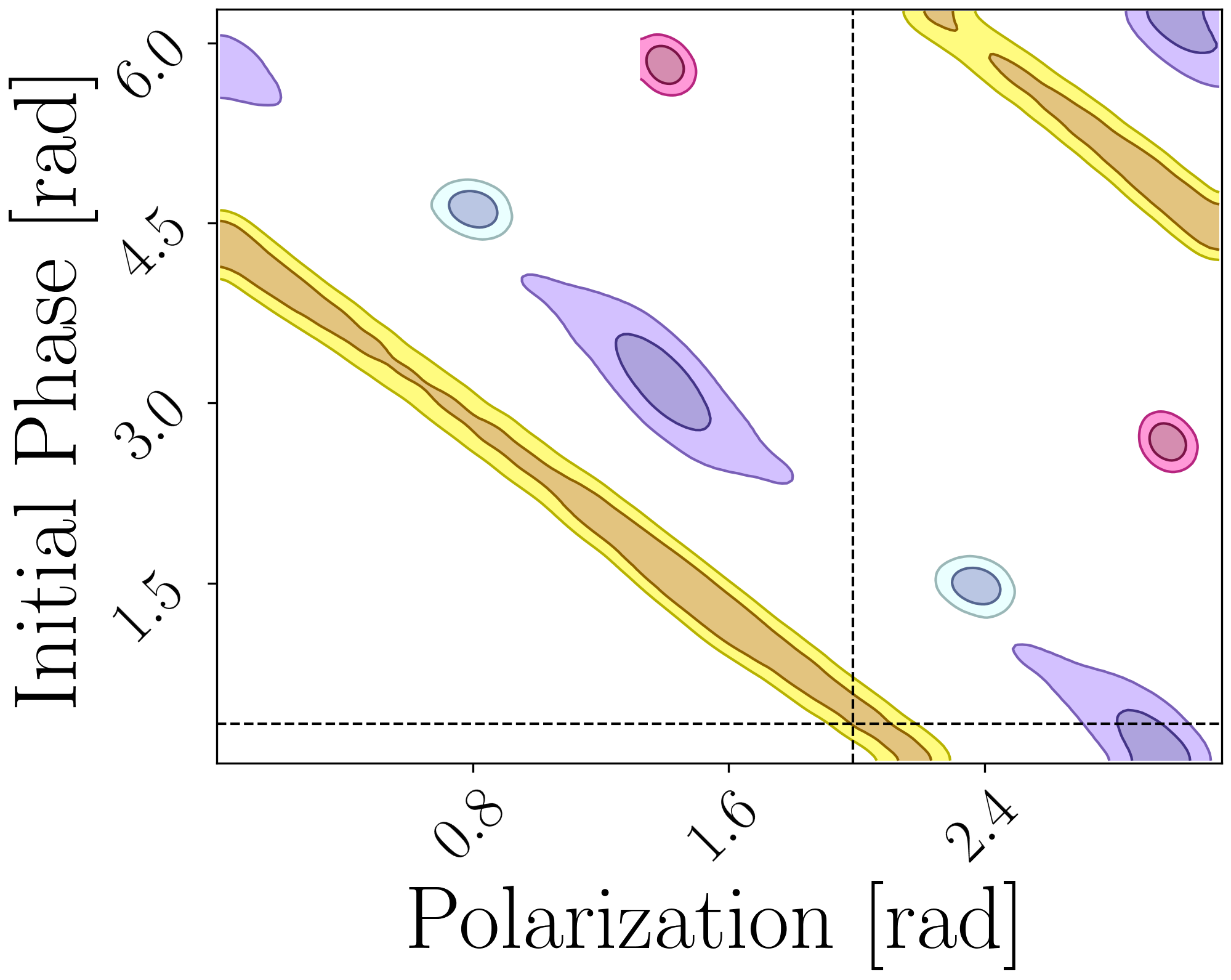}{0.15\textwidth}{{\vul}}
          \fig{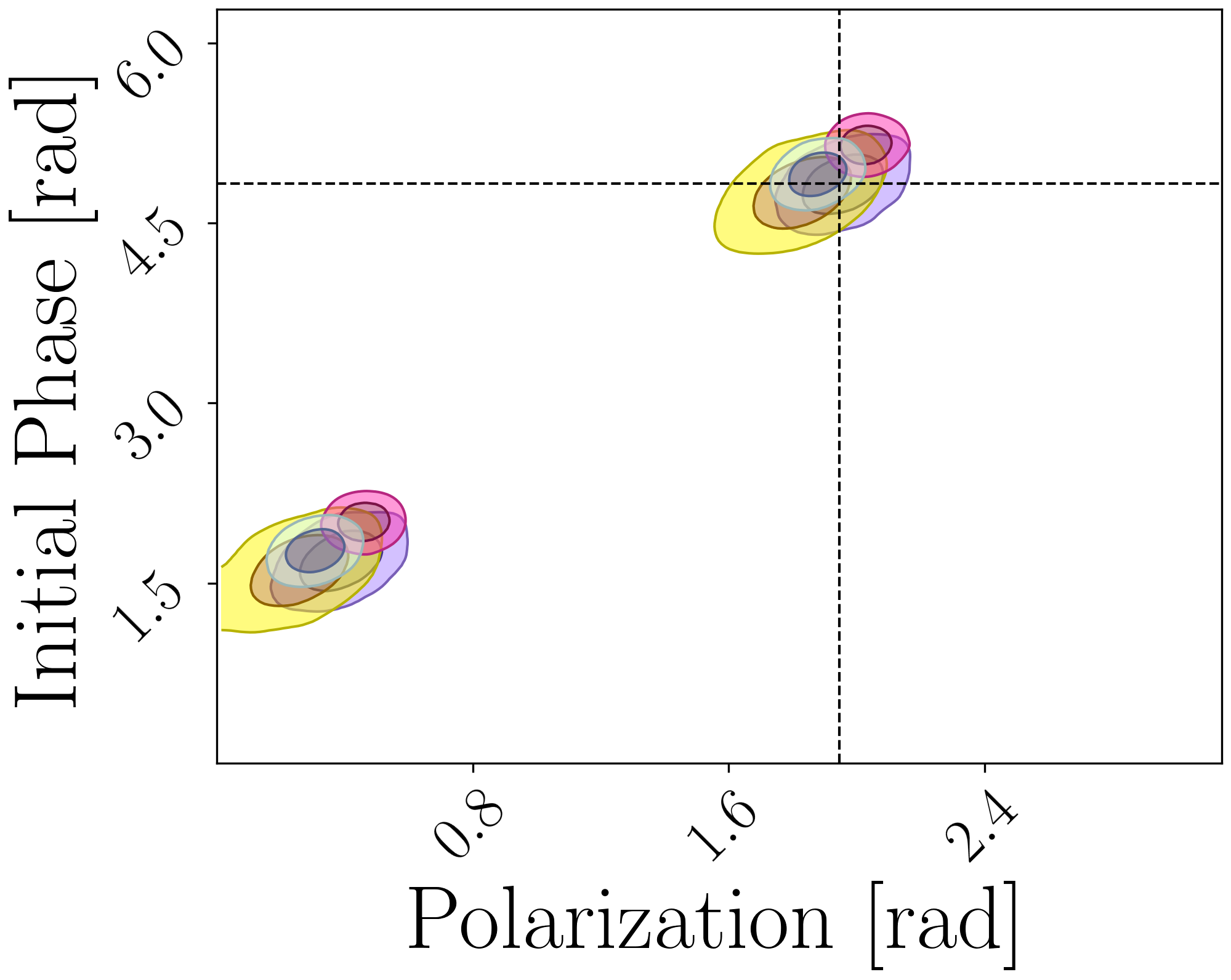}{0.15\textwidth}{{\ztf}}}
\caption{Same as Fig~\ref{fig:vgbonly_by_week_phase} but for {\vbgalaxy} model.}
\label{fig:by_week_phase}
\end{figure}

As before, performing the same analysis but in data which includes the rest of the galactic binary population ({\vbgalaxy}) results in egregious failures of the test due to contamination from other sources not accounted for by the data model. 
In this example all six binaries show biases larger than the statistical error in either the amplitude (Fig.~\ref{fig:by_week_amplitude}) or phase (Fig.~\ref{fig:by_week_phase}) parameters, even for the examples which manage to include the true values in at least some of the recoveries (e.g., {\amcvn}, {\sdss}, {\escet}, and {\ztf}).  
The biases are particularly catastrophic for {\hmcnc} and {\vul}.

\section{Discussion}

In this set of numerical experiments we have demonstrated that the highest S/N currently known LISA binaries cannot be robustly recovered from the LISA data without simultaneously considering the other galactic sources.
Any study of currently known binaries must be performed in context with a global LISA analysis considering other galactic sources and the resulting uncertainty in the effective noise level in the data, not to mention the other source types expected to be present in the data which were ignored in this work (e.g. massive black hole mergers).

Our analyses took a maximalist approach to using the available electromagnetically-determined source parameters, fixing the sky location and orbital period of the known binaries in our search pipeline to their known values.
The ``failures'' demonstrated in this work would be exacerbated by allowing the model for the known binaries to include those extra degrees of freedom, and to quantify whether the source was even detectable.  

This result calls into question the utility of the currently known binaries as verification sources for LISA.  
We conclude that the common nomenclature calling these sources ``verification binaries'' for LISA is, at present, misleading.  
Use of this term to refer to the known LISA galactic binaries should be accompanied by a quantitative prescription for how these particular sources can be exploited for instrument/data characterization.

On the other hand, population synthesis models predict an abundance of UCBs at sufficiently high gravitational wave frequency that the sources are well separated from the confusion-limited foreground, and one another, to make for unambiguous detection and monitoring without requiring an electromagnetic discovery before LISA operations. 
\begin{figure}
\plotone{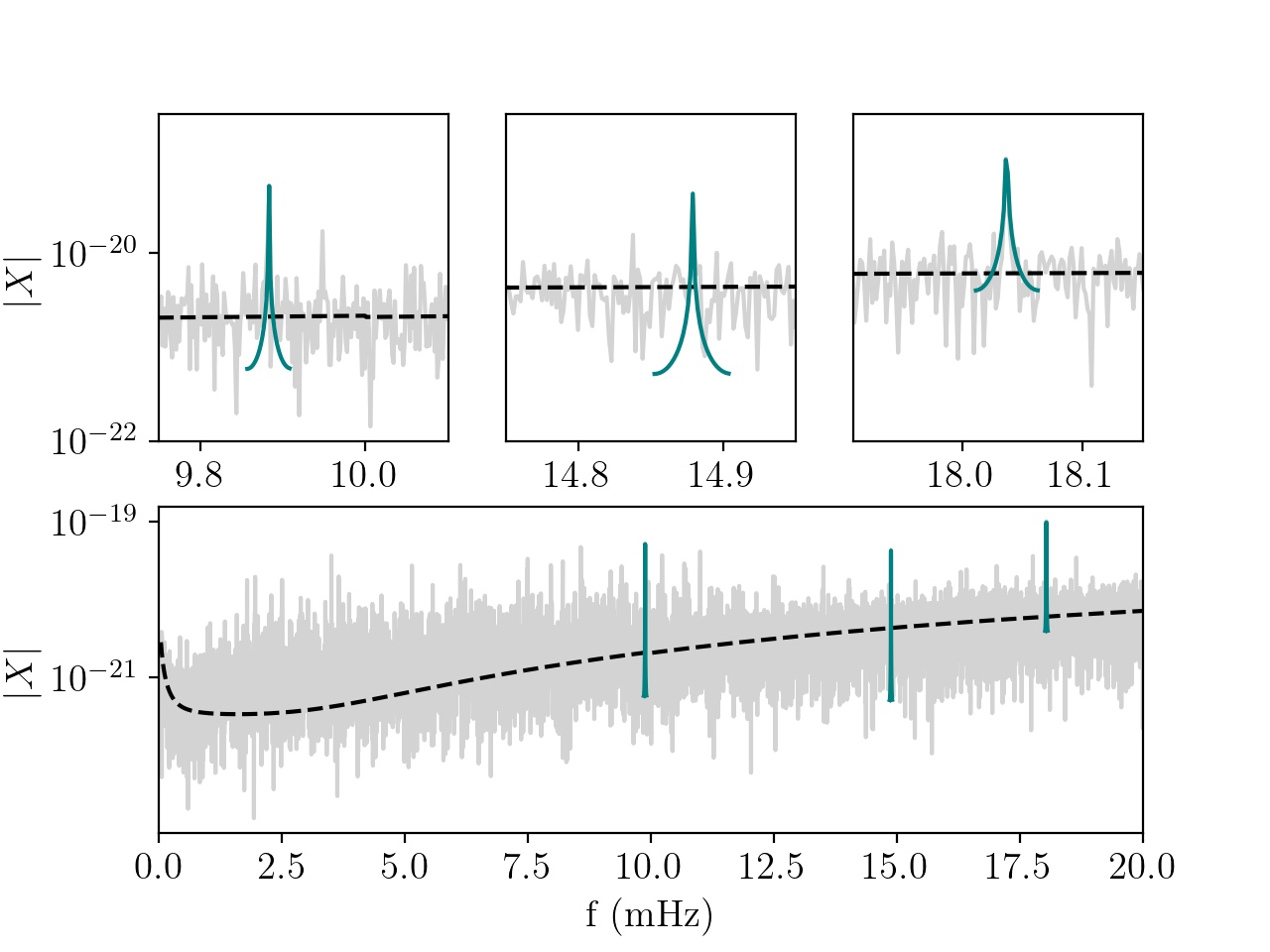}
\caption{Power spectrum of the TDI X channel for the {\sangria} dataset using one week of observing time.  The data are in gray while the black dashed line shows the inferred instrument noise level (not including a fit to the confusion noise) from a global analysis of the week-long data. The teal lines are three of the highest S/N binaries used for a jackknife test like Scenario 3 without the benefit of an electromagnetic counterpart.}
\label{fig:high_f}
\end{figure}

For example, the power spectrum of the {\sangria} dataset after one week of observing time is shown in Fig.~\ref{fig:high_f}.  The data are in gray while the black dashed line shows the inferred instrument noise level (not including a fit to the confusion noise) from a global analysis of the week-long data. Above ${\sim}8$ mHz the UCBs begin to appear as isolated spectral lines in the data.  Highlighted in teal are three of the highest S/N binaries in the low-source-density region.

\begin{figure}
\gridline{\fig{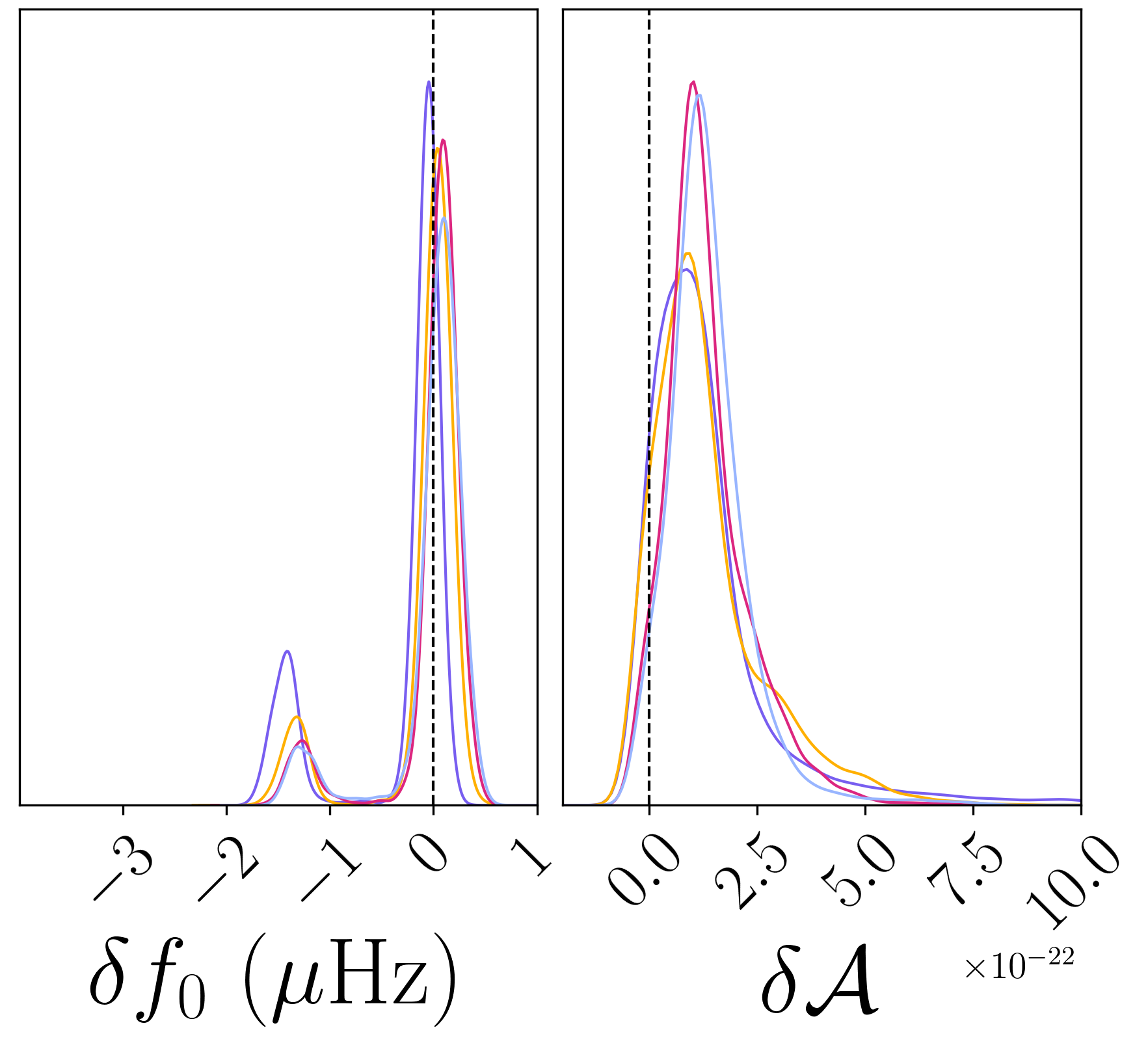}{0.15\textwidth}{{10 mHz}}
	\fig{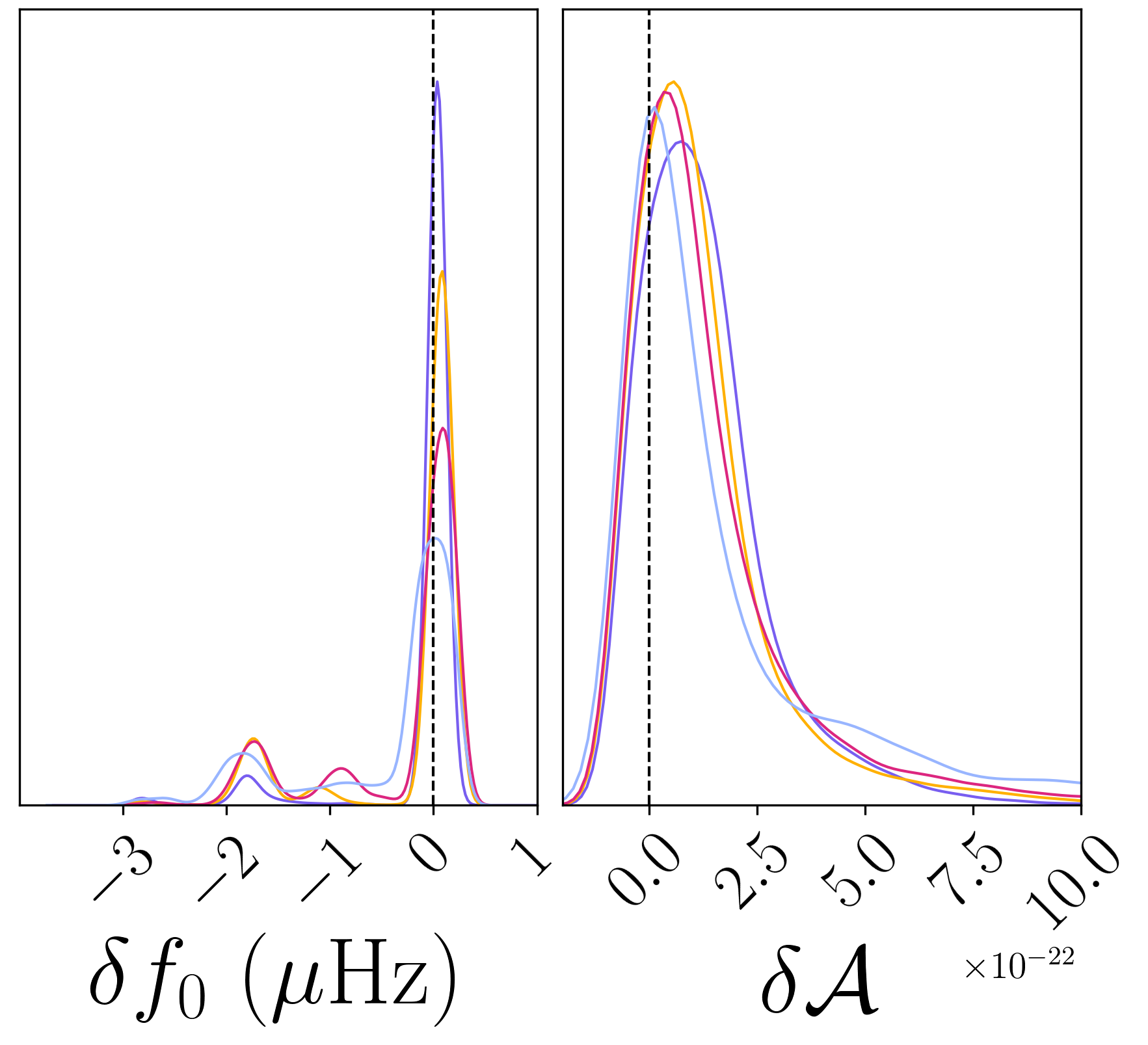}{0.15\textwidth}{{15 mHz}}
	\fig{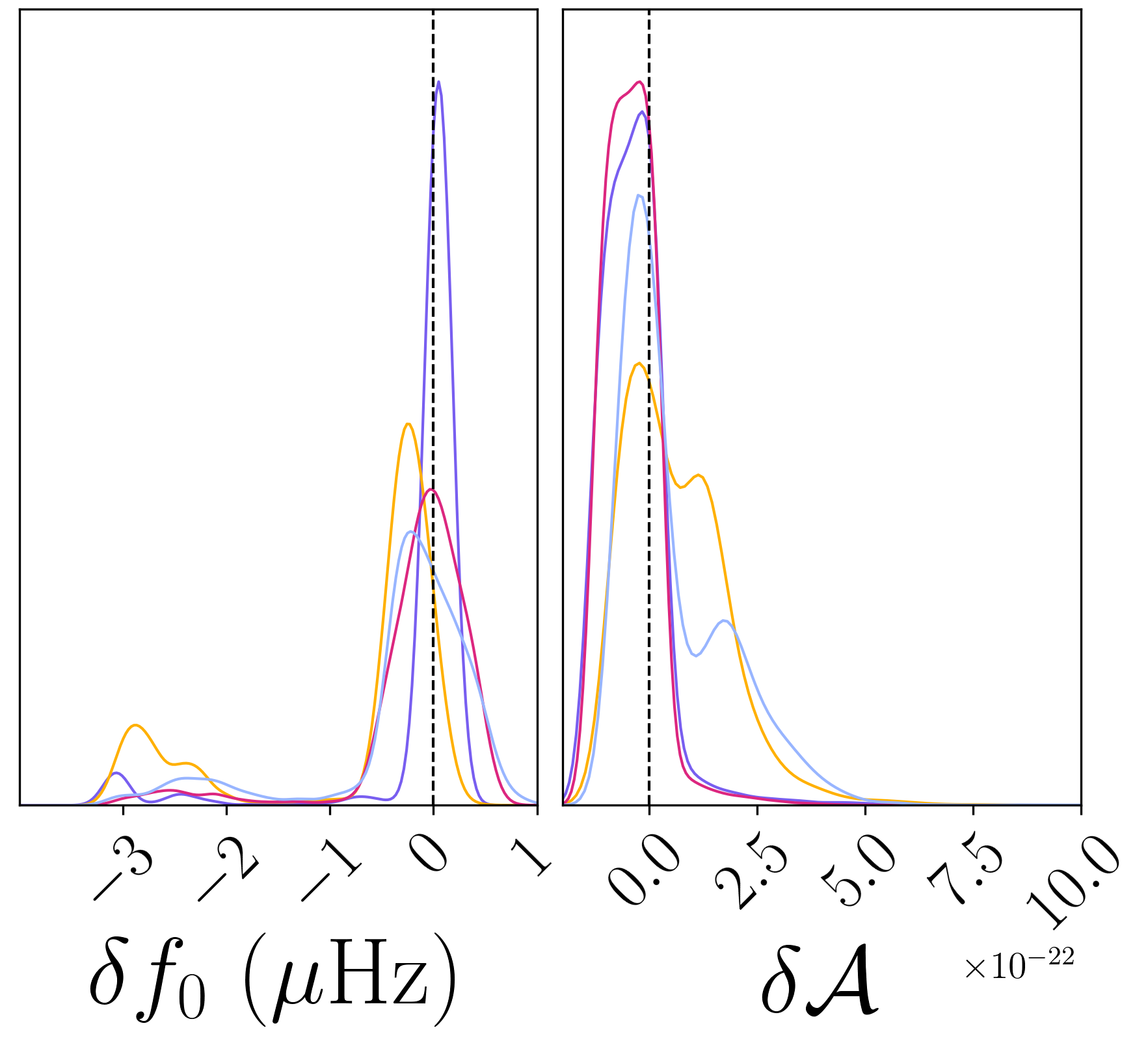}{0.15\textwidth}{{18 mHz}}}
\caption{Frequency and amplitude jack-knife tests for high frequency examples. For each example the frequency and amplitude are relative to the simulated value.  The sources are identified consistently from week-to-week with $\mathcal{O}(\mu{\rm Hz})$ precision in frequency and parts in $10^{22}$ in amplitude.}
\label{fig:high_f_freq_amp}
\end{figure}

As a proof of concept, we perform a similar jackknife test as Scenario 3 for the three highlighted sources. 
Considering that they will likely not be identified electromagnetically ahead of LISA operations, the analysis has to also search over the orbital period and sky location of the source.
The analyses use a restricted frequency range limited to a 100 $\mu{\rm Hz}$ band centered on the targeted source, taking a few minutes to run on a commercial-grade desktop computer.
Fig.~\ref{fig:high_f_freq_amp} shows the inferred gravitational wave frequency (left panel) and amplitude (right panel) for sources at 10, 15, and 18 mHz. 
The distributions are centered at the true value of the simulated source.  
The different week-long analyses are all self consistent at the ${\sim}\mu{\rm Hz}$ level in frequency, and $\mathcal{O}(10^{-22})$ in amplitude. Is this level of precision monitoring on a week-by-week cadence useful?
Just like with the analyses of the currently known binaries, it is unclear without a specifically articulated question about the data that we wish to answer using astrophysical sources as gravitational wave standards.

A more specific example was shown in \cite{PhysRevD.98.043008} which considered a toy problem of using the highest-frequency binaries as timing standards for LISA data by demanding phase coherence of the UCB signal across a gap in the LISA data to measure its duration.
While that scenario is unlikely to be necessary during LISA operations, it is perhaps a useful example exercise to begin considering how to exploit UCBs in our galaxy as gravitational wave standard sources without restricting the discussion to the already-known binaries which currently occupy regions of the overall parameter space prone to source contamination.

Whether or how galactic UCBs are to be utilized for understanding the LISA data is a question that is ripe for detailed study.
Regardless of the outcome, the scientific value of UCB observations is a linchpin of the LISA survey, be it from precision measurement of individual sources, multimessenger observations of systems detectable in both the GW and EM bands, or population-level inferences.
 
\section{Acknowledgements}

\emph{Software:} Results presented here used v2.0 of \hyperlink{github.com/tlittenberg/ldasoft}{\tt ldasoft}, a public C library which includes the noise, UCB, verification binary, and global fit samplers. Post-processing and visualization tools for the source catalogs are available in the python package \hyperlink{github.com/tlittenberg/lisacattools}{\tt lisacattools} which in turn depends on {\tt numpy}~\citep{harris2020array}, {\tt pandas}~\citep{reback2020pandas,mckinney-proc-scipy-2010}, {\tt matplotlib}~\citep{Hunter:2007}, {\tt astropy}~\citep{2022ApJ...935..167A}, {\tt seaborn}~\citep{Waskom2021}, and {\tt ChainConsumer}~\citep{Hinton2016}.

The authors thank J. Baker, C. Cutler, M. Katz, D. Mukherjee, R. Rosati, J. Slutsky, J.I. Thorpe, and M. Vallisneri for their useful discussions and suggestions during preliminary stages of this project, and the LISA Data Challenge group for providing and supporting the simulated data. The authors are supported by the NASA LISA Study Office.


\bibliography{references}{}
\bibliographystyle{aasjournal}



\end{document}